\documentclass[aps,preprint,pre,showpacs]{revtex4-1}

\usepackage{color}
\usepackage{subfigure}
\usepackage{amsmath}
\usepackage{amssymb}
\usepackage{graphicx}
\usepackage{dcolumn}
\usepackage{bm}

\usepackage{psfrag}

\newcommand\etal{{\it et~al.\/} }
\newcommand\ie{{i.e.}\ }

\renewcommand\Re{{\rm Re}}

\def\Ca{{C\!a}}
\def\r{{\ti r}}
\def\x{{\ti x}}
\def\s{{\ti s}}
\def\A{{\ti A}}
\def\V{{\ti V}}

\def\P{\partial}

\def\G{\Gamma}
\def\D{\delta}
\def\e{\eta}
\def\l{\lambda}
\def\kap{\kappa}

\def\eps{\epsilon}
\def\vareps{\varepsilon}
\def\ti{\tilde}

\def\q0{q_0}

\def\F{{F}}
\def\o{\omega}
\def\hN{h_{\rm N}}
\def\bhN{\bar h_{\rm N}}

\def\qN{q_{\rm N}}
\def\bqN{{\bar q}_{\rm N}}
\def\kap{\kappa}
\def\R{R\!e}

\def\Bo{B\!o}
\def\aN{\ti \alpha}
\def\Go{G\!o}
\def\Ga{G\!a}

\def\ux{u_x}

\def\a{\alpha}
\def\G{\Gamma}

\def\UI{{\bf U}_{\rm I}}
\def\UII{{\bf U}_{\rm II}}

\def\GG{G}

\def\I{I}

\def\J{J}
\def\K{K}
\def\L{L}
\def\M{M}

\def\BC{\begin{center}}
\def\EC{\end{center}}
\def\BM{\begin{displaymath}}
\def\EM{\end{displaymath}}
\def\BE{\begin{equation}}
\def\EE{\end{equation}}
\def\BEN{\begin{equation}}
\def\EEN{\end{equation}}
\def\BAN{\begin{eqnarray*}}
\def\EAN{\end{eqnarray*}}
\def\BA{\begin{eqnarray}}
\def\EA{\end{eqnarray}}
\def\NNM{\nonumber}

\def\BSE{\begin{subequations}}
\def\ESE{\end{subequations}}
\def\BPS{\begin{psfrags}}
\def\EPS{\end{psfrags}}

\def\CCW{\color{white}}

\begin{document}

\title[Wavy regimes of film flow down a fibre]
{Wavy regimes of film flow down a fibre}

\author{Christian Ruyer-Quil}
\affiliation{FAST -- UMR CNRS 7608, Universit\'e
Pierre et Marie Curie (UPMC),\\
 Campus universitaire,
91405 Orsay, France}

\author{Serafim Kalliadasis}
\affiliation{Department of Chemical Engineering, Imperial College London,\\
London SW7 2AZ, United Kingdom}

\date{\today}

\begin{abstract}
We consider axisymmetric traveling waves propagating on the
gravity-driven flow of a liquid down a vertical fibre. Our starting
point is the two-equation model for the flow derived in the study by
Ruyer-Quil \emph{et al.} [\emph{J. Fluid Mech.} {\bf 603}, 431
(2008)]. The speed, amplitude and shape of the traveling waves are
obtained for a wide range of parameters by using asymptotic analysis
and elements from dynamical systems theory. Four different regimes
are identified corresponding to the predominance of four different
physical effects: Advection by the flow, azimuthal curvature,
inertia and viscous dispersion. Construction of the traveling-wave
branches of solutions reveals complex transitions from one regime to
another. A phase diagram of the different regimes in the parameter
space is oferred.
 \end{abstract}


\maketitle

\section{Introduction}
\label{S-Intro}

A liquid film flowing down a vertical fibre is often encountered in
a wide variety of technological applications such as condensers,
emergency cooling of nuclear fuel rods and optical fibre coating. It
can also serve as a simple prototype for the study of wave
instabilities and transitions in open flow hydrodynamic and other
nonlinear systems. As a consequence, this problem has been an active
topic of both experimental and theoretical research, especially over
the past two decades.

Experimental studies have revealed a complex wave dynamics dominated
by axisymmetric and localised tear-drop-like structures which
continuously interact with each
other~\cite[]{Que90,Kli01,Cra06,Dup07,Dup09a}. These structures are
robust as they propagate over long distances without changing their
speed or shape significantly; they are also separated by portions of
nearly flat film, and they can be referred to as traveling waves.
When the portions of nearly flat film between these structures are
much longer than their characteristic length, the structures can be
referred to as solitary waves. The formation of solitary/traveling
waves results from the interplay between two different instability
mechanisms: (i) The classical instability of a liquid film flowing
down an inclined planar substrate prompted by inertia effects. This
mode was initially characterised experimentally and theoretically by
Kapitza and his son~\cite[]{Kap49}; (ii) The interfacial instability
of a liquid cylinder, considered first in experiments by Plateau and
was theoretically explained by Lord Rayleigh \cite[]{Pla73,Ray78}.
These two mechanisms are hereinafter referred to as the K and
RP~modes of instability, respectively.

The experimental study by Duprat \etal \cite{Dup09a}, in particular,
 provided
reports on the traveling waves' characteristics, namely shape, speed
and amplitude, for very viscous fluids. The parameter values in the
experiments were chosen in order to investigate the interplay of the
K and RP modes on the waves; thus the study by Duprat
\etal~\cite{Dup09a} completed the flow regime portrait obtained by
Kliakhandler \cite{Kli01} for very viscous fluids but in the
inertialess limit. Regular wavetrains were generated by means of a
periodic forcing at the inlet. Two flow regimes were identified. For
thin fibres and/or small flow rates, the RP mode is dominant and the
solitary waves resemble beads or sliding drops whose shape is
affected by gravity. In fact, when rescaled by the amplitude and the
tail length, the profiles are nearly superimposed. At the same time,
the flow field in a frame moving with the beads is characterised by
recirculation zones within the beads. When closed streamlines exist
in the moving frame, a fluid particle is trapped in both moving and
laboratory frames. Hence, the beads transport the trapped fluid mass
downstream. For these reasons, we refer to the flow regime observed
for thin fibres and/or small flow rates, as the {\sl drop-like
regime\/}.

For thick fibres and/or large flow rates, the K mode is dominant and
a steepening of the wave front is observed with an increase of the
wave amplitude. Mass transport is not observed in this regime except
for a few cases corresponding to the largest waves. We refer to this
regime as the {\sl wave-like regime\/}. We can conjecture that the
onset of recirculation zones in the wave-like regime is a signature
of the increased prevalence of the K mode. Eventually, we observe a
transition from the {\sl drag-gravity\/} regime, where inertia plays
a perturbative role, to the {\sl drag-inertia\/} regime, where
inertia effects become dominant. Similar regimes and a transition
between the two as inertia effects increase were first observed in
the planar case~\cite[]{Oos99,Ruy05b}.

Noteworthy is that the drop-like regime is similar to the one
observed by Qu\'er\'e~\cite{Que90} on a fibre or wire being pulled out form a
liquid bath and results from the same physical mechanisms. The thin
annular film that coats the wire can break up into drops. This drop
formation process occurs when the typical time of growth of the RP
instability is much smaller than the typical time of advection of a
structure by the flow, \ie for small fibre radii and/or small flow
rates.

At the theoretical front, a number of modelling approaches have been
proposed within the framework of the long-wave approximation of the
Navier-Stokes equations and associated wall and free-surface
boundary
conditions~\cite[]{Fre92,Tri92,Kal94,Kli01,Roy02,Cra06,Rob06,Nov09}:
The basic assumption of this approximation is that of slow spatial
and time modulations of the film thickness motivating the
introduction of a formal perturbation parameter, the `long-wave' or
`film parameter' $\eps$ measuring such modulations. Perturbation
expansions in terms of this parameter then lead to substantial
simplifications of the governing equations and boundary conditions.
Additional assumptions lead to further simplifications, i.e. small
film thickness $h$ in comparison to the fibre radius $R$ or
negligible inertia. The resulting models are either single evolution
equations for the film thickness $h$, e.g. the model by
Frenkel~\cite{Fre92} based on the long-wave approximation only, or
systems of two coupled evolution equations for the film thickness
$h$ and streamwise flow rate $q$ which combine the long-wave
approximation and other approaches, e.g. the model by~Trifonov
\cite{Tri92} based on the `integral-boundary-layer' approximation
and the more recent model by Novbari and Oron~\cite{Nov09} based on
an `energy integral' method.

It should be noted that all the above studies neglected the
second-order viscous terms originating from the stream-wise momentum
equation (streamwise viscous diffusion) and tangential stress
balance (second-order contributions to the tangential stress at the
free surface). The recent study by Ruyer-Quil \etal~\cite{Ruy08}
formulated a two-evolution equation model for $h$ and $q$ that took
into account the second-order viscous terms but also included
inertia and was not limited to small aspect ratios $h/R$. The model
was based on a combination of the long-wave approximation and a
weighted-residuals approach using appropriate polynomial test
functions for the velocity field -- a `weighted residuals integral
boundary layer' (WRIBL) model following the terminology introduced
by Oron~\cite{Oro08}. It should be noted that the WRIBL model is
consistent~\footnote{Following the terminology introduced
 in~\cite{Ruy98,Ruy00,Ruy02}, a film flow model is
 {\it consistent}
at $O(\eps^n)$ when all neglected terms are of higher order, or
equivalently no terms of $O(\eps^n)$ or smaller have been omitted,
and hence a gradient expansion of the model up to $O(\eps^n)$ agrees
exactly with the single evolution equation for $h$ at $O(\eps^n)$
obtained with just the long-wave approximation.} at $O(\eps)$ for
the inertial terms and at $O(\eps^2)$ for the remaining
contributions, whereas the models obtained by Trifonov~\cite{Tri92}
and Novbari and Oron~\cite{Nov09} are not consistent at $O(\eps)$.
Furthermore, the study by Duprat \etal~\cite{Dup09b} compared the
wavetrains generated experimentally by periodic forcing at the inlet
to the traveling-wave solutions of the WRIBL model showing
remarkable agreement in all cases, thus validating experimentally
the model. Experimental validation was done in the study by
Ruyer-Quil \emph{et al.} where the traveling-wave solutions of the
WRIBL model were favorably compared to the experiments by
Kliakhandler~\emph{et al.}~\cite{Kli01} while the spatio-temporal
dynamics of the film computed numerically with the WRIBL model was
shown in be in good agreement with the experiments by Duprat
\emph{et al.}~\cite{Dup07}

As was shown in \cite{Ruy08}, the second-order viscous terms have a
dispersive effect on the speed of the linear waves (they introduce a
wavenumber dependence on the speed) and hence we shall refer to this
effect as {\sl viscous dispersion\/}. Viscous dispersion influences
the shape of the capillary ripples in front of a solitary hump, more
specifically, the amplitude and frequency of the capillary ripples,
an effect which is amplified as the Reynolds number is increased.
Hence, viscous dispersion is in fact a linear effect, but
interestingly it can have some crucial consequences on the nonlinear
dynamics of the film and the wave-selection process in the
spatio-temporal evolution. After all, solitary pulses interact
through their tails which overlap, i.e. the capillary ripples and
their amplitude and frequency will affect the separation distance
between the pulses. For example, smaller-amplitude ripples will
allow for more overlap between the tails of neighboring pulses, thus
decreasing their separation distance. These points have been
analyzed in detail in the recent work by Pradas \etal~\cite{Pra11}
on coherent structures interaction on falling films on planar
substrates and the influence of viscous dispersion on interaction.
Their analysis was based on the weighted-residuals models obtained
in~\cite{Ruy98,Ruy00,Ruy02}.

The main aim of the present study is to characterize theoretically
the solitary/traveling waves propagating down the fibre within the
framework of the WRIBL model. A first stab to the investigation of
the traveling wave solutions of the WRIBL model was the recent study
by Ruyer-Quil~\emph{et al.}~\cite[Sect. 6]{Ruy08}. In this study
traveling wave solutions were constructed numerically and were
favorably compared to the experiments by Kliakhandler~\emph{et
al.}~\cite{Kli01}, as noted earlier. Here we undertake an asymptotic
analysis of the governing equations for solitary/traveling waves in
various limiting cases, i.e. in the limits of small/large values of
the pertinent parameters. We also obtain both numerically and
asymptotically static drops in the drop-like regime. Furthermore, by
using elements from dynamical systems theory, we provide a detailed
and systematic parametric study of their speed, shape and amplitude,
i.e. we construct bifurcation diagrams for their speed as a function
of pertinent parameters, as well as obtain ranges in the parameter
space for which the K~or RP~modes of instability, prompted by
inertia and azimuthal curvature, respectively, are dominant. We
scrutinize the four different regimes described earlier (drop-like,
wave-like, drag-gravity, drag-inertia), and provide detailed phase
diagrams and corresponding regime maps for very viscous and less
viscous fluids, thus providing a deeper understanding of the problem
as well as new insights and also completing the flow regime diagram
provided in~\cite{Dup09a} for viscous fluids.

Section~\ref{Formulation} introduces the pertinent
non-dimensionalization and the WRIBL model. Solitary wave solutions
are constructed in \S~\ref{S-dyn}. Asymptotic limits for small/large
values of the pertinent parameters are analyzed in \S~\ref{S-asymp}.
Traveling waves corresponding to the experimental conditions
considered in \cite{Kli01,Dup07} are next discussed in
\S~\ref{S-TW}. Section~\ref{S-Phase} presents a phase diagram of the
different regimes. Finally, a summary of our findings and concluding
remarks are offered in \S~\ref{S-concl}.

\section{Formulation} \label{Formulation}
\subsection{Natural set of parameters}
\label{S-Natural}
 Consider a film flowing down a vertical cylinder under the
action of gravity. The liquid has dynamic and kinematic viscosity,
$\mu$ and $\nu$, respectively, density $\rho$ and surface tension
$\sigma$. The flow is assumed to remain axisymmetric. $\bar r$,
$\bar x$, $\bar u$ and $\bar t$ denote the radial, pointing outwards
from the fibre centreline coordinate, the axial coordinate oriented
along gravity, the axial velocity distribution and time,
respectively (bars are used to distinguish dimensional from
dimensionless quantities unless the distinction is unnecessary).
From simple physical considerations and without prior knowledge of
the specific details of the system, the following scales can be
readily identified: The fibre radius $\bar R$, the Nusselt thickness
$\bhN$ of the uniformly coated film, the length and time scales,
$l_\nu=\nu^{2/3} g^{-1/3}$ and $t_\nu = \nu^{1/3} g^{-1/3}$, based
on gravity and viscosity (making explicit the balance between
gravity and viscous forces giving rise to the Nusselt flat-film
solution) and the capillary length $l_c= \sqrt{\sigma/(\rho g)}$.

A first set of pertinent dimensionless groups arises from these
scales. The aspect ratio
\BSE
\label{natural-params}
\BE
\aN \equiv \bhN/{\bar R}
\EE
 which assesses azimuthal curvature effects at the scale of the film, the
 Goucher number \cite[]{Que99},
\BE
\Go\equiv{\bar R}/l_c\,,
\EE
 that compares azimuthal and axial surface tension effects and the
 Kapitza number,
\BE
\G\equiv\sigma/(\rho \nu^{4/3} g^{1/3}) = (l_c/l_\nu)^2\,,
\EE
\ESE
 comparing surface tension and viscosity. Useful combinations of these
parameters are $\hN \equiv \bhN/l_\nu$ and $\bhN/l_c$. The former
compares the film thickness to the gravity-viscous length scale and,
indirectly, inertia and viscosity since the Nusselt base flow is the
result of the balance of gravity and viscosity. The latter,
$\bhN/l_c$ is related to the Bond number $\Bo=\rho g \bhN^2/\sigma =
(\bhN/l_c)^2$ comparing surface tension and gravity at the scale of
the film.

The advantage of the set of parameters $\aN$, $\Go$ and $\G$ is that
when the geometry and the working fluid are fixed, the Goucher and
the Kapitza numbers $\Go$ and $\G$ are constant and the only free
parameter is $\aN$. From an experimental point of view, $\aN$ can be
varied independently by varying the inlet flow rate. The Kapitza
number $\G$ is entirely defined by the fluid properties
independently of the flow characteristics, whereas the Goucher
number $\Go$ can be easily varied by replacing the fibre. Hence, the
parameters $\aN$, $\Go$ and $\G$ can therefore be viewed as
`natural' for the fibre problem, and we will systematically recast
our results in terms of these parameters $\aN$, $\Go$ and $\G$.

Table~\ref{table2} gives the physical properties of four different
fluids of increasing viscosities commonly used in experiments and
corresponding to a wide range of Kapitza numbers together with the
corresponding capillary lengths and viscous-gravity length scales.
For simplicity, our results will be presented for the Kapitza
numbers listed in table~\ref{table2}.
\begin{table*}
\begin{center}
\begin{tabular}{l|ccc|ccc}
&$\nu$ (${\rm mm}^2s^{-1}$) & $\rho$ (${\rm kg}{\rm m}^{-3}$) &
 $\sigma$ (${\rm mN} {\rm m}^{-1}$) & $l_c$ (mm) & $l_\nu$ (mm) & $\G$
\\
water                      &  1  & 998 & 72.5 &   2.7 & 0.047 &  3376 \\
Rhodorsil silicon oil v50  & 50  & 963 & 20.8 &   1.5 & 0.63  &  5.48 \\
castor oil                 & 440 & 961 & 31   &   1.8 & 2.7   &  0.45\\
Rhodorsil silicon oil v1000 & 1000 & 980 & 21.1 & 1.5 & 4.7   &  0.10 \\
\hline
\end{tabular}
\end{center}
\caption{Fluid properties, capillary length $l_c$, gravity-viscous
length $l_\nu$ and Kapitza number used in the present study. The
data for silicon oil v50 and castor oil have been taken from
\cite{Dup07,Kli01}. \label{table2} }
\end{table*}

\subsection{WRIBL model}
\label{S-WRIBL} We now adapt Shkadov's scaling~\cite[]{Shk77} and
introduce different length scales for the streamwise (axial) and
radial directions. The length scale in the radial direction is the
Nusselt thickness $\bhN$, whereas the length scale in the
streamwwise direction is chosen as $\kap \bhN$ defined by the
balance of the streamwise pressure gradient induced by capillarity,
$\propto \sigma \P_{xxx} h$, and gravity acceleration, $\rho g$,
which gives $\kap =[\sigma/(\rho g \bhN^2)]^{1/3} =
(l_c/\bhN)^{2/3}$. The time scale is defined with reference to the
Nusselt solution of uniform thickness (a result of the balance of
gravity and viscosity). The volumetric flow rate per unit length of
circumference, $\qN =R^{-1} \int_{R}^{R + \hN} {u} \,{r} d{r}$,  of
a film of constant thickness $\bhN$ is given by \BE \label{bqN} \bqN
\equiv \frac{g\bhN^3}{3\nu} \phi(\aN)\,, \EE where $\phi$ is a
geometric factor defined in (\ref{def-phi}) and measures the
deviation of the flow-rate-to-thickness relation from the cubic
dependency corresponding to the planar geometry ($\phi(0)=1$).
Similarly to the streamwise  length scale, the time scale is stretched by a
factor $\kap$ and thus defined as
$3\kap\bhN^2/\bqN=\nu\kap/[g \bhN \phi(\aN)]$.

Our basic equations for the analysis to follow are the WRIBL model
obtained in~\cite{Ruy08}, a set of two evolution equations for the
local film thickness $h(x,t)$ and the local flow rate $q(x,t) \equiv
R^{-1} \int_{R}^{R + h(x,t)} {u} \,{r} d{r}$. For the sake of
clarity and completeness we re-write the WRIBL model here,
 \BSE
\BA
\label{q} \P_t h &=& -\frac{1}{1 + \aN h} \P_x q\,,
\\\NNM
\D \P_t q &=& \D\left[- \F (\aN h)\,\frac{q}{h} \P_x q
+ \GG(\aN h) \,\frac{q^2}{h^2} \P_x h   \right]
+ \frac{\I(\aN h)}{\phi(\aN)} \left[
-\frac{3\phi(\aN)}{\phi(\aN h)} \frac{q}{h^2}
\right. \\&&  \left. \NNM
+  h \left\{1 + \P_{xxx} h +
 \frac{\beta}{(1 +\aN h)^2} \P_x h
-  \frac{1}{2} \P_x \left(\frac{\aN}{1 +\aN h} (\P_x h)^2  \right)
\right\}  \right]
\\ && + \e\left[ \J(\aN h) \frac{q}{h^2} (\P_x h)^2
- \K(\aN h) \frac{\P_x q \P_x h}{h} - \L(\aN h) \frac{q}{h} \P_{xx}
h + \M(\aN h) \P_{xx} q \right] \,, \label{mom-shk} \EA
\label{model2shk} \ESE in terms of the Shkadov scaling, where
$\phi$, $\F$, $\GG$, $\I$, $\J$, $\K$, $\L$, and $\M$ are functions
of the aspect ratio $\aN$ given in Appendix~\ref{S-Coeffs}.
Equation~(\ref{q}) is the (exact) dimensionless mass balance whereas
(\ref{mom-shk}) is the streamwise momentum equation averaged across
the film with a weighted-residuals approach.
It should be emphasized that the WRIBL model has been validated in
\cite{Ruy08,Dup09a} through direct comparisons to the experiments in
\cite{Kli01,Dup07,Dup09a} as noted in \S~\ref{S-Intro} [for both
very viscous and less viscous liquids (castor oil and silicon oil
V50, see table~\ref{table2}) and a wide range of the parameters
($0.15\le\Go\le1$ and $0.5\le\aN\le4.5$)].

Shkadov's scales introduce three new dimensionless groups besides
the aspect ratio $\aN=\hN/R$, a `reduced Reynolds number', \BSE
\label{Shk-param} \BE
 \D \equiv 3 \bqN /(\nu \kap) =  \left(\aN \Go \right)^{11/3} \phi(\aN)
 \G^{3/2},
\EE which compares inertia and the viscous drag at the scale $\kap
\bhN$ introduced by the balance of gravity and capillarity, a
streamwise `viscous dispersion parameter', \BE \e \equiv 1/\kap^2 =
({\bhN}/l_c)^{4/3} = \left(\aN \Go \right)^{4/3}, \EE and the
dimensionless group, \BE \beta\equiv\aN^2/\e = \aN^{2/3} \Go^{-4/3},
\EE \ESE which is a combination of $\aN$ and $\e$ and compares
azimuthal to axial surface tension effects. We have made explicit in
(\ref{Shk-param}) the relations of $\D$, $\e$ and $\beta$ to the
`natural' parameters $\aN$, $\Go$ and $\G$. Notice that all
second-order/viscous-dispersion terms are gathered in the last row
of (\ref{mom-shk}) and are multiplied by $\e$.  Finally, we recall
the usual definition of the Reynolds number based on the flow rate,
$\R=\bqN/\nu=\hN^3\phi(\aN)/3$ where again $\hN=\bhN/l_\nu$.

The advantage of Shkadov's scaling stems from (i) the direct
reference to the Nusselt uniform film flow that simplifies the
comparisons between solutions, with the Nusselt solution
corresponding to constant values of the film thickness and flow
rates $h=1$ and $q=1/3$; (ii) the association of a single parameter
to each physical effect affecting the balance of the different
forces: Inertia ($\D$), azimuthal surface tension ($\beta$), viscous
dispersion ($\e$) and geometry ($\aN$).

As noted in \S~\ref{S-Intro} the WRIBL model was derived with the
long-wave approximation (i.e. under the assumption of slow space and
time modulations, $\P_{x,t} \sim \eps$, where $\eps$ is the
long-wave/film parameter) and a weighted-residuals approach in which
the velocity field is expanded on an appropriately chosen set of
test functions. This expansion takes into account the (small)
deviations of the velocity field from the uniform-thickness
solution. As also noted in \S~\ref{S-Intro}, contrary to the model
obtained by Trifonov~\cite{Tri92} -- see also~\cite{Sis06} -- and to the
model by Novbari and Oron \cite{Nov09}, (\ref{model2shk}) is consistent up to
$O(\eps)$ for the inertia terms and up to $O(\eps^2)$ for the
remaining contributions (and accounts for viscous dispersion).
Indeed, both Trifonov's and Novbari and Oron's approaches assume a
self-similar velocity distribution and do not account for the
deviations of the velocity profile induced by the free-surface
deformations. For this reason, their two-equation formulations lack
consistency even at first order in the film parameter. Furthermore,
we note that the energy-integral approach employed by Novbari and Oron
 \cite{Nov09}
is not consistent
 with the kinetic energy balance of the flow.
Indeed, writing formally the axial momentum equation as ${\cal
M}(u)=0$, Novbari and Oron's averaged momentum equation reads
$\int_R^{R+h} {\cal M}(u)u  \,dr=0$ whereas the kinetic energy
balance of a section of the liquid corresponds to $\int_R^{R+h}
{\cal M}(u) u\, d(r^2)=0$. Truncating then ${\cal M}(u)$ at
$O(\eps^2)$ is equivalent to the Galerkin approach that can be used
to reduce the algebra leading to (\ref{mom-shk}) \cite[]{Ruy08}.
Noteworthy is that the two-equation model (\ref{model2shk}) is not
limited to small aspect ratios unlike, e.g. the model by Roberts and
Li~\cite[]{Rob06}.

To end this section let us point out one apparent drawback of the
Shkadov scaling, namely the divergence of the
 dimensionless parameter $\beta=\aN^2/\e$ as the viscous dispersion parameter
 goes to zero for $\aN=O(1)$.
The divergence of $\beta$ signals that the typical length of a wave
is not determined by the balance of gravitational and axial
capillary forces, as assumed in Shkadov's scaling, but rather by the
balance of axial and azimuthal capillary forces, in which case the
typical curvatures of the beads in the azimuthal and axial
directions must coincide. The beads have thus a drop-like rounded
shape, the long-wave approximation starts to be violated and the
viscous dispersion effects cannot be a priori discarded as in the
planar case. Considering drop-like beads (see \S-\ref{S-drop}) it
can thus be useful to adopt a scaling based on the radius ${\bar R}$
of the fibre, which gives the time scale $\nu/(g {\bar R})$.
This scaling introduces a Galilei number, $\Ga \equiv g {\bar
R}^3/\nu^2= \Go^3 \G^{3/2}$.

\subsection{Surface equations and saturation numbers}
\label{S-saturation}
Neglecting inertia and viscous dispersion, Craster and Matar \cite{Cra06}
 formulated a single
evolution equation for the film thickness $h$, the Craster and Matar
 (CM) equation,
 \BE
\label{KDB} \P_t \left( h + \frac{\aN}{2} h^2 \right) + \P_x \left[
\frac{h^3}{3} \frac{\phi(\aN h)}{\phi(\aN)} \left( 1 +
\frac{\beta}{(1+\aN h)^2} \P_x h + \P_{xxx} h \right) \right] =0\,.
\EE
 For sufficiently thin films, that is $\aN \to 0$, we obtain from
(\ref{KDB}) \BE \label{Serafim} \P_t  h + \P_x \left[ \frac{h^3}{3}
\left( 1 + \beta \P_x h + \P_{xxx} h \right) \right] =0\,, \EE which
is the equation derived initially by Frenkel \cite{Fre92} (see
also~\cite{Kal94}). Equations~(\ref{KDB}) and~(\ref{Serafim}) are
the simplest evolution equations balancing all relevant physical
effects, gravity, viscous drag, surface tension and the fibre
curvature. They are equations for the free surface $h(x,t)$ only and
following the terminology introduced by Ooshida~\cite{Oos99}, we
shall refer to them as {\sl surface equations\/}.

The CM equation offers a simple prototype for the flow, easily
amenable to mathematical and numerical scrutiny. It can be obtained
asymptotically from (\ref{model2shk}) in the limit $\D\to0$ and
$\e\to0$ when the nonlinear term $-  1/2 \P_x \left(\aN/(1 +\aN h)
(\P_x h)^2 \right)$ of the azimuthal curvature gradient is omitted.
Yet, as already stated, $\aN=O(1)$ implies that $\beta$ diverges to
infinity and therefore that second-order viscous dispersive terms
cannot be a priori discarded and that the long-wave assumption is
invalid. By contrast, the derivation of the Frenkel's equation
(\ref{Serafim}) can be obtained in the distinguished limit
$\aN\to0$, $\e\to 0$ and $\beta=\aN^2/\e=O(1)$. Nevertheless,
comparisons of the solutions to the CM equation to the experiments
show good agreement~\cite[]{Ruy08,Dup09b}.

Using the surface equation (\ref{Serafim}), Kalliadasis and
Chang~\cite{Kal94}, and Chang and Demekhin~\cite{Cha99} analysed the
mechanism of the formation of drops observed by
Qy\'er\'e~\cite{Que90} when a fibre or wire is drawn out of a liquid
bath. Qu\'er\'e observed suppression of the RP mode for sufficiently
small coating films. More specifically, below a certain critical
thickness the film deposited on the wire develops small-amplitude
interfacial waves with the flow preventing their growth into drops
(such drops would always be observed when the wire is horizontal, as
the suppression of the RP mode induced by the flow is absent in this
case). On the other hand, beyond the critical film thickness growth
of the interfacial waves into drops was observed. Kalliadasis and
Chang~\cite{Kal94} found that the amplitude of the
 solitary-wave solutions
to (\ref{Serafim}) diverges/`blows-up' for $\beta$ larger than a
critical value $\beta_c=1.413$, which closely corresponds to the
formation of drops in Qu\'er\'e's experiments. These authors also
performed an analytical construction of the solitary waves for
$\beta \rightarrow \beta_c$ using matched asymptotic expansions.
They showed that drop formation results from the inability of the
flow advection to saturate the instability growth through a
nonlinear saturation mechanism.

The advection and the growth of the instability can be compared
through the definition of the advection time $\tau_a$ of an
interfacial structure over its length and the definition of a
typical time of growth of the structure $\tau_g$ as the inverse of
the maximum growth rate.
Based on the Fenkel equation (\ref{Serafim}), the stability analysis
of the uniform film leads to the dispersion relation
\BE
\label{disp-Frenk}
\o = k+ \frac{{\rm i} k^2}{3} (\beta - k^2)
\EE
which governs infinitesimal perturbations around the base Nusselt flow
of wavenumber $k$ and angular frequency $\o$.
Based on (\ref{disp-Frenk}) the ratio  $\tau_a/\tau_g$ reads:
\BE \label{tauKDB} \tau_a/\tau_g =\o_i/\o_r|_{k=\sqrt{\beta/2}}
\propto \beta^{3/2} \EE
(see \cite{Ruy08} for details).
 Therefore, $\beta$ compares $\tau_a$ to
$\tau_g$. For, $\beta<\beta_c$, the instability growth is slower and thwarted
than the advection by the flow. The same mechanism is also in play in
the saturation of the drops though it is then strongly nonlinear.
For these reasons, we refer to $\beta$ as a {\sl saturation
number\/}, a term that was first introduced in~\cite{Dup09a}.

From the CM equation (\ref{KDB}), we get
 $\tau_a/\tau_g\propto (\beta^\star)^{3/2}$  where the composite parameter
 $\beta^\star$ is defined as \cite[]{Dup07,Ruy08}:
\BE \label{def-bstar} \beta^\star= \beta c_k^{-2/3}(1+\aN)^{-8/3}\,.
\EE $c_k$ is the speed at which infinitesimal deformations of the
free surface are kinematically transported by the flow, or {\sl
kinematic wave speed\/}: \BE \label{ck} c_k = \frac{1}{1 + \aN}
\left[  1+ \frac{\aN \phi'(\aN)}{3 \phi(\aN)} \right] = \frac{8
(b-1) \left(2 \log (b) b^2-b^2+1\right)}{3 \left(4 \log (b) b^4-3
b^4+4 b^2-1\right)}\,, \EE with $b = 1 + \aN$.

Since $c_k(\aN=0) = 1$, $\lim_{\aN\to0}\beta^\star = \beta$. In
this limit the Frenkel equation~(\ref{Serafim}) limited to very thin
films ($\aN \ll 1$) applies and $\beta^\star$ is a generalisation of
the saturation number $\beta$ to film flows with
thicknesses comparable to the fibre radius.

\section{Solitary waves and dynamical systems theory}
\label{S-dyn} Experimental studies \cite{Kli01, Cra06, Dup07,Dup09a}
 reported the formation of axisymmetric
traveling waves (TWs) propagating without deformations and at
constant speed over long distances. Solitary waves, \ie TWs
separated by constant-thickness layers of fluid, or substrates, much
longer than the characteristic length of the waves, were commonly
observed sufficiently far downstream. Theoretically, solitary waves
can be viewed as periodic TWs with an infinitely long wavelength.
The aim of this section is to investigate infinite-domain solitary
waves using elements from dynamical systems theory.

In a frame of reference moving with the speed $c$ of the waves, $\xi
= x -c t$, the flow is stationary and the set of partial
differential equations (\ref{model2shk}) can be converted into a set
of ordinary differential ones. The mass balance (\ref{q}) can be
integrated once, \BE \label{q0} q -c\left(h + \frac{\aN}{2} h^2
\right) \equiv q_0\,, \EE where $q_0$ is the rate at which the fluid
flows under the waves. The averaged momentum balance (\ref{mom-shk})
next reads, \BA \NNM
 \D\left[c\, q' - \F (\aN h)\,\frac{q}{h} q'
+ \GG(\aN h) \,\frac{q^2}{h^2} h'   \right] + \frac{\I(\aN
h)}{\phi(\aN)} \left[ -\frac{3\phi(\aN)}{\phi(\aN h)} \frac{q}{h^2}
\right. &&\\ \left. \NNM +  h \left\{1 +  h''' + \frac{\beta}{(1
+\aN h)^2} h' + \frac{\aN}{1 +\aN h} h' \left[ h'' -
\frac{1}{2}\frac{\aN}{1 +\aN h} (h')^2 \right] \right\} \right] &&\\
+ \e\left[ \J(\aN h) \frac{q}{h^2} (h')^2 - \K(\aN h) \frac{q'
h'}{h} - \L(\aN h) \frac{q}{h}  h'' + \M(\aN h) q'' \right] &=&0 \,,
\label{MB_ssdyn} \EA where the primes denote differentiation with
respect to the moving coordinate $\xi$.
 Using (\ref{q0}), (\ref{MB_ssdyn}) can be recast into
$h''' = f(h,h',h'')\,,$
 where $f$ is a function of the thickness $h$, its first and second
derivatives and the parameters $\D$, $\aN$, $\e$ and $c$. We thus
end up with a dynamical system of dimension three, \BE \label{ssdyn}
\frac{d}{d\xi} {\bf U} = (U_2, U_3, f(U_1,U_2,U_3))^t, \EE where
${\bf U} = (U_1,U_2, U_3)^t \equiv (h,h',h'')^t$.

\begin{figure}
\begin{center}
\BPS
\psfrag{h}{$h$}
\psfrag{hI}{$h_{\rm I}$}
\psfrag{hII}{$h_{\rm II}$}
\psfrag{c/ck}{$c/c_k$}
\includegraphics[width=0.5\textwidth]{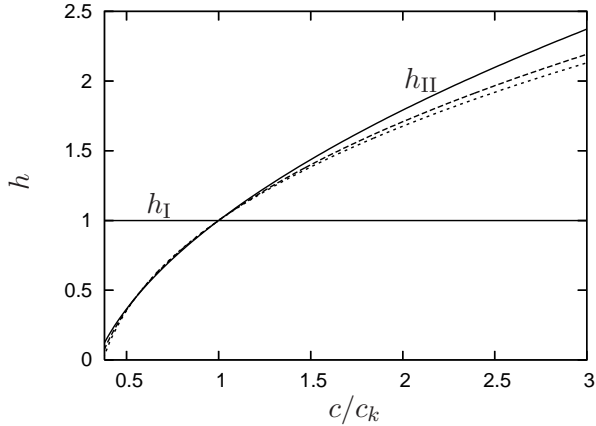}
\EPS
\end{center}
\caption{Film thicknesses $h_{\rm I}=1$ and  $h_{\rm II}$
corresponding to the location of the fixed points as function of the
ratio of the normalised wave speed $c$ to the kinematic wave speed
$c_k$ defined in (\ref{ck}). Solid, dashed and dotted lines refer to
$\aN=0$ (planar case), $\aN=0.5$ and $\aN=1$, respectively.}
\label{fig-fp}
\end{figure}

Solitary waves correspond to homoclinic orbits in the phase space
connecting a fixed point to itself. Here we restrict our attention
to single-loop homoclinic orbits corresponding to single-hump
solitary waves in real space. The fixed points of the dynamical
system (\ref{ssdyn}) satisfy $h'=h''=0$ and, \BE \label{fp}
 \frac{h^3}{3} \frac{\phi(\aN h)}{\phi(\aN)}
-c\left(h + \frac{\aN}{2} h^2 \right) = q_0\,. \EE Requiring that
$h=1$ is a solution to (\ref{fp}) gives: \BE \label{q0_h1} q_0 =
\frac{1}{3} - c \left(1+ \frac{\aN}{2} \right)\,. \EE In addition to
the solution  $h=1$, there is one more real positive solution to
(\ref{fp}) with (\ref{q0_h1}) and the dynamical system (\ref{ssdyn})
admits two fixed points $\UI = (1,0,0)^t$ and $\UII = (h_{\rm
II},0,0)^t$ whose positions are displayed in Fig.~\ref{fig-fp} as
functions of the wave speed $c$ and aspect ratio $\aN$. The two
fixed points coincide when the wave speed $c$ is equal to the speed
of linear waves of infinitesimal amplitude and infinite length that
are neutrally stable. This situation corresponds to the definition
of the linear kinematic waves in the zero-wavenumber limit, whose
speed is given in (\ref{ck}). In the limit $\aN =0$, we recover the
expression $h_{\rm II} =  -1/2 + \sqrt{3(c-1/4)}$ corresponding to a
film flowing down an inclined plane \cite[]{Pum83,Ruy05b}. We note
that extending this expression to the axisymmetric geometry by
replacing $c$ with $c/c_k$, gives a reasonable approximation $h_{\rm
II} \approx  -1/2 + \sqrt{3[(c/c_k)-1/4]}$ for the position of the
second fixed point even for $\aN = O(1)$ (see Fig.~\ref{fig-fp}).
Finally, we note that it is sufficient to consider homoclinic orbits
around only one of the two fixed points because of the presence of a
continuous family of Nusselt flat-film solutions parameterized by
the reduced Reynolds number $\D$ (or $\hN = \bhN/l_\nu$) when $\Go$
and $\G$ are held constant. Indeed, homoclinic orbits connecting
$\UII$ correspond to phase-space trajectories connecting $\UI$
through the transformation $\hN \to \hN h_{\rm II}$.

The shape of the tail and front of a solitary wave can be determined
by considering how the corresponding homoclinic orbit in the phase
space approaches and leaves the fixed point it connects to. Let us
consider the linear stability of the fixed point $\UI$. The
dispersion relation governing infinitesimal perturbations $\sim
\exp(\l \xi)$ is \BE \label{dispU1} \l^3 + \l^2 \e D_\e + \l \D D_\D
- 3 (1+\aN) \left( c -c_k \right) =0\,, \EE where \BE D_\e =
\frac{\phi}{\I} \left[- \frac{\L}{3} + c (1+\aN) \M \right] \quad
\hbox{and}\quad D_\D = \left\{ \frac{\phi}{\I} \left[(1 + \aN)
\left( c^2 - \frac{\F}{3} c \right) + \frac{\GG}{9} \right]
+\frac{\beta}{\D(1+\aN)^2} \right\}\,. \EE Equation~(\ref{dispU1})
can be reduced to the canonical form $P(y)=y^3 + py + q=0$ by the
change of variable $\l=y - \e D_\e/3$, where \BE p = \D D_\D -
\frac{\e^2 D_\e^2}{3} \quad \hbox{and}\quad q = -3(1+\aN)(c-c_k) -
\frac{\D\e}{3} D_\D D_\e + \frac{2\e^3}{27} D_\e^3\,. \EE Using the
Cardan formulae, (\ref{dispU1}) admits a real eigenvalue and a
complex conjugate pair when the discriminant $\Delta=4p^3+27q^2$ is
$>0$. When $\Delta <0$,  (\ref{dispU1}) admits three real
eigenvalues. Therefore $\UI$ changes from a saddle to a saddle-focus
at $\Delta=0$. When $\UI$ is a saddle, the homoclinic orbit departs
and returns to the fixed point monotonically along the two
eigenspaces corresponding to the eigenvalues of smallest absolute
value, and the corresponding tail and front of the solitary wave are
monotonic. Conversely, when $\UI$ is a saddle-focus, the homoclinic
orbit leaves monotonically along the eigenspace spanned by the
eigenvector corresponding to the real eigenvalue and returns to the
fixed point by spiralling on the eigenspace spanned by the
eigenvectors corresponding to the complex pair. This spiral
corresponds to capillary ripples at the front of the solitary wave
while the tail remains monotonic.

\begin{figure}
\begin{center}
\BPS
\psfrag{D}{$\D$}
\psfrag{c/ck}{$c/c_k$}
\psfrag{h}{$h_{\rm max}$}
\subfigure[]{\label{sol-b}%
\includegraphics[width=0.48\textwidth]{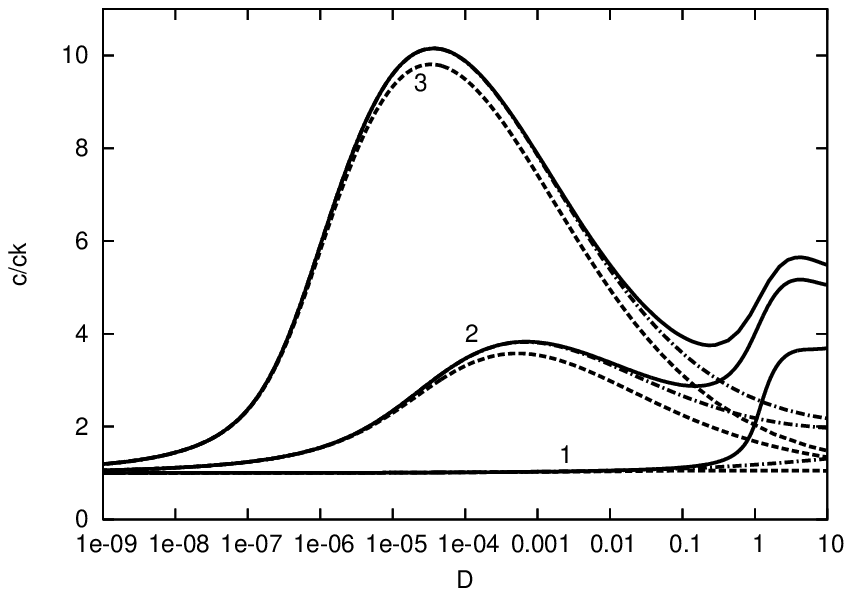}}\hfill%
\subfigure[]{\label{sol-d}%
\includegraphics[width=0.48\textwidth]{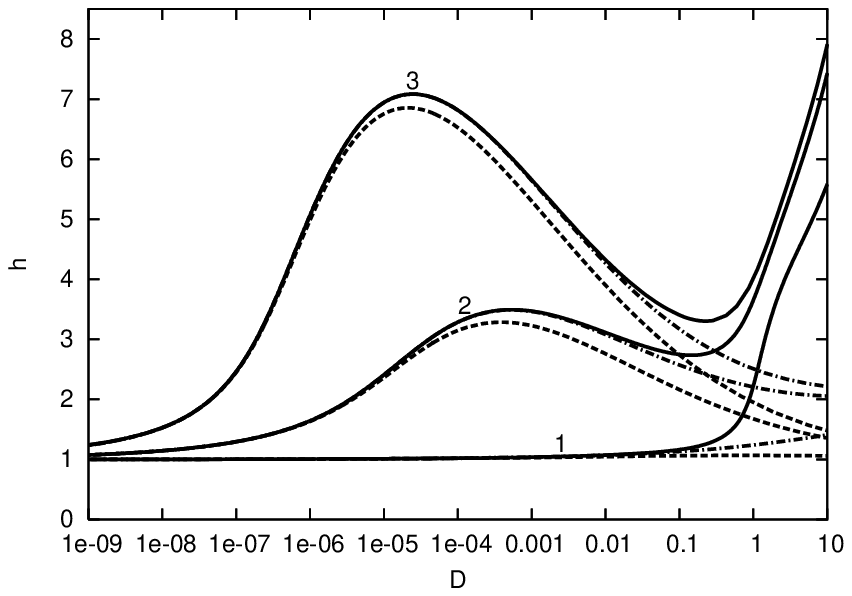}}
\EPS
\end{center}
\caption{Speed $c$ (left) and maximum height $h_{\rm max}$ (right)
of the solitary waves as function of the reduced Reynolds number
$\D$ for different fibre radii: $R=1.5$~mm (Curves~1), $0.35$~mm (2)
and $0.25$~mm (3). Solutions to (\ref{model2shk}) (solid lines) are
compared to the solutions of the CM~equation (\ref{KDB}) and of the
model (\ref{model2shk}) when inertial terms are set to zero (dashed
and dashed-dotted lines). The fluid properties correspond to
Rhodorsil silicon oil v50 (see table~\ref{table2}).}
 \label{fig-sol}
\end{figure}

Figure~\ref{fig-sol} displays the speed and maximum amplitude of the
solitary waves corresponding to homoclinic orbits around
$\UI=(1,0,0)^t$. Parameters correspond to the fluid properties of
silicon oil v50 ($\G=5.48$, see table~\ref{table2}) and to different
fibre radii, $R=1.5$~mm, $0.35$~mm and $0.25$~mm ($\Go=1$, $0.24$
and $0.17$). The solutions have been computed by continuation using
{\sc Auto97} together with its subroutine {\sc HomCont}
\cite[]{auto97}. In all cases $\Delta>0$, i.e. $\UI$ is a
saddle-focus and the solitary waves are characterised by monotonic
tails and oscillatory fronts.

Two different behaviours can be observed for small and large
thicknesses, or for $\D\ll1$ and $\D \gg 1$.
\footnote{Even though, strictly speaking $\D$
is not allowed to tend to infinity, the question of the behaviour of
different quantities of interest for large $\D$ is a valid one
within the context of the WRIBL model as model equations.}
For small thicknesses ($\D \ll 1$), the maximum height $h_{\rm max}$
and speed of the solitary waves exhibit local maxima which strongly
increase for thin fibres corresponding to a stronger RP~instability.
In Figs.~\ref{sol-b} and \ref{sol-d}, the characteristics of the
solitary wave solutions of the WRIBL model (\ref{model2shk}) and of
the CM~equation (\ref{KDB}) are compared showing reasonable
agreement. As the CM~equation is parameterized by the aspect ratio
$\aN$ and $\beta$, the local maxima are the result of the balance of
curvature effects and the advection by the flow. This is reminiscent
of the sharp increase, or `blow-up', of speed and amplitude observed
by Kalliadasis and Chang~\cite{Kal94} at
 $\beta=\beta_c \approx 1.413$ in their study of
the solitary wave solutions of the Frenkel equation (\ref{Serafim}).

\begin{figure}
\begin{center}
\BPS
\psfrag{hM}{$h_{\rm max}$}
\psfrag{betaM}{$\beta^\star$}
\includegraphics[width=0.6\textwidth]{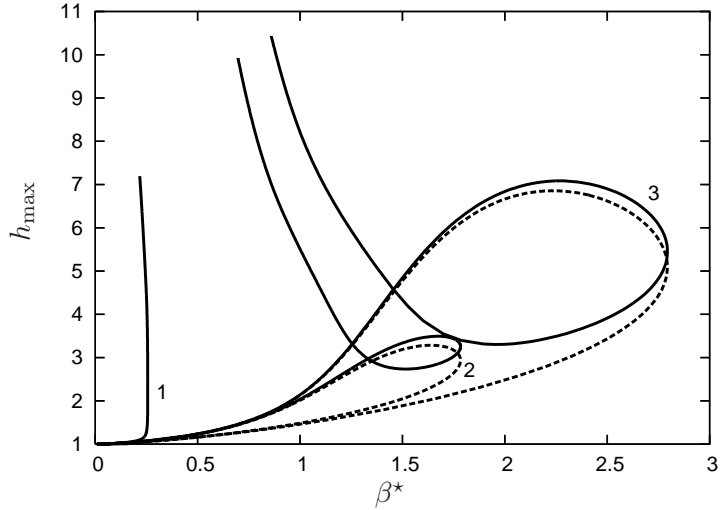}
\EPS
\end{center}
\caption{Maximum height of solitary waves versus saturation
parameter $\beta^\star$. See also caption of Fig.~\ref{fig-sol}.}
\label{fig-beta-star}
\end{figure}

One might expect that the sharp increase of the local maxima of the
speed and amplitude observed by lowering the fibre radius is related
to the nonlinear saturation mechanism of the instability by the
advection discussed earlier and, therefore, it should be correlated
with the saturation number $\beta^\star$. The validity of this
hypothesis is checked in Fig.~\ref{fig-beta-star} where the maximum
height of the solitary waves is depicted as a function of
$\beta^\star$. At a given value of the Goucher number $\Go$,
$\beta^\star$ reaches a maximum for $\aN \approx 0.44$ and tends to
zero for $\aN \to 0$ and $\aN \to \infty$. For a fixed radius $R$,
$\D$ and $\aN$ have the same trend. This explains why $\beta^\star$
tends to zero when $\D \to 0$ and $\D \to \infty$ and justifies the
shapes of the curves. The local maximum of $h_{\max}$ occurs at
$\beta^\star$ close to the maximum reached by this parameter as $\D$
is varied. The increase of the local maximum of $h_{\max}$ is
related to the increase of the maximal value of $\beta^\star$
achieved as the ratio $\Go$ is lowered.

We have also computed solutions of model (\ref{model2shk}) when the
inertial terms are cancelled ($\D\to0$). The results are shown as
dashed-dotted lines in Figs.~\ref{sol-b} and \ref{sol-d}. As they
asymptote to the solid lines corresponding to (\ref{model2shk}) in
the limit $\D \ll 1$, we can conclude that the difference in speed
and amplitude between solutions to (\ref{model2shk}) and to the
CM~equation result from the viscous dispersion effects that
contribute to the increase of the speed and amplitude of the
solitary waves.

\begin{figure}
\begin{center}
\BPS
\psfrag{D}{$\D$}
\psfrag{c/ck}{$c/c_k$}
\psfrag{h}{$h_{\rm max}$}
\subfigure[]{\label{solv440-a}%
\includegraphics[width=0.48\textwidth]{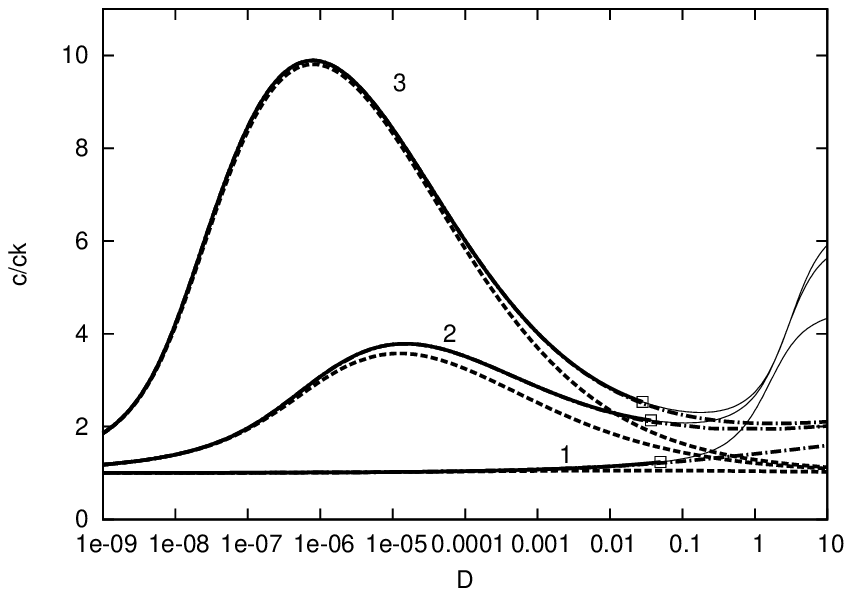}}\hfill%
\subfigure[]{\label{solv440-b}%
\includegraphics[width=0.48\textwidth]{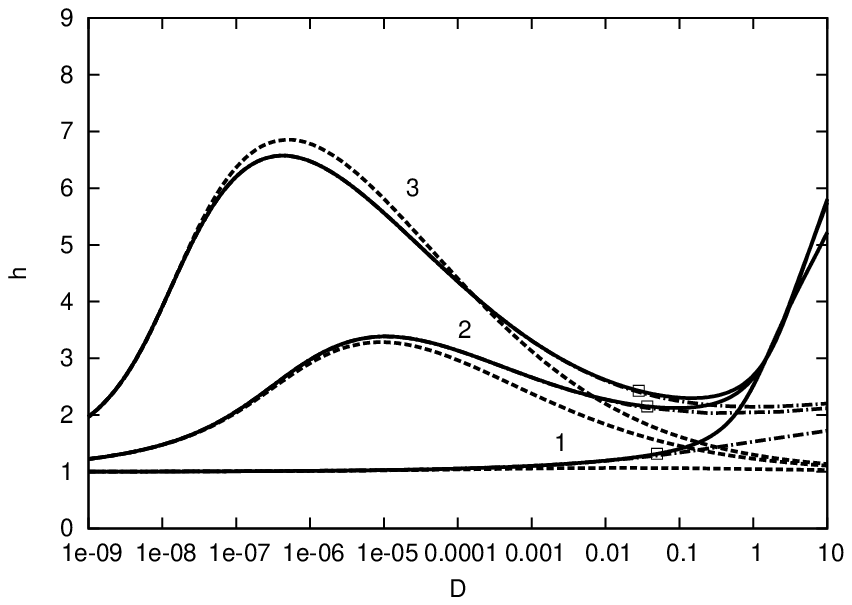}}
\EPS
\end{center}
\caption{Speed $c$ (left) and maximum height $h_{\max}$ (right) as
function of the reduced Reynolds number $\D$ for different fibre
radii: $R=1.83$~mm (Curves~1), $0.43$~mm (2) and $0.31$~mm (3).
Solutions to (\ref{model2shk}) (solid lines) are
compared to the solutions of the CM~equation (\ref{KDB}) and of the
model (\ref{model2shk}) when inertial terms are set to zero (dashed
and dashed-dotted lines).
Homoclinic orbits connecting a saddle-focus (saddle) fixed point to
itself are shown in thick (thin) solid lines. Squares indicate the
loci of saddle to saddle-focus transitions.  The fluid properties
correspond to castor oil ($\G=0.45$).} \label{fig-solv440}
\end{figure}

Figure~\ref{fig-sol} reveals a sharp increase of the maximum height
and speed of the waves above $\D\approx 2$ corresponding to the
predominance of the K mode (as the planar case is approached, i.e.
when the fibre radius increases, this sharp increase corresponds
precisely to the transition between the drag-gravity and
drag-inertia regimes). The characteristics of the waves in this
region will be considered in detail later on in \S\,\ref{S-DI}. The
separation between the local maxima for the speed and amplitude
corresponding to the RP-dominated waves ($\D\ll1$) and the
K-dominated waves at large $\D$ increases for even more viscous
fluids like the castor oil used by Kliakhandler \cite{Kli01} (cf.
Table~\ref{table2}) as can be observed from Fig.~\ref{solv440-b}
where the solutions of the model (\ref{model2shk}) with and without
inertia, and of the CM~equation are compared for fibre radii
corresponding to the same ratios $\Go=1$, $0.24$ and $0.17$ as in
figure~\ref{fig-sol}. This increase can be understood by considering
the definition of $\beta^\star$, which is a function of the aspect
ratio $\aN$ and $\Go$, and the definition of $\D=(\aN \Go)^{11/3}
\phi(\aN) \G^{3/2}$. Since the Kapitza number $\G$ decreases with
the kinematic viscosity, the maximum of $\beta^\star$ for a given
value of the Goucher number $\Go$ corresponds to smaller and smaller
values of the reduced Reynolds number $\D$ as the viscosity of the
fluid is increased. In contrast with the results for silicon oil v50
(cf. Fig.~\ref{fig-sol}), a transition from a saddle-focus to a
saddle fixed point has been observed corresponding to the
disappearance of the capillary ripples at the front of the solitary
wave at values of $\D$ above $0.1$. The precise location of this
transition varies with the fibre radius $R$ and is indicated by
squares in figure~\ref{solv440-a}.

\begin{figure}
\begin{center}
\BPS
\psfrag{D}{$\D$}
\psfrag{c/ck}{$c/c_k$}
\psfrag{h}{$h_{\rm max}$}
\psfrag{hmax, bstar}{$h_{\rm max}$, $\beta^\star$}
\subfigure[]{\label{sol2-b}%
\includegraphics[width=0.48\textwidth]{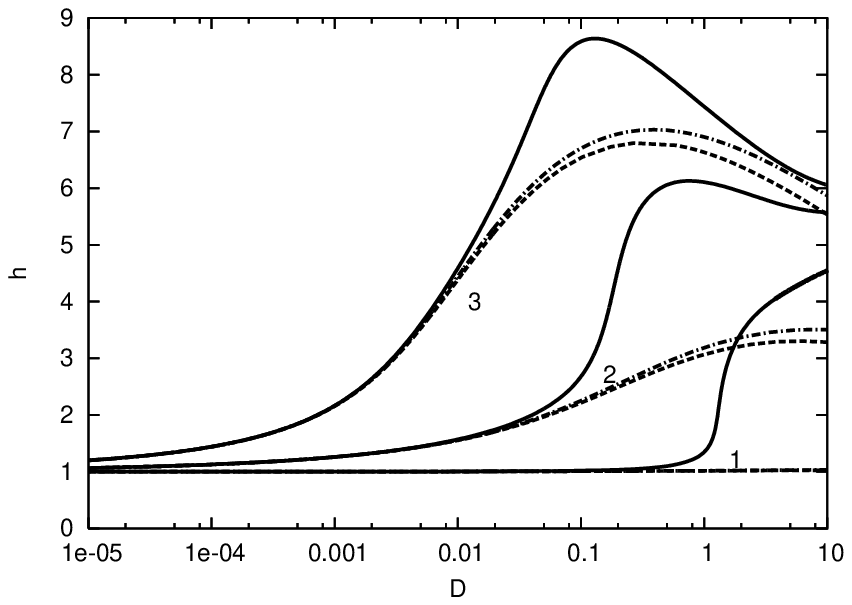}}\hfill%
\subfigure[]{\label{sol2-c}%
\includegraphics[width=0.48\textwidth]{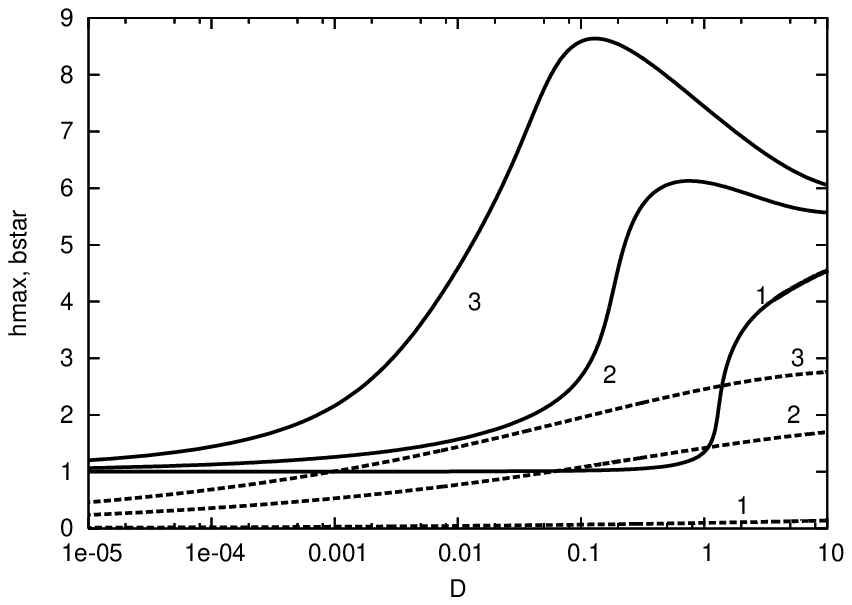}}
\EPS
\end{center}
\caption{(a) Maximum height $h_{\rm max}$ of the solitary waves as
function of $\D$. Solutions of (\ref{model2shk}) (solid lines) are
compared to the solutions of the CM~equation (\ref{KDB}) (dashed
lines) and solutions of (\ref{model2shk}) where the inertia terms
are set to zero (dashed-dotted lines). (b) Maximum height $h_{\rm
max}$ (solid line) and parameter $\beta^\star$ (dashed lines).
Parameters correspond to water ($\G=3376$) and different fibre
radii: $R=2.75$~mm (Curves~1), $0.64$~mm (2) and $0.46$~mm (3).}
\label{fig-sol2}
\end{figure}

Conversely, the separation of the solitary wave characteristics,
such as speed and maximum height, as a function of $\D$ into two
distinct regions, at low and high reduced Reynolds numbers, vanishes
at low viscosities. Indeed at low viscosities, or equivalently, high
Kapitza numbers, the RP~mode occurs at relatively high values of
$\D$ where the K~mode already takes over. In Fig.~\ref{sol2-b} we
have redrawn Fig.~\ref{sol-d} for water ($\G=3376$, see
Table~\ref{table2}), which is fifty times less viscous than silicon
oil v50. The figure compares the maximum height of the solitary
waves for a ratio of the Goucher number $\Go$ equal to the ones
chosen for the computations shown in Figs.~\ref{fig-sol}, to the
amplitude of the solutions of the CM~equation (\ref{KDB}). At small
values of $\D$, the amplitude of the waves is significantly larger
than the amplitude of the solutions of the CM~equation, which
signals the influence of the K~mode on the RP~instability. Notice
that this effect cannot arise from the second-order viscous terms as
the cancellation of the inertial terms leads to results comparable
to the solutions of the CM~equation. On the other hand, the
influence of the RP instability on the K mode is illustrated in
Fig.~\ref{sol2-c} where $\beta^\star$ and $h_{\rm max}$ are plotted
versus $\D$. At small fibre radii $R=0.64$~mm and $R=0.46$~mm
($\Go=0.24$ and $0.17$), the characteristic sharp increase of the
solitary-wave amplitude is displaced to values of $\D$ smaller than
$\D \approx 2$  corresponding to a generalised saturation number
$\beta^\star \gtrapprox 1$.

\section{Limiting cases}
\label{S-asymp}

In this section, we focus on the different regions of the
wave-characteristics' diagrams displayed in Figs.~\ref{fig-sol},
~\ref{fig-solv440} and \ref{fig-sol2}, and consider all possible
asymptotic limits.

\subsection{Small Goucher number limit: the drop-like regime}
\label{S-drop} Let us first consider the local maxima of the
amplitude $h_{\rm max}$ and speed $c$ with respect to $\D$ for given
values of the Goucher number $\Go$, or equivalently $\aN$, observed
for viscous fluids like silicon oils in the inertia-less limit
($\D\ll1$). Table~\ref{table1} depicts these quantities as obtained
from the CM equation.
 As $\Go$ tends to zero, the RP instability mechanism becomes more and more
efficient and we observe a sharp increase of the local maxima of the
amplitude $h_{\rm max}$ and of the maxima of the speed of the waves.
For such waves, the amplitude can be several orders of magnitude
larger than the Nusselt flat film on which they stand, and
variations of the Nusselt thickness should only slightly modify the
wave characteristics (except perhaps when the film becomes so thin
that the corner dissipation at the wave front dominates over the
dissipation in the bulk).
\begin{table}
\begin{center}
\begin{tabular}{c|cccccc}
\hline $\Go$ & $\aN$ & $h_{\rm max}$ & $c$ & $\bar h_{\rm max}/{\bar
R}$ & $ c \,\nu/(g {\bar R}^2)$ & $c_{\rm drops} \,\nu/(g {\bar
R}^2)$
\\
\hline
0.236   & 0.23  & 3.3        & 3.1   & 0.77 &  0.22   & 0.25\\
0.168   & 0.15  & 6.9        & 8.8   & 1.05 &  0.25   & 0.30\\
0.110   & 0.075 & 18         & 38.5  & 1.36  & 0.24   & 0.29\\
0.055   & 0.022 & 81.2       & 383   & 1.80  & 0.19   & 0.22\\
0.044   & 0.014 & 129        & 773   & 1.85  & 0.16   & 0.18\\
\hline
\end{tabular}
\end{center}
\caption{Local maxima of the speed $c$ and amplitude $h_{\rm max}$
with respect to $\D$ for given values of the Goucher number $\Go$,
or equivalently $\aN$, obtained from the CM~equation (\ref{KDB}).
$c_{\rm drops}$ refers to the speed of quasi-static drops sliding
downwards coated fibres (see text and
Appendix~\ref{S-Static-Drops}). \label{table1} }
\end{table}

Since $\bar R\ll l_c$, azimuthal surface tension effects dominate
over gravity and the typical length scale and height of a wave
should correspond to the radius $\bar R$ of the fibre as already
pointed out in \S\,\ref{S-WRIBL}. The wave speed should then be
determined by the balance of viscosity and gravity at the scale
$\bar R$ which gives a typical velocity of $g {\bar R}^2/\nu$ for
viscous-gravitational drainage.
 Justification of the
neglect of the inertial terms demands that the Galilei number
$\Ga=\Go^3\G^{3/2}$ is small, which is satisfied for all tested
fluids except for water ($\G=3376$). Our computations of the
solutions to the CM~equation (\ref{KDB}) confirm these scaling
arguments as shown in table~\ref{table1}.

\begin{figure}
\begin{center}
\BPS
\psfrag{x/R}{$\xi/R$}
\psfrag{r/R}{$r/R$}
\psfrag{0.044}{$\Go=0.044$}
\psfrag{0.110}{$0.110$}
\psfrag{0.168}{$0.168$}
\psfrag{0.236}{$0.236$}
\includegraphics[width=\textwidth]{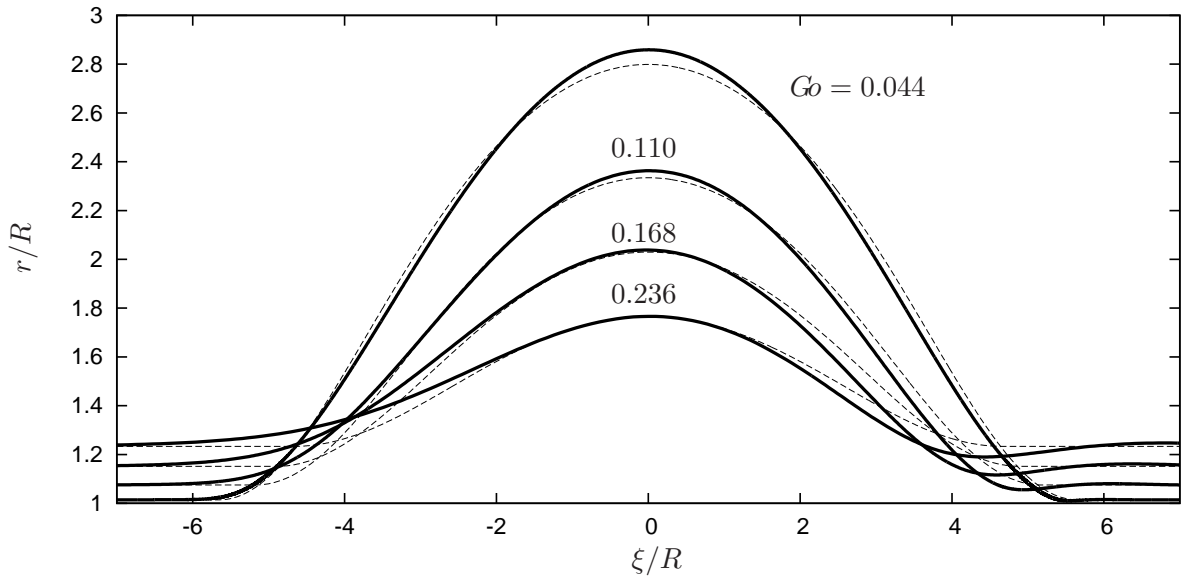}
\EPS
\end{center}
\caption{Wave profiles corresponding to the solutions to the CM
 equation (solid lines) and static drop shape (thin dashed lines).
Labels correspond to the Goucher number. Values of the other parameters
 are listed in table~\ref{table1}.
\label{fig-prof-tab2}
}
\end{figure}

In Fig.~\ref{fig-prof-tab2}, we contrast the  wave profiles rescaled
with respect to $\bar R$ corresponding to Table~\ref{table1}. Except
from the front and back of the waves corresponding to the return to
the fixed point, the wave profiles are rather symmetric. This
front-to-back symmetry shows that gravity does not affect the wave
profile, as expected, since the typical size of the wave $\bar R$ is
much smaller than the capillary length $l_c$. Therefore, solitary
waves in this regime resemble isolated drops sliding under the
action of gravity on a wettable fibre, which is precisely why we
refer to this regime as the drop-like regime. [When $\bar R$ is
larger and approaches $l_c$, on the other hand, the drops will feel
the effect of gravity as they grow and they will eventually resemble
falling pendant drops.] It corresponds to the observation by
Qu\'er\'e~\cite{Que90,Que99} in the coating of wires or thin fibres
drawn out of a bath of viscous liquids that the thin annular film
deposited on the wires/fibres breaks up into drops.

We have checked this analogy by computing the shape of static drops
with zero contact angles sitting on a fibre coated with a substrate
film of the same liquid (details of the calculation are given in
Appendix~\ref{S-Static-Drops}). The agreement of the wave shapes to
the symmetrical static-drop shapes is remarkable even in the case of
nearly spherical drops such as those obtained at $\Go=0.044$, where
the long-wave assumption is no longer valid. We can therefore
conclude that the CM equation (\ref{KDB}), and by extension the
WRIBL model (\ref{model2shk}), are accurate in the drop-like regime
where surface tension is dominant and even if the long-wave
approximation does not hold. Besides, the remarkable agreement
between the solutions to the CM~equation (\ref{KDB}) and the WRIBL
model (\ref{model2shk}) already noted in Fig.~\ref{fig-sol}, shows
that the second-order streamwise viscous terms do not affect the
amplitude and speed of the drops.

Following \cite{Kal94}, an analytical estimate of the amplitude and
speed of the drop-like waves in the limit $\Go\to0$ may be obtained
via matched asymptotic expansions. The appropriate small parameter
is the dimensionless speed of the drops. By balancing viscous and
capillary forces at the back of the waves, one can easily extend to
sliding drops the Landau-Levich-Derjaguin law obtained by
Qu\'er\'e~\cite{Que99} in the case of fibres drawn out of a bath.
The speed of sliding drops is thus governed by Eq. (\ref{Landau})
which compares favorably to the results from the CM equation in
table~\ref{table1}. As a matter of fact, this agreement shows that
the thickness of the residual film on which the drops slide is
determined by the balance of surface tension and viscous dissipation
in the meniscus region. To estimate the speed and amplitude of
drop-like waves, one would have to take into account the gravity
acceleration and higher-order corrections in the outer region,
namely the viscous dissipation in the drop. A task which is
difficult, as (a) with the Frenkel equation resolving fully the
leading-order outer region requires matching up to third
order~\cite[]{Kal94} and quite likely this is the case here, (b) a
single `composite equation' for the whole domain, i.e. for both the
drop and residuals films, as e.g. in the `drag-out' problem in
coating theory~\cite[]{Wil82}, does not exist.

\subsection{Large $\D$ limit: the drag-inertia regime}
\label{S-DI} 
\begin{figure}
\begin{center}
\BPS
\psfrag{h}{$h$}
\psfrag{hx}{$h'$}
\psfrag{xi}{$\xi$}
\subfigure[]{\label{fig-profhom_a}
\includegraphics[width=0.45\textwidth]{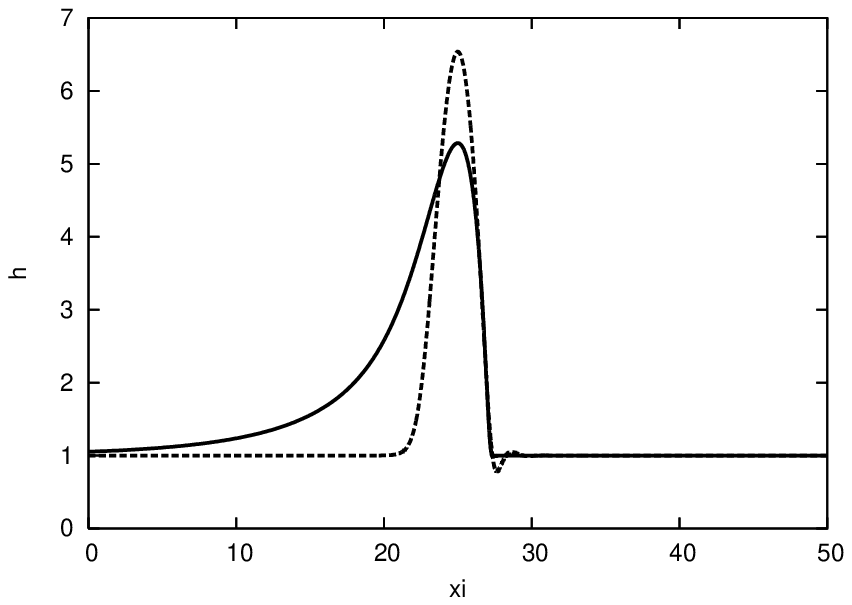}}\hfill%
\subfigure[]{\label{fig-profhom_b}
\includegraphics[width=0.45\textwidth]{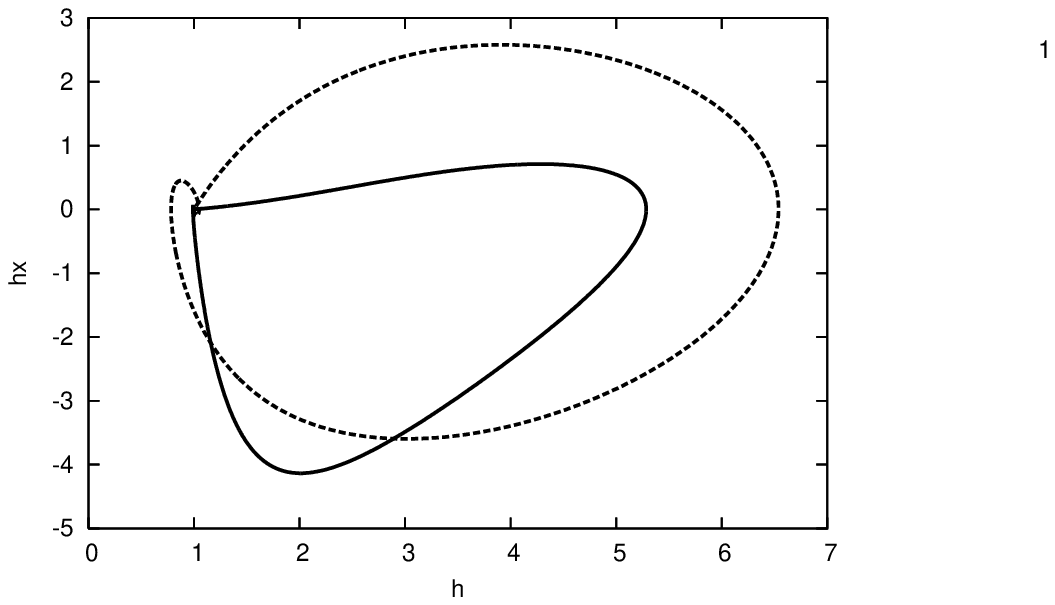}}
\EPS
\end{center}
\caption{Profile (a) and projected trajectory (b) onto the plane
($h$, $h'$) of single-loop homoclinic solutions to (\ref{model2shk})
corresponding to single-hump solitary waves. Parameters correspond
to Rhodorsil silicon oil v50 ($\G=5.48$) and $R=0.25$~mm. Solid and
dashed lines refer to $\D=5$ and $\D=3\,10^{-5}$, respectively.}
\label{fig-profhom}
\end{figure}

To understand the change of behaviour of the
solitary-wave characteristics for $\D\gg1$, we look at the wave
profiles and their projections on the plane ($h$, $h'$).
Figure~\ref{fig-profhom} compares two single-loop homoclinic orbits
for a small and a large value of $\D$. For $\D\ll 1$, solitary waves
have a relatively symmetric shape. Except from the neighbourhood of
the fixed point $\UI$, where the escape from $\UI$ is monotonic and
the return to it is oscillatory, a symmetry between the front and
back of the waves is observed with a steep front and a steep back.
On the contrary, at large values of $\D$, the front and back of the
solitary waves present radically different shapes, with a gentle
sloping back edge and a steep front edge.

The observed difference between front and back of the solitary waves
in the large-$\D$ case can be explained by examining the linearised
dynamics around $\UI$. The dispersion relation governing
infinitesimal perturbations varying as $\exp(\l \xi)$ is given in
(\ref{dispU1}). At a given radius $R$ and capillary length $l_c$,
large thickness, $\hN \gg 1$, corresponds to $\D \sim\hN^{11/3}
\phi(\aN) \gg 1$, $\beta/\D = (l_c /R)^2 /(3 \R) \ll 1$ and possibly
large viscous dispersion since $\e \sim \hN^{4/3}$. Let $\l_1$ be
the positive eigenvalue corresponding to the unstable manifold of
$\UI$. The eigenvalues of the tangent subspace to the unstable
manifold satisfy $\Re(\l_3) \le \Re(\l_2) < 0$. When $\UI$ is a
saddle-focus ($\Delta>0$), we further  denote $\l_{2,3}$ by $-
\Sigma \pm{\rm i} \Omega$  with $\Sigma >0$ and $\Omega > 0$. From
(\ref{dispU1}) we obtain the estimate $\l_1 \sim \D^{-1} \ll 1$.
Since $\l_1 + \l_2 + \l_3= - \e D_\e$, we immediately get an
estimate of the mean value, $(\l_2 + \l_3)/2 \sim -\e$, so that
$\Sigma \sim \e$ when $\Delta>0$. As a consequence, at the back of a
solitary wave, the monotonic escape from the fixed point is slow,
whereas at the front, the return to $\UI$ is fast. The above
estimates have been confirmed by computations of $\l_1$ and $\Sigma$
for the solitary waves shown in Fig.~\ref{fig-sol} corresponding to
silicon oil v50.


Focusing now at the back of the solitary wave and defining a slow
variable ${\ti \xi} = \xi/\D$, (\ref{MB_ssdyn}) reads,
 \BA
\NNM
&& \left\{ (1 + \aN h) \left[ c^2 - c \F(\aN h)
\frac{q}{h} \right] + \GG(\aN h)  \frac{q^2}{h^2}
+ \frac{\beta \I(\aN h)}{\D (1+\aN h)^2 \phi(\aN)}
\right\}
\frac{{\rm d} h }{{\rm d} {\ti \xi}}
\\
&& \qquad \qquad \qquad \qquad =
\frac{\I(\aN h)}{\phi(\aN)}
\left[ \frac{3 \phi(\aN)}{\phi(\aN h)} \frac{q}{h^2} - h \right]
 +O (\e/\D^2,\D^{-3})\,,
\label{MB_tixi}
\EA
 where $q$ is given by (\ref{q0}) and
 (\ref{q0_h1}). Equation~(\ref{MB_tixi}) can be formally rewritten as,
\BE
 \label{MB_tixi_formal}
{\cal G}(h,c;\aN,\beta/\D)  \frac{{\rm d} h }{{\rm d} {\ti \xi}} = - {\cal
H}(h,c;\aN)
 + O(\e/\D^2,\D^{-3})\,,
\EE
expressing the balance at the back of the solitary waves of inertia (at the
left-hand side), viscous drag and gravity acceleration (at the
right-hand side). As a consequence, the roots of the right-hand side
of (\ref{MB_tixi}) correspond to the fixed points of (\ref{ssdyn}).

As the homoclinic orbit departs from $\UI$, $h$ increases up to
$h_{\rm II}$ which is larger than unity since $c > c_k$ (cf.
Fig.~\ref{fig-fp}). At this point, $h$ goes through a maximum if
${\cal G}$ is nonzero. The resulting orbit is thus a heteroclinic
one linking $\UI$ and $\UII$ which contradicts the fact that we are
considering a single-loop homoclinic orbit \cite[]{Ruy05b}. As a
consequence, ${\cal G}$ must go to zero at $h=h_{\rm II}$ which
signals a `critical film thickness' $h_c$ at which inertial terms
must go to zero. In the limit $\D\to\infty$, we thus
 obtain the condition
\BE
\label{def-crit}
 h_{\rm II}(c;\aN) = h_c(c;\aN,\beta/\D)
\EE which gives a unique solution $c_{\rm crit}$ and then $h_{\rm
crit}$ as function of
 $\aN$ and $\beta/\D$. As the limit speed
$c_{\rm crit}$ is governed by the balance of inertia, wall friction and
gravity acceleration, we may refer to this situation as the {\it
drag-inertia} regime \cite[]{Oos99}.

Having shown that ${\cal G}$ possesses at least one root, ${\cal G}$
can be factorised into ${\cal G} = (1 + \aN h) [c-c_{d+}(\aN h,
q/h,\beta/\D)][c-c_{d-}(\aN h, q/h,\beta/\D)]$ where, \BSE
 \label{cdpmahq/h}
\BA
c_{d\pm}(\aN h, q/h) &=& \frac{q}{h} \frac{\F(\aN h)}{2}  \pm
\sqrt{\Delta_{\aN h,q/h,\beta/\D}}
\\\hbox{and} \quad
 \Delta_{\aN h, q/h,\beta/\D} &=&
\left(\frac{q}{h} \right)^2 \left[\frac{\F(\aN h)^2}{4} -
 \frac{\GG(\aN h)}{1 + \aN h} \right] -
 \frac{\beta \I(\aN h)}{\D (1+\aN h)^3\phi(\aN)}
\,,
\EA
\ESE
 are the speeds of linear dynamic waves for a uniform layer of
thickness $h$ and averaged speed $q/h$ \cite[]{Ruy08}. In other words,
 the position of the second fixed point must coincide with the critical layer
$h_c$ at which the speed of the solitary wave $c$ is equal to the
speed of one of the dynamic waves $c_{d\pm}$ -- in fact the fastest
one with speed $c_{d+}$ --, which separates the flow into a
`subcritical region' ($c<c_{d+}$) and a `supercritical region'
($c>c_{d+}$). The condition $h_{\rm II} = h_{\rm crit}$ is similar
to the `Thomas condition' derived in the mathematical treatment of
periodic bores on steep slopes, or {\it roll-waves}
\cite[]{Tho39,2Liu05} made of the regular succession of laminar
flows and hydraulic jumps. For $\D \gg 1$, the front of the solitary
waves, where surface tension arrests the breaking of the waves,
plays a role similar to that of a hydraulic jump connecting the
subcritical and supercritical regions of a roll wave. For this
reason, we may also refer to this regime as the {\it roll-wave
regime\/}.

Considering now large but finite values of $\D$, condition
(\ref{def-crit}) is not verified and one has to go back to
(\ref{MB_tixi_formal}). At the critical point $h=h_c$ we have ${\cal
G}(h_c,c;\aN) =  0$ and a Taylor expansion close to criticality
gives \BE \label{Taylor-G} h_c - h_{\rm crit} \approx - \left[\P_c
{\cal G}/\P_h {\cal G}\right] ( h_{\rm crit}, c_{\rm crit}) \left(c
- c_{\rm crit}\right) \EE whenever $\P_h {\cal G}( h_{\rm crit},
c_{\rm crit}) \ne 0$. Proceeding next to a Taylor expansion of
${\cal H}(h,c)$, (\ref{MB_tixi_formal}) yields \BE \label{Taylor-H}
\left[ \P_c {\cal H} -\frac{\P_h {\cal H} \P_c {\cal G}}{\P_h {\cal
G}} \right] (c -c_{\rm crit}) \approx
 K_{\e} \frac{\e}{\D^{2}}
+ \frac{K_{\rm st}}{\D^{3}}\,,
\EE
where the constants $K_{\e}$ and $K_{\rm st}$ are functions
 of $h_{\rm crit}$, $c_{\rm crit}$ and the derivatives  $h_{\rm crit}'$,
  $h_{\rm crit}''$ and   at $h_{\rm crit}'''$ at
the critical thickness of the asymptotic solution. Since for $\hN
\gg 1$, $\D \sim \hN^{11/3}$ and $\e \sim \hN^{4/3}$ the
asymptotically dominant term in the right-hand-side of
(\ref{Taylor-H}) are the viscous terms
 $K_{\e} \e \D^{-2}$. Consequently, the convergence of the speed of
the waves to the asymptotic value $c_{\rm crit} (\aN)$ satisfies \BE
\label{conv-rate} c - c_{\rm crit} (\aN,\beta/\D) \propto \e/\D^2
\sim 1/\R^2 \EE
in agreement with our numerical computations.
%

\begin{figure}
\begin{center}
\BPS
\psfrag{D}{$\D$}
\psfrag{c/ck}{$c/c_k$}
\subfigure[water ($\G=3376$)]{\label{DI-sol2-a}%
\includegraphics[width=0.48\textwidth]{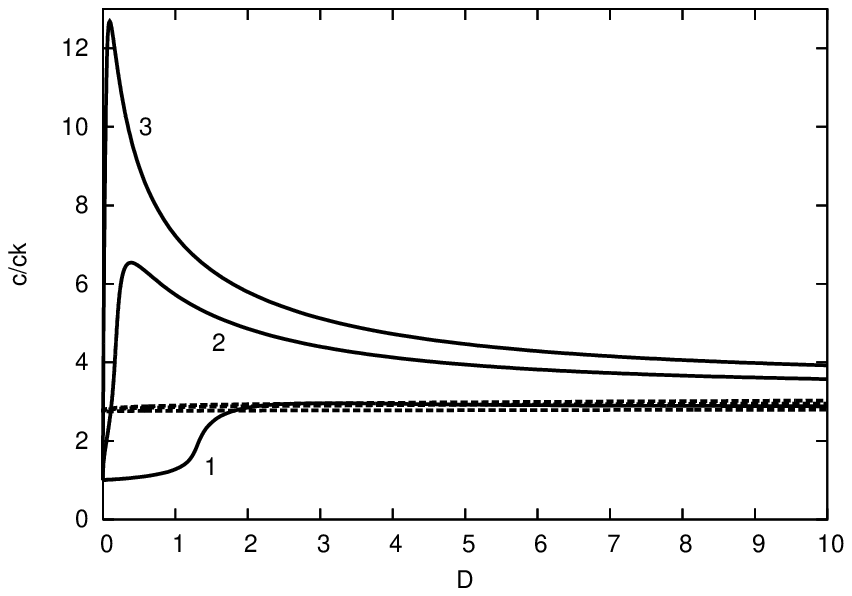}}\hfill%
\subfigure[silicon oil v50  ($\G=5.48$)]{\label{DI-sol-a}%
\includegraphics[width=0.48\textwidth]{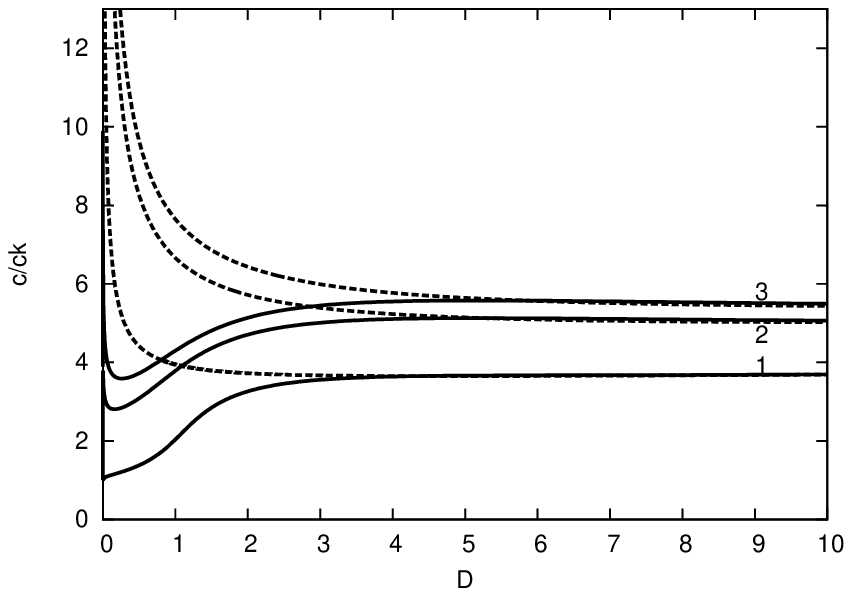}}
\subfigure[castor oil ($\G=0.45$)]{\label{DI_solv440-a}%
\includegraphics[width=0.48\textwidth]{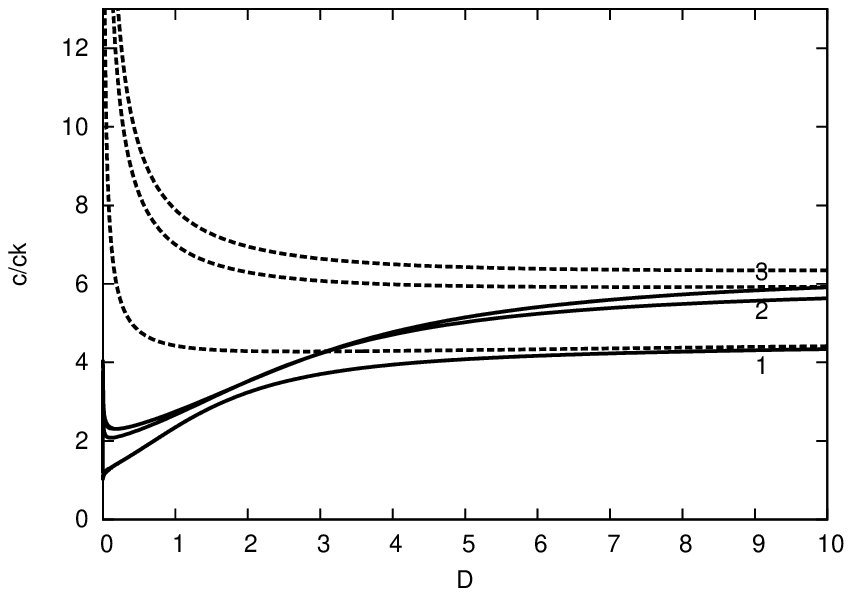}}\hfill%
\subfigure[silicon oil v1000 ($\G=0.10$)]{\label{DI-solv1000-a}%
\includegraphics[width=0.48\textwidth]{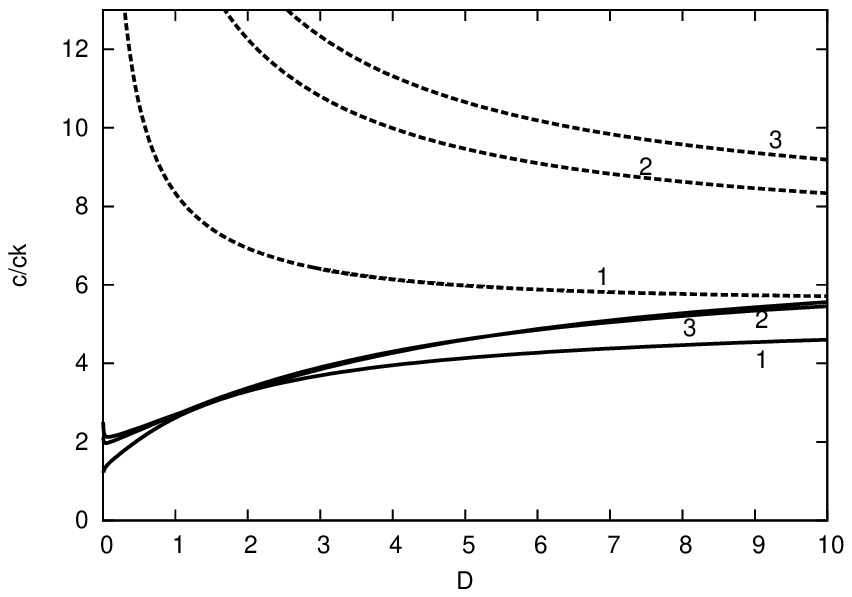}}
\EPS
\end{center}
\caption{Speed $c$ of the solitary waves as function of the reduced
Reynolds number $\D$ for three values of the Goucher number
$\Go=1.01$ (curves 1), $0.236$
 (curves 2) and $0.168$ (curves 3) for fluids of decreasing Kapitza numbers.
 Solutions to (\ref{model2shk}) (solid lines) are compared to
the asymptotic predictions $c_{\rm crit}$ (dashed lines).
} \label{fig-DI}
\end{figure}

Figure~\ref{fig-DI} compares the phase speed $c$ of the solitary
waves to the asymptotic limit $c_{\rm crit}$ as function of $\D$ for
the four different fluids of increasing viscosity that we considered
(fluid properties are gathered in Table~\ref{table1}). For weakly
viscous flows, the RP and K mode reinforce each other and the speed
$c$ of the waves is much higher than the asymptote $c_{\rm crit}$
(see Fig.~\ref{DI-sol2-a}).
At larger viscosities (Figs.~\ref{DI-sol-a} and \ref{DI_solv440-a})
the K mode ceases to be affected by the RP instability and a rapid
convergence to $c_{\rm crit}$ is observed as $\D$ is increased.
Whereas for very viscous fluids like silicon oil v1000 (one thousand
times more viscous than water, cf.~Table~\ref{table2}), the curves
start again to separate signalling the arrest of the convergence to
the asymptote
 $c_{\rm crit}$ by the axial viscous effects.

\subsection{Weakly nonlinear analysis}
\label{S-weakNL} Let us now consider the regions of the speed and
maximum height diagrams in Figs.~\ref{fig-sol}, \ref{fig-solv440}
and \ref{fig-sol2} corresponding to the transition between the
drop-like regime and the drag-inertia regime. They correspond to
situations where neither the K nor the RP~instability mechanisms are
strong enough to promote large amplitude waves, \ie when
$\D\lessapprox 1$ and $\beta^\star \lessapprox 1$. We can then
proceed to a weakly nonlinear analysis to characterize the shape,
speed and amplitude of the waves. Since $\D\ll1$, we substitute $h =
1+H$ and $q= 1/3 + Q$ where $H= O(\vareps)$ and $Q=O(\vareps)$ are
small deviations from the base state $(h,q)=(1,1/3)$, (\ie $\vareps
\ll 1$). Taking the distinguished limit $\vareps\ll \eps^2$, we are
led from the WRIBL (\ref{model2shk}) model to a single evolution
equation for the deviation $H$: \BA \P_t H + c_k \P_x H +  \Phi(\aN)
H \P_x H +
 \frac{\beta}{3(1+\aN)^3} \P_{xx} H
+ \frac{1}{3(1+\aN)}\P_{xxxx} H
\NNM&&\\
+\D \frac{\phi(\aN)}{3 \I(\aN)}
\left[\P_{tt} H + \frac{\F(\aN)}{3} \P_{tx} H
 + \frac{\GG(\aN)}{9(1+\aN)} \P_{xx} H
\right]
\NNM&&\\
+\e  \frac{\phi(\aN)}{3 \I(\aN)} \left[\M(\aN) \P_{xxt} H -
\frac{\L(\aN)}{3(1+\aN)} \P_{xxx} H \right] &=& O(\vareps
\eps^5,\vareps^2 \eps^2)\,, \label{H} \EA where \BEN \Phi(\aN) =
\frac{3(2+\aN)\phi + \aN[(6+5\aN)\phi' + \aN(1+\aN)\phi'']}
{3(1+\aN)^2\phi} \EEN is a function of the aspect ratio $\aN$.
Noteworthy is that a stability analysis of the base state
$(h,q)=(1,1/3)$ based on (\ref{H}) leads to the same dispersion
relation as for (\ref{model2shk}).

\subsubsection{Drag-gravity regime}
We consider here the limit of a thin film compared to the fibre radius,
$\aN\ll 1$, when the radius of the fibre is constant or equivalently
 $\Go$  is constant.
Since $\e = (\bhN/l_c)^{4/3} = \aN^ {4/3}\Go^{4/3}$, viscous
dispersion is negligible in this limit and (\ref{H}) reduces to \BE
\label{H6} \P_t H + \P_x H + 2  H \P_x H + \frac{\beta}{3} \P_{xx} H
+\frac{2}{5}\D \left[\P_{tt} H + \frac{17}{21} \P_{tx} H
+\frac{1}{7} \P_{xx} H \right] + \frac{1}{3}\P_{xxxx} H = 0\,. \EE
By making use of the first-order equivalence of the time- and
space-derivatives of $H$, $\P_t H \approx- \P_x H$, (\ref{H6})
reduces to the Kuramoto-Sivashinsky (KS) equation: \BE \label{H6b}
\P_t H + \P_x H + 2  H \P_x H + \frac{1}{3}\left[\frac{2}{5}\D  +
\beta \right] \P_{xx} H + \frac{1}{3}\P_{xxxx} H = 0\,. \EE To look
for the TW solutions to (\ref{H6b}) in their moving frame, $\xi = x-
ct$, we rescale the velocity as $c= 1+ C$, and the amplitude as $H =
\vareps A$ and stretch the moving coordinate as
 $\xi=B X$:
 \BE
\label{H7}
 -3 C B^3 A + 3 \vareps B^3  A^2
+ B^2 \Upsilon \frac{d}{dX} A
+ \frac{d^3}{dX^3} A  = 0\,,
\EE
where $\Upsilon = \frac{2}{5}\D + \beta$ and the condition
 $\lim_{\xi \to \pm \infty} H =0$ has been enforced yielding a vanishing integration
constant. Balancing each term in (\ref{H7}) gives $C B^3  \sim
\vareps B^3 \sim B^2 \Upsilon\sim 1$ so that
 $B \sim \Upsilon^{-1/2}$, $\vareps \sim \Upsilon^{3/2}$ and
 $C \sim \Upsilon^{3/2}$.

Writing $B = \Upsilon^{-1/2}$, $\vareps = \frac{2}{3} \Upsilon^{3/2}$ and
$C = \mu \Upsilon^{3/2}/3$ then leads to
\BE
\label{H5}
 -\mu A +  2A^2 + \frac{d}{dX} A
+ \frac{d^3}{dX^3} A  = 0\,, \EE which is an ordinary-differential
equation governing the TW solutions to the KS equation propagating
at speed $\mu$. Equation~(\ref{H5}) admits a one-hump solitary-wave
solution for a particular value of $\mu = \mu_0\approx 1.216$ and an
amplitude $A_{\rm max} \approx 0.784$. Consequently, in the limit
$\aN \to 0$, the speed $c$ and the amplitude $h_{\rm max}$ of the
one-hump solutions to (\ref{model2shk}) follow power laws of the
form: \BE \label{power-laws-bis} c \approx 1 + 0.405\,
{\Upsilon}^{3/2} \qquad h_{\rm max} \approx 1 +
0.523\,{\Upsilon}^{3/2}\,. \EE Recall that the relations
(\ref{power-laws-bis}) have been obtained under the assumption of
small amplitude, $h_{\rm max} - 1 \ll 1$, hence $\Upsilon$ must be
small, which implies that both inertia and azimuthal curvature
effects must be small. Viscous drag and gravity acceleration are the
dominant physical effects. Following \cite{Oos99}, we may refer to
this regime as the {\it drag-gravity regime\/.}

It is possible to interpret this regime as one characterized by the
arrest of the instability growth by the flow advection. The
dispersion relation corresponding to the KS~equation (\ref{H6b}) and
governing infinitesimal perturbations around the base Nusselt flow
of wavenumber $k$ and angular frequency $\o$ is identical to
(\ref{disp-Frenk}) when $\beta $ is substituted with $\Upsilon$. One
can then follow the same line of reasoning in going from
(\ref{disp-Frenk}) to (\ref{tauKDB}) and define  the ratio of the
typical time of advection of the structure over its length, $\tau_a=
k^{-1}$, to the typical time of growth, $\tau_g=[\max(\o_i)]^{-1}$,
of the instability: \BE \label{tauDG} \tau_a/\tau_g
=\o_i/\o_r|_{k=\sqrt{\Upsilon/2}} \propto \Upsilon^{3/2}\,. \EE
Therefore $\Upsilon \ll 1$ corresponds to $\tau_a \ll \tau_g$. The
flow advection is much faster than the instability and the ratio
$\tau_a/\tau_g$ controls the amplitude and speed  of the observed
structures as reflected by (\ref{power-laws-bis}).

\subsubsection{Soliton-like regime}
We consider in this section the limit of a thick film, $\aN\gg 1$,
though in practice, if $\aN$ is large, the axisymmetry of the flow
might be difficult to achieve. We thus have $\phi(\aN) \sim (3/4)
\aN \log(\aN)$, $c_k \sim 4/3 \aN^{-1}$ and $\beta^\star \sim
(3/4)^{2/3} \left(\aN \Go\right)^{-4/3}$, $\D \sim
\frac{3}{4}\aN\log(\aN) \G^{3/2} \left(\aN \Go\right)^{11/3}$ while
in all cases $\e =(\aN\Go)^{4/3}$. Therefore the conditions $\D \ll
1$  and $\beta^\star \lessapprox 1$, necessary to sustain the weakly
nonlinear analysis, can be justified if e.g. $\aN \Go = O(1)$, i.e.
$\e=O(1)$ and $\G^{3/2} \ll 1$. For silicon oil v1000, $\G=0.1$ so
that $\G^{3/2}=0.03$. [We note that strictly speaking the derivation
of the WRIBL model (\ref{model2shk}) demands that $\G$ is at least
$O(1)$ since the long-wave approximation is sustained only when
surface tension effects are strong \cite[]{Ruy08}.]

\begin{figure}
\begin{center}
\BPS
\psfrag{c/ck-1}{$c/c_k - 1$}
\psfrag{aN}{$\aN$}
\psfrag{xi}{$\xi$}
\psfrag{h}{$h$}
\subfigure[]{\label{fig-KDB_KaRb_aN_log_c_e}
\includegraphics[width=0.45\textwidth]{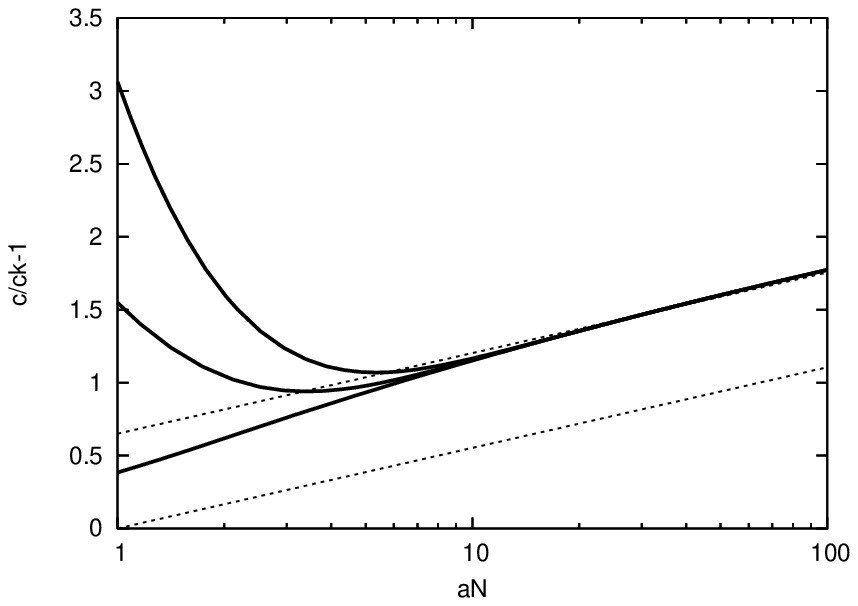}}\hfill%
\subfigure[]{\label{fig-prof_14e_KDBev440_23}
\includegraphics[width=0.45\textwidth]{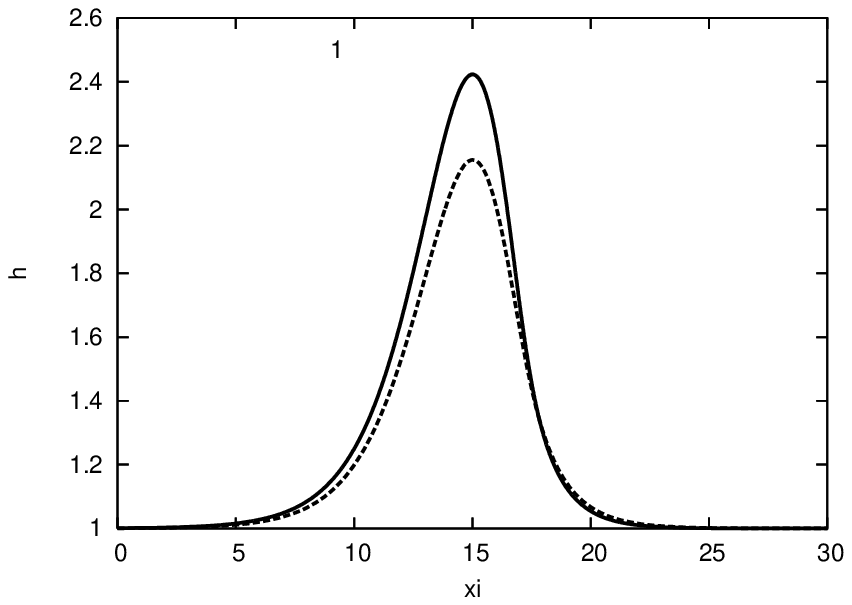}}
\EPS
\end{center}
\caption{(a) Speed of solitary wave solutions to  (\ref{model2shk})
when inertia is neglected ($\D\to0$) as function of the aspect ratio
$\aN$. Dotted lines refer to (\ref{c_ck_aN}) and to $c/c_k \approx
1.65 + 0.24\log(\aN)$. (b) Profile of the solitary wave solutions to
(\ref{model2shk}) with (solid line) and without (dashed line)
inertial terms (castor oil $\G=0.45$, $R =0.31$~mm, $\D =0.5$,
$\e=0.45$ and $\aN=4.06$). } \label{fig-KDB_KaRb_e}
\end{figure}

Neglecting inertial effects and looking for the dominant terms in (\ref{H})
 leads to
\BA \NNM \P_t H + \frac{4}{3 \aN} \P_x H + \frac{8}{3\aN} H \P_x H +
\frac{1}{3\aN\e} \P_{xx} H + \frac{1}{3 \aN} \P_{xxxx} H  &&\\
-  \e\log(\aN)\left[\frac{3}{2} \P_{xxt} H
 + \frac{1}{\aN}\P_{xxx} H \right] &=& 0\,.
\label{H2}
\EA
The first-order equivalence of the time and space derivatives of
$H$, \ie $\P_t H = -(4/(3\aN) \P_x H +
O(\eps\vareps^2,\eps^2\vareps)$, can again be utilized. Equation
(\ref{H2}) is then simplified into \BE \P_t H + \frac{4}{3 \aN} \P_x
H + \frac{8}{3\aN} H \P_x H + \frac{1}{3\aN\e} \P_{xx} H +
\frac{\e\log(\aN)}{\aN} \P_{xxx} H + \frac{1}{3 \aN} \P_{xxxx} H
  = 0\,,
\label{H10} \EE the `Kawahara' equation that was scrutinized
numerically by Kawahara and coworkers~\cite[]{Kaw83,Kaw88}. The
equation is also often referred to as the `generalized KS'
equation~\cite[]{Dup09b,Tse10a,Tse10b}, and it is the KS equation
appropriately extended to include dispersion ($\P_{xxx} H$). As
before, we look for TW solutions in their moving frame, $\xi = x-
ct$. Stretching the moving coordinate as $\xi = \sqrt{\e} X$, the
amplitude as $H= \e^{-3/2} A/2$, and the speed as $c=(4/3)\aN^{-1}
(1+ \mu\, \e^{-3/2}/4)$, then gives \BE \label{H9}
 -\mu A +  2 A^2 + \frac{d}{dX} A +\delta_K  \frac{d^2}{dX^2} A
+ \frac{d^3}{dX^3} A  = 0\,, \EE with $\delta_K = 3\e^{3/2}
\log(\aN)$ is a parameter that measures the relative importance of
dispersion. Equation~(\ref{H9}) governs TW solutions of the Kawahara
equation propagating at speed $\mu$. Since $\aN \gg 1$ and
$\e=O(1)$, the dispersion parameter $\delta_K$ is large, in which
case the speed of the one-hump solitary waves solutions to
(\ref{H9}) is given by $\mu \approx 0.3256 \delta_K$.
Solitary wave solutions of (\ref{model2shk}) in the limit $\aN \gg
1$ and $\D \ll 1$ should therefore satisfy: \BE \label{c_ck_aN}
\frac{c}{c_k} \approx 1 + 0.24 \log(\aN). \EE
The speed of the one-hump solitary wave solutions to (\ref{model2shk}) is
 compared to the asymptotic limit (\ref{c_ck_aN}) in
 Fig.~\ref{fig-KDB_KaRb_aN_log_c_e}.
However, our computations show that for moderate values of $\aN$
the wave speed is in fact closer to $c/c_k \approx 1.65 +   0.24 \log(\aN)$.
The reason for this discrepancy can be traced in the violation of the
 assumption of small amplitude deviations from the base state ($h-1\ll 1$)
 that was necessary to obtain (\ref{H10}) from (\ref{model2shk}). Indeed, TW
solutions to the Kawahara equation have a speed and an amplitude
that diverge as $\delta_K$ becomes large.

\subsection{Speed to amplitude relation}
\label{S-eed-amp} The weakly nonlinear analysis presented in
subsection~\ref{S-weakNL} provides us with the dependencies
(\ref{power-laws-bis}) of the speed and maximum height as functions
of $\Upsilon = \frac{2}{5}\D + \beta$ for small film
thicknesses.
 Combining these two relations give a linear dependence of the deviation
 amplitude $h_{\rm max} - 1$ to the deviation speed $c/c_k - 1$, where the wave
speed $c$ has been normalised with the kinematic wave speed $c_k$:
\BE \label{c-hm-asymp} h_{\rm max} - 1 \approx 1.29 \left(
\frac{c}{c_k} - 1\right) \EE A linear dependence of the speed as a
function of amplitude was initially found by  Chang~\cite{Cha86} \BE
\label{speed-Chang} h_{\rm max} -1 \approx \frac{c}{c_k} - 1 \EE by
utilizing a normal form analysis of the TW solutions of the KS
equation (\ref{H6b}). This linear dependence is a characteristic of
the drag-gravity regime and must be contrasted with the experimental
relation \BE \label{c-hm-Tihon} h_{\rm max} - 1 \approx 1.67 \left(
\frac{c}{c_k} - 1\right) \EE obtained by Tihon \etal~\cite{Tih06}
for solitary waves running down a plane inclined at an angle
$5^\circ$. A linear dependence of $h_{\rm max}$ with respect to
$c/c_{k}$ has also been found experimentally in the recent study by
\cite{Dup09a} for solitary waves on a relatively thick fibre.

\begin{figure}
\begin{center}
\BPS
\psfrag{c/ck-1}{$c/c_k -1$}
\psfrag{hmax-1}{$h_{\rm max} - 1$}
\psfrag{D>>1}{$\D\gg1$}
\psfrag{D<<1}{$\D\ll1$}
\subfigure[silicon oil v50 ($\G=5.48$)]{
\label{fig-KaRb-c-hmv50}
\includegraphics[width=0.45\textwidth]{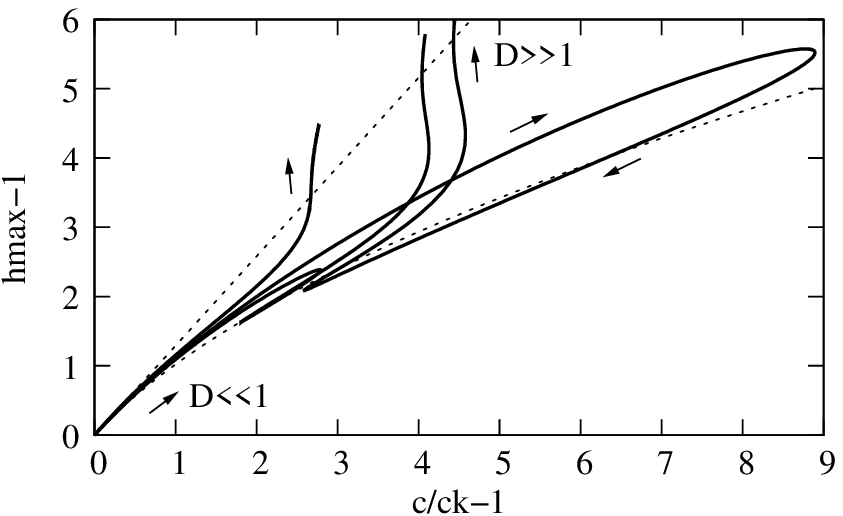}}\hfill
\subfigure[water ($\G=3376$)]{
\label{fig-KaRb-c-hmv1}
\includegraphics[width=0.45\textwidth]{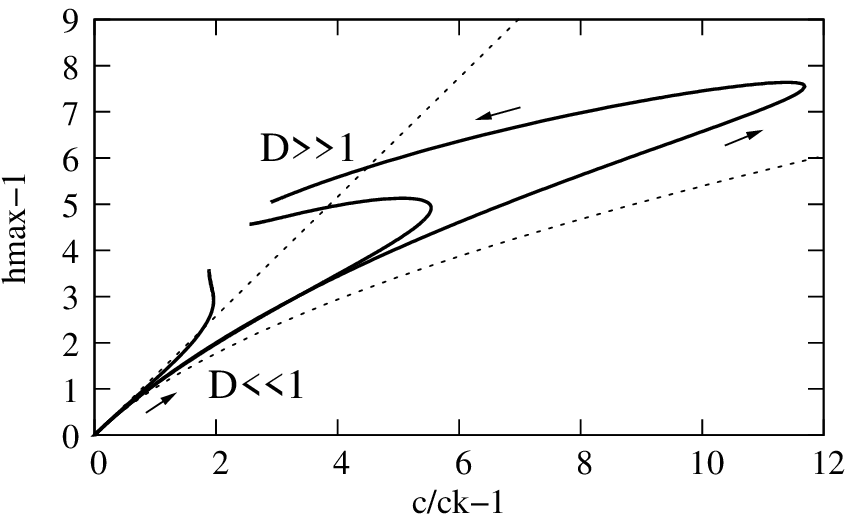}}
\EPS
\end{center}
\caption{Deviation wave amplitude $h_{\rm max}- 1$ versus relative
wave speed $c/c_k - 1$ for the solitary-wave solutions to
(\ref{model2shk}) for $\Go=1.01$, $0.236$ and $0.168$. The dashed
lines correspond to the estimates (\ref{c-hm-asymp}) and
(\ref{c-hm-asymp-2}). \label{fig-KaRb-c-hm} }
\end{figure}

Similarly, in the drop-like regime, a law relating $c$ and $h_{\rm
max}$ can easily be obtained by recognizing that the wave amplitude
$h_{\rm max} -1$ should be proportional to the distance  $h_{\rm II}
- 1$ separating the fixed points. The constant of proportionality is
determined by continuity with (\ref{c-hm-asymp}) in the limit
$h_{\rm II} - 1 \ll 1$ for which $h_{\rm II} - 1 \approx c/c_k-1$
(see Fig.~\ref{fig-fp}). Next, by approximating $h_{\rm II}$ with
$-1/2 + \sqrt{3[c/c_k-1/4]}$ we arrive at: \BE \label{c-hm-asymp-2}
h_{\rm max} - 1 \approx 1.29 \left(-\frac{3}{2} +
\sqrt{3\left[\frac{c}{c_k} - \frac{1}{4}\right]} \right). \EE
Figure~\ref{fig-KaRb-c-hmv50} presents the dependency of $h_{\rm
max}-1$ on the relative speed $c/c_k -1$ for the results in
figure~\ref{fig-sol} corresponding to the WRIBL model
(\ref{model2shk}) and to silicon oil v50. The direction of
increasing reduced Reynolds number $\D$ along the curves is
indicated by arrows. There is good agreement with relation
(\ref{c-hm-asymp}) in the limit $c \approx c_k$. A good agreement
with relation (\ref{c-hm-asymp-2})
 is also found in the case of the fast waves ($c/c_k
> 6$), typical of the drop-like regime observed at small Goucher
numbers. In the drag-inertia regime, however, at large values of
$\D$, the speed of the waves saturates to the limit $c_{\rm
crit}(\aN,\beta/\D)$ and no univocal mapping of the amplitude to the
wave speed is observed.

When the drop-like regime is affected by inertia for weakly viscous
liquids, the linear relation (\ref{c-hm-asymp-2}) is no longer
applicable reflecting the influence of the K mode on the drop-like
waves and in particular the steepening of the wave fronts (see
Fig.~\ref{fig-KaRb-c-hmv1}).

It is then clear from the above, that equation~(\ref{c-hm-asymp-2})
provides a good approximation to the speed-to-amplitude dependence
for solitary waves both in the drop-like and the drag-gravity
regime.

\section{Traveling waves}
\label{S-TW} In this section, we compare periodic TW solutions of
the WRIBL model~(\ref{model2shk}) ---\ie limit cycles of the
dynamical system (\ref{ssdyn})--- with ($\e \neq 0$) and without
viscous dispersion ($\e\to0$) and of the CM~equation (\ref{KDB})
(both $\D\to0$ and $\e\to0$), for the experimental conditions
considered by~\cite{Kli01} and~\cite{Dup07}.
We revisit the description of the TWs branches of solutions given in
\cite{Ruy08} with reference to the characteristics of TWs of
infinite extension, i.e. as TWs approach homoclinicity. In the
experimental set-ups of~\cite{Kli01,Dup07} the flow rate was
controlled and maintained at a constant value at the inlet.
Amplification of the ambient noise resulted in a wavy dynamics with
waves traveling with constant shapes and speeds over long distances.
Periodic TWs can be produced experimentally by applying periodic
perturbations at the inlet \cite[]{Liu94,Dup09a}. If the signal
remains periodic in time at each location in space, an integration
in time of the mass balance (\ref{q}) shows that the time average of
the flow rate,
 $T^{-1} \int_0^T q \,dt$, where $T$ is the period,
is conserved all along the fibre and is equal to its value at the
inlet which gives the condition $T^{-1} \int_0^T q \,dt = 1/3$
\cite[]{Sch05}. Periodic TW solutions of the model equations aiming
to describe the experimental conditions must therefore verify the
condition $\langle q \rangle\equiv k/(2\pi) \int_0^{2\pi/k} q\,d\xi
=1/3$, where $k$ again denotes the wavenumber.

\begin{figure}
\begin{center}
\BPS
\psfrag{c/ck}{$c/c_k$}
\psfrag{k/kRP}{$k/k_{\rm RP}$}
\psfrag{h}{$h$}
\psfrag{10x}{(iii)}
\psfrag{19}{(ii)}
\psfrag{20}{(i)}
\psfrag{16}{(iv)}
\psfrag{14}{(v)}
\psfrag{12}{(vi)}
\psfrag{9}{(vii)}
\psfrag{10}{(viii)}
\psfrag{11}{(ix)}
\psfrag{g1}{$\gamma_1$}
\psfrag{g2}{$\gamma_2$}
\subfigure[]{\label{Kliac_a}
\includegraphics[width=0.45\textwidth]{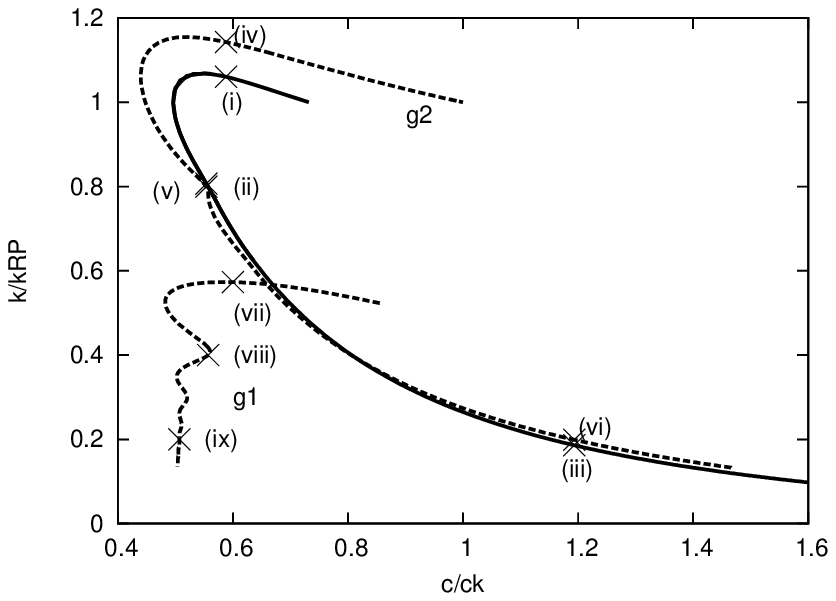}}\hfill
\subfigure[]{\label{Kliacha}
\includegraphics[width=0.45\textwidth]{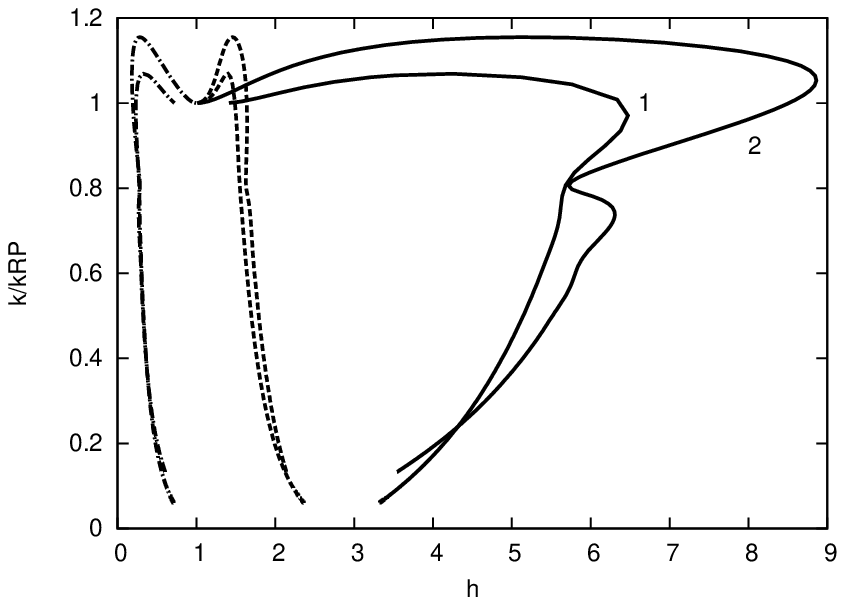}}
\subfigure[]{\label{Kliac_b}
\begin{minipage}[c]{0.6\textwidth}
\includegraphics[width=0.32\textwidth]{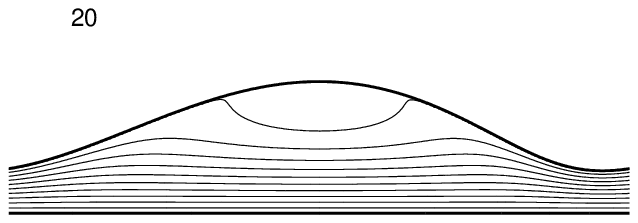}
\includegraphics[width=0.32\textwidth]{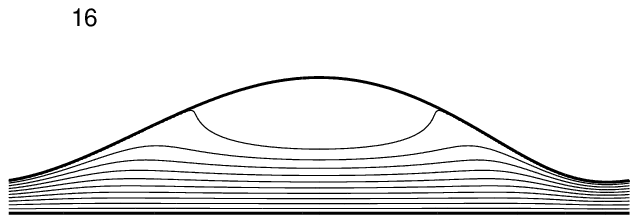}
\includegraphics[width=0.32\textwidth]{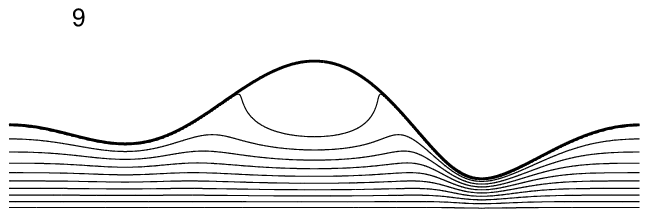}\\
\includegraphics[width=0.32\textwidth]{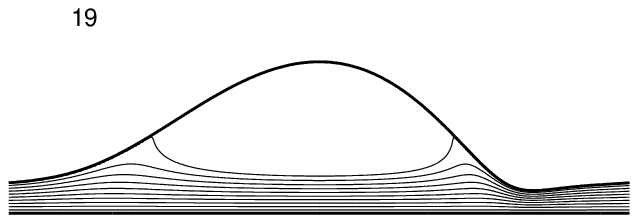}
\includegraphics[width=0.32\textwidth]{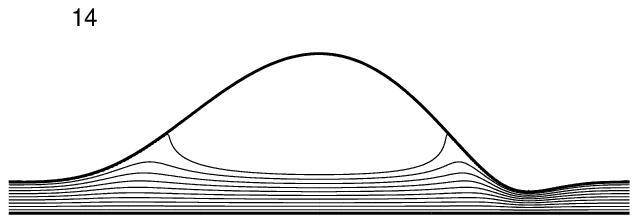}
\includegraphics[width=0.32\textwidth]{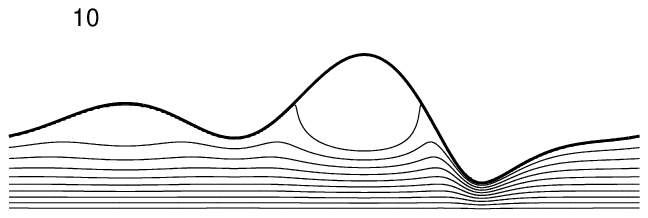}\\
\includegraphics[width=0.32\textwidth]{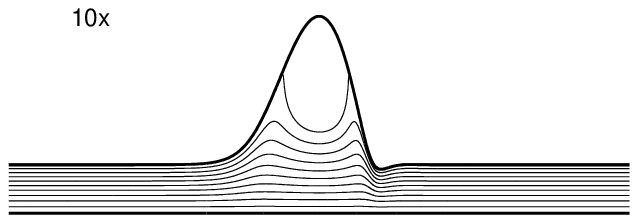}
\includegraphics[width=0.32\textwidth]{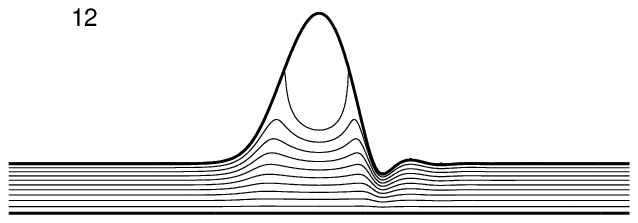}
\includegraphics[width=0.32\textwidth]{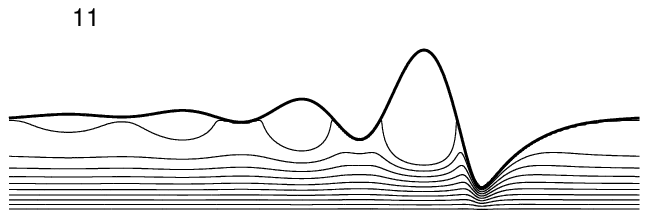}\\
\end{minipage}}
\EPS
\end{center}
\caption{TW branch of solutions bifurcating from the marginal
stability conditions.(a) Normalised speed $c/c_k$ as function of the
normalised wavenumber $k/k_{\rm RP}$ where $k_{\rm RP} =
\sqrt{\beta}/(1+\aN)$ (see text). Solid and dashed lines refer to
(\ref{model2shk}) and to the CM~equation (\ref{KDB}), respectively.
 (b) Maximum
thickness $h_{\rm max}$ (dashed line) and substrate thickness $h_s$
(dashed-dotted line) as functions of the normalised wavenumber
$k/k_{\rm RP}$. The solid line refers to the relative maximum thickness $h_{\rm
max}/h_s$.  Labels~1 refer to model (\ref{model2shk}), whereas
labels~2 refer to the CM~equation (\ref{KDB}).
(c) Wave profiles and streamlines in the moving frame for solutions
indicated by crosses in panel~a; left: solutions to
(\ref{model2shk}); right: solutions to (\ref{KDB}).
Parameters correspond
to the experimental conditions of \cite{Kli01}: $\qN=5.3$~mg/s and
$R=0.25$~mm ($\Go=0.138$, $\aN=2.04$, $\D=0.01$, $\e=0.18$ and
$\beta^\star = 2.1$). \label{Kliac}}
\end{figure}

Let us first consider the experimental conditions corresponding to
the very viscous fluid considered by \cite{Kli01} (castor oil,
$\G=0.45$; see Table~\ref{table2}). The bifurcation diagram of
TW~solutions of the WRIBL model (\ref{model2shk}) and of the
CM~equation (\ref{KDB}) are compared in Figure~\ref{Kliac_a}. The
parameters in the figure correspond to regime `c' reported by
Kliakhandler {\it et al.} ($\qN=21$~mg/s and $R=0.25$~mm). We have
normalised the wavenumber $k$ with the reference $k_{\rm RP} \equiv
\sqrt{\beta}/(1+\aN)$ corresponding to the marginal stability
condition (real $k$ ad $\o$) for linear waves solutions to the
CM~equation
$k_{\rm RP}$ corresponds to a dimensional wavelength $2\pi(\bar R +
\bhN)$ proportional to the diameter of the liquid cylinder.

Figure~\ref{Kliac_b} shows corresponding wave profiles with
regularly spaced streamlines in the moving frame, i.e. iso-contours
of the function $\psi = \int_R^r \ux\,rdr +c (R^2-r^2)/2$ at levels
$n \q0/N$, $1\le n \le N$ with $N=10$. The surface of the fibre
corresponds to $\psi=0$ and the free surface to $\psi = \q0$.
In the case of the WRIBL model (\ref{model2shk}), only one branch of
TW~solutions has been found emerging from the marginal linear
stability conditions  at $k \approx k_{\rm RP}$ through a Hopf
bifurcation, whereas a secondary branch has also been found by
period doubling for the CM equation (\ref{KDB}). We note that weakly
nonlinear TW~solutions of~(\ref{model2shk}) travel at a smaller
speed than the TW~solutions of the CM~equation (\ref{KDB}) since the
speed of linear kinematic waves is significantly affected by the
second-order viscous effects~\cite[]{Ruy08}. At small wavenumbers,
TW solutions of the branch emerging from $k \approx k_{\rm RP}$
accelerate, become more and more localized and terminate into the
single-hump solitary waves discussed in~\S\,\ref{S-dyn}. The speed,
amplitude and shape of the solutions of (\ref{model2shk}) and
(\ref{KDB}) are comparable in this limit. Following the terminology
introduced by Chang and co-workers for falling films on a planar
substrate~\cite[]{Cha93a,Cha02}, we refer to this branch of
solutions as the `$\gamma_2$ waves'. The branch of solutions to
(\ref{KDB}) emerging through period doubling terminates by slow
waves with a shape made of a trough followed by capillary ripples.
Following again \cite{Cha93a,Cha02} for planar substrates, we refer
to these TWs as the `$\gamma_1$ waves'.

For the conditions of Fig.~\ref{Kliac}, the fibre radius is small
compared to the capillary length ($\Go =0.138$) and the Nusselt film
thickness is comparable to the fibre radius ($\aN=2.04$). As a
consequence, the RP instability mechanism is strong and TWs have
amplitudes comparable to the fibre radius, which corresponds to the
typical situation of the `drop-like' regime of the solitary waves
discussed in \S~\ref{S-drop}. Yet, a direct transposition of the
results obtained in \S~\ref{S-drop} to solitary-like wavetrains in
the limit $k\to0$ is not possible since the reference thickness of
the substrates of the solitary-like waves is not constant (recall
that in the treatment of solitary wave solutions, the constant film
thickness far from the solitary waves was the reference thickness).
 We have computed the parameter $\beta_s^\star$ based on the substrate
thickness $h_s$ defined by the position of the fixed point in the
phase space when the corresponding limit cycle approaches
homoclinicity.  The maximum of the relative thickness $h_{\rm
max}/h_s$, $6.5$ here, is reached for a value of the local number
$\beta_s^\star$ close to its maximum, namely $3.65$, as expected
since the waves' characteristics are piloted by the balance between
the advection by the flow and the RP instability in the drop-like
regime. Figure~\ref{Kliacha} shows the evolution of the maximum
thickness, $h_{\rm max}$, of the substrate thickness $h_s$ and of
the ratio of the two as function of the normalised wavenumber
$k/k_{\rm RP}$. At small $k$, a TW is solitary-like and travels on a
portion of nearly flat substrate of increasing length. The mass
transported by the wave decreases in comparison to the mass carried
by the substrate and $h_s$ asymptotes to the Nusselt film thickness.
As a consequence $h_s$ increases as $k$ tends to zero. This trend is
followed by the maximum height $h_{\rm max}$. As a TW gets more and
more localized, it tends to have bigger and bigger size. However,
the maximum relative thickness, $h_{\rm max}/h_s$, is a rapidly
decreasing function of $k$ at small wavenumbers. Indeed, larger
substrate thicknesses imply weaker RP instability mechanisms and
stronger advection: $\beta_s^\star$ (not shown) follows a trend
similar to $h_{\rm max}/h_s$. We therefore obtain a rather
paradoxical picture: Despite the weakening of the RP instability ---
thus the lowering of $h_{\rm max}/h_s$---, a TW has its absolute
amplitude augmented as $k$ tends to zero.

\begin{figure}
\begin{center}
\BPS
\psfrag{c/ck}{$c/c_k$}
\psfrag{k/kRP}{$k/k_{\rm RP}$}

\psfrag{h}{$h$}
\psfrag{i}{(i)}
\psfrag{ii}{(ii)}
\psfrag{iii}{(iii)}
\psfrag{iv}{(iv)}
\psfrag{v}{(v)}
\psfrag{vi}{(vi)}
\psfrag{5}{(vii)}
\psfrag{4}{(viii)}
\psfrag{4b}{(ix)}
\psfrag{g1}{$\gamma_1$}
\psfrag{g2}{$\gamma_2$}
\subfigure[]{\label{KliacR0.5a}
\includegraphics[width=0.45\textwidth]{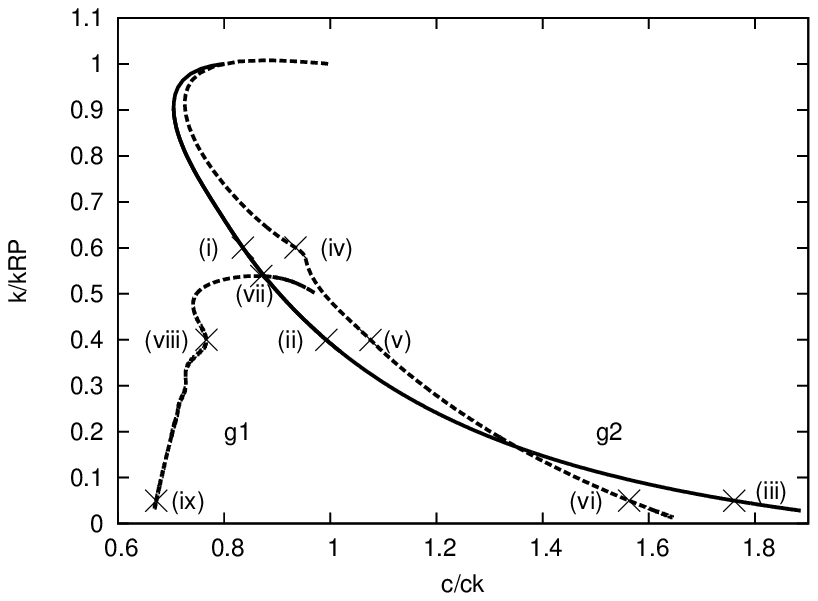}}\hfill%
\subfigure[]{\label{Kliachb}
\includegraphics[width=0.45\textwidth]{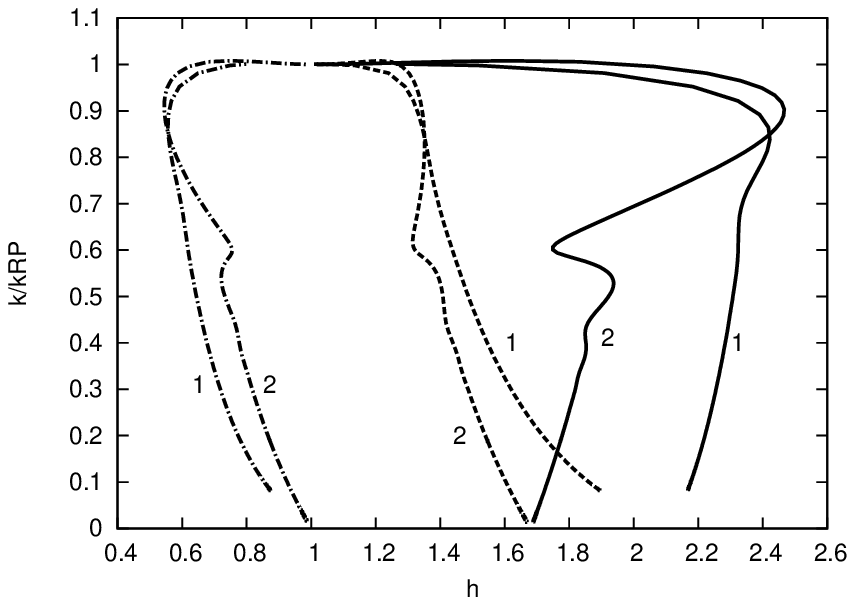}}
\subfigure[]{\label{KliacR0.5b}
\begin{minipage}[c]{0.6\textwidth}
\includegraphics[width=0.32\textwidth]{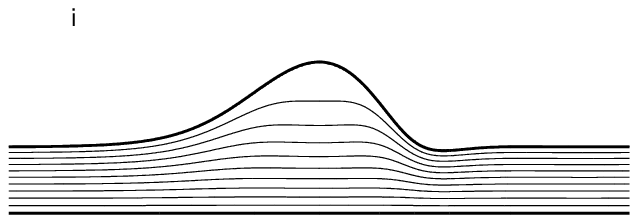}
\includegraphics[width=0.32\textwidth]{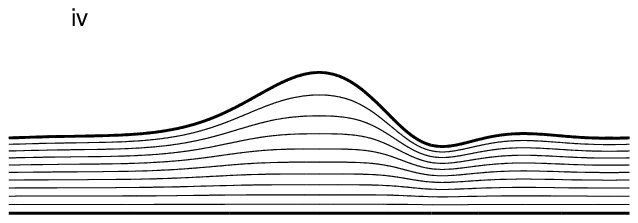}
\includegraphics[width=0.32\textwidth]{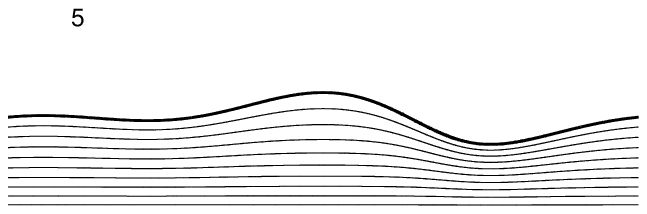}\\
\includegraphics[width=0.32\textwidth]{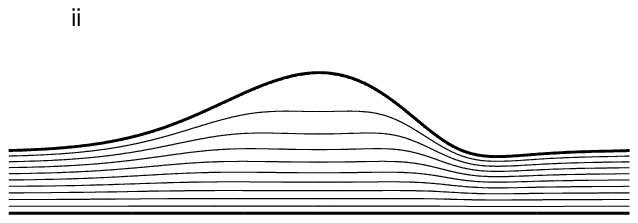}
\includegraphics[width=0.32\textwidth]{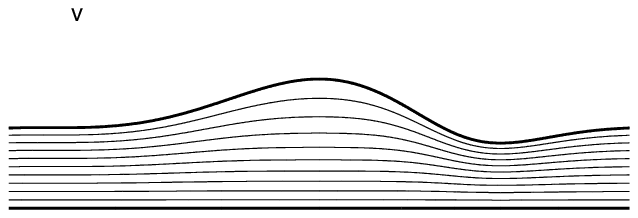}
\includegraphics[width=0.32\textwidth]{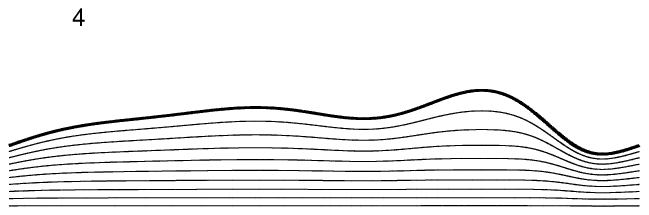}\\
\includegraphics[width=0.32\textwidth]{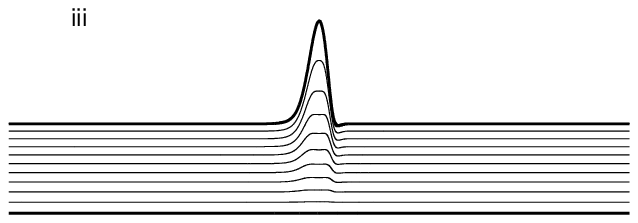}
\includegraphics[width=0.32\textwidth]{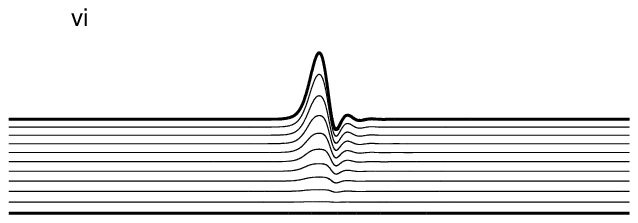}
\includegraphics[width=0.32\textwidth]{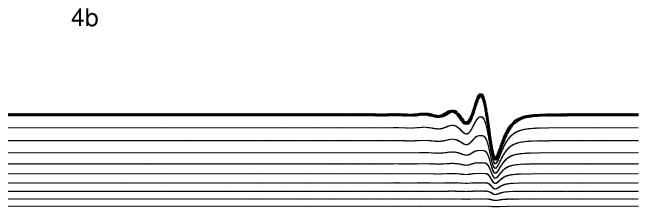}
\end{minipage}}
\EPS
\end{center}
\caption{TW branch of solutions bifurcating from the marginal
stability conditions. Parameters correspond to castor oil
($\G=0.45$), $\qN=10$~mg/s and $R=0.5$~mm ($\D=0.01$, $\e=0.22$,
$\aN=1.15$ and $\beta^\star=1.16$). See caption of
figure~\ref{Kliac}.} \label{KliacR0.5}
\end{figure}

For the experiments by Kliakhandler {\it et al.}, the RP instability
mechanism is strong while inertia and viscous dispersion are weak
($\D=0.01$, $\e=0.18$) which explains the good agreement obtained
with the results from the CM~equation (\ref{KDB}). However, as the
RP~instability is weakened by raising the Goucher number, streamwise
viscous dispersion should play an increasingly important role. We
have checked this by computing the TW~solutions of~(\ref{model2shk})
and~(\ref{KDB}) corresponding to a flow rate $\qN=10$~mg/s and a
larger radius $R=0.5$~mm, i.e. for $\D=0.01$, $\beta^\star=1.16$ and
$\e=0.22$, parameters that are comparable to those corresponding to
the regime `c' discussed by Kliakhandler {\it et al.} ($\D=0.01$,
$\beta^\star=2,1$ and $\e=0.18$) but with a value of $\beta^\star$
$\sim$ two times smaller. Branches of solutions are compared in
figure~\ref{KliacR0.5a} in the $(k/k_{\rm RP},c/c_k)$-plane.
$\gamma_2$ fast TW solutions of (\ref{KDB}) are again observed to
emerge from the marginal conditions $k\approx k_{\rm RP}$ whereas
$\gamma_1$ solutions are obtained through a period doubling
bifurcation of the $\gamma_2$ branch. We note again that only the
$\gamma_2$ branch of solutions can be found for the WRIBL model
(\ref{model2shk}), with no period doubling bifurcation being
detected in this case. We therefore conclude that the inclusion of
second-order viscous effects reduces the number of wave families
and, as a consequence, it may drastically simplify the complex
sequence of bifurcations and topological structure of associated
bifurcation diagrams.
In fact, such a reduction of the number of wave families by the
second-order viscous terms was also evidenced in the planar case
\cite[]{Ruy99}.

In comparison to Fig.~\ref{Kliac_a}, the $\gamma_2$ branches
portrayed in figure~\ref{Kliac_a} now deviate significantly from
each other even at small wavenumbers. The TW solutions of
model~(\ref{model2shk}) have a larger amplitude than the TW
solutions of equation~(\ref{KDB}) as indicated in
Fig.~\ref{KliacR0.5} which compares the wave profiles of waves of
equal wavenumber. When approaching homoclinicity, this effect is
enhanced, and solutions of~(\ref{model2shk}) travel with a notably
higher speed and larger amplitude than solutions of~(\ref{KDB}).
They are also preceded by smaller capillary ripples.

Figure~\ref{Kliachb} compares $h_s$ and $h_{\rm max}$. Again, the
substrate thickness and the maximum thickness have the same trend as
$k$ tend to zero whereas the maximum relative thickness $h_{\rm
max}/h_s$ evolves in the opposite direction. However, a comparison
of Fig.~\ref{Kliachb} to Fig.~\ref{Kliacha} reveals that the maximum
reached by $h_{\rm max}/h_s$ is $\sim$ three times smaller than with
the experimental conditions considered by \cite{Kli01}. Computations
of the local values of the parameters based on the substrate
thickness show that $\beta_s^\star$ does not exceed $1.41$ and
decreases with $k$ whereas $\eta_s$ increases. The local parameters
are $\beta_s^\star=1.2$ and $\e_s=0.19$ for the wave profile
labelled (iii) in Fig.~\ref{KliacR0.5} so that viscous dispersion
effects start to be comparable with the RP instability mechanism. We
conclude that viscous dispersion has a non-trivial amplifying effect
on the RP~instability mode.

\begin{figure}
\begin{center}
\BPS
\psfrag{k/kRP}{$k/k_{\rm RP}$}
\psfrag{h}{$h$}
\psfrag{D}{$\D_s$}
\psfrag{bs}{$\beta^\star_s$}
\psfrag{D,bs}{$\D_s$, $\beta^\star_s$}
\subfigure[]{\label{q0.151Rb0.23h}%
\includegraphics[width=0.45\textwidth]{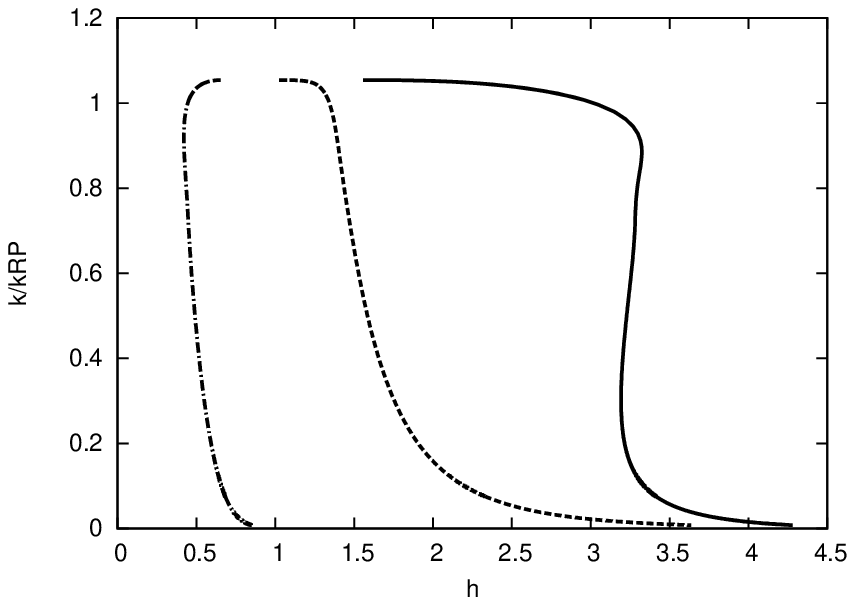}}\hfill%
\subfigure[]{\label{q0.151Rb0.23Dbs}%
\includegraphics[width=0.45\textwidth]{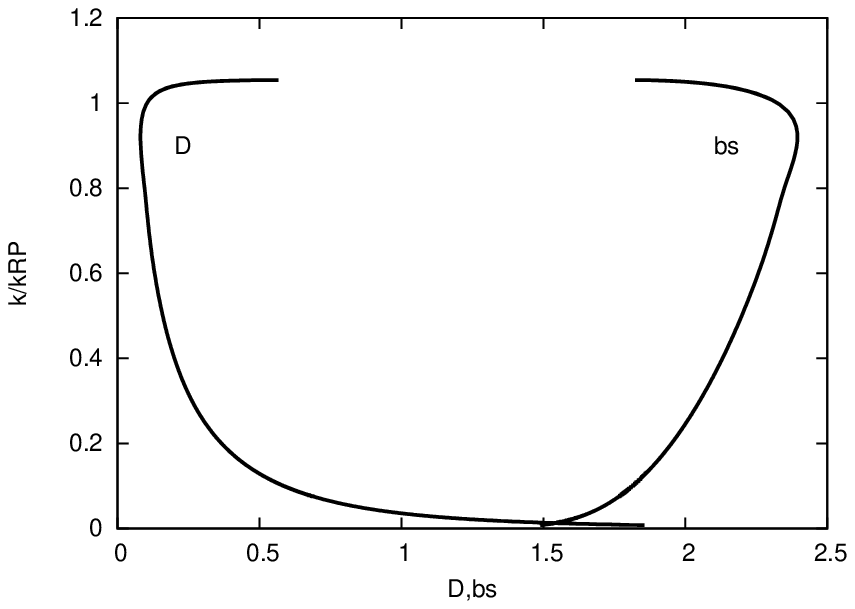}}
\EPS
\end{center}
\caption{(a) Maximum thickness $h_{\rm max}$ (dashed line) and
substrate thickness $h_s$ (dashed-dotted line) versus the normalised
wavenumber $k/k_{\rm RP}$. The solid line refers to the relative
thickness $h_{\rm max}/h_s$. (b) Local reduced Reynolds number
$\D_s$ and $\beta^\star_s$ versus $k/k_{\rm RP}$. Parameters
correspond to Rhodorsil v50 silicon oil ($\G=5.48$), $\qN=151$~mg/s
and $R=0.23$~mm ($\Go=0.155$, $\D=3.9$, $\aN=3.0$ and $\e=0.36$). }
\label{q0.151Rb0.23-h-Dbs}
\end{figure}

For less viscous fluids like Rhodorsil silicon oil v50, one may
expect a more significant effect of inertia and of the K~mode of
instability.
In Fig.~\ref{q0.151Rb0.23h}, we compare the substrate, maximum and
relative maximum thicknesses, $h_s$, $h_{\rm max}$ and $h_{\rm
max}/h_s$, respectively, corresponding to the TW branch of solutions
to (\ref{model2shk}). Parameters are chosen to correspond to one of
the experimental conditions considered by \cite{Dup07} ($R=0.23$~mm,
$\qN=151$~mg/s) and to a Goucher number $\Go=0.155$ close to the
corresponding one for the experiments in \cite{Kli01} ($\Go=0.137$).
As $k$ is lowered and the TWs approach homoclinicity, $h_s$ and
$h_{\rm max}$ grow, a trend similar to what is observed in
figures~\ref{Kliacha} and \ref{Kliachb} for a more viscous fluid.
However, the relative amplitude $h_{\rm max}/h_s$ follows a
surprising non-monotonic behaviour with an S-shape curve in the
plane ($h$,$k/k_{\rm RP}$). This behaviour can be understood by
examining the local values of $\D$ and $\beta^\star$ based on the
substrate thickness $h_s$ and presented in
figure~\ref{q0.151Rb0.23Dbs}. As $k$ is lowered to zero,
$\beta_s^\star$ decreases whereas $\D_s$ varies in the opposite
direction. As a result, the two curves ultimately cross and very
long waves correspond to a relatively stronger K instability mode
than the initially dominant RP instability. We thus recover the
usual trend observed with the solitary waves propagating on a planar
film: Larger substrates imply larger amplitudes $h_{\rm max}$ and
also larger relative amplitude $h_{\rm max}/h_s$
\cite[]{Ale94,Cha02}.

\begin{figure}
\begin{center}
\BPS
\psfrag{k/kRP}{$k/k_{\rm RP}$}
\psfrag{h}{$h_{\rm max}$}
\psfrag{D}{$\D_s$}
\psfrag{bs}{$\beta^\star_s$}
\psfrag{D,bs}{$\D_s$, $\beta^\star_s$}
\subfigure[]{\label{R0.475q0.296h}%
\includegraphics[width=0.45\textwidth]{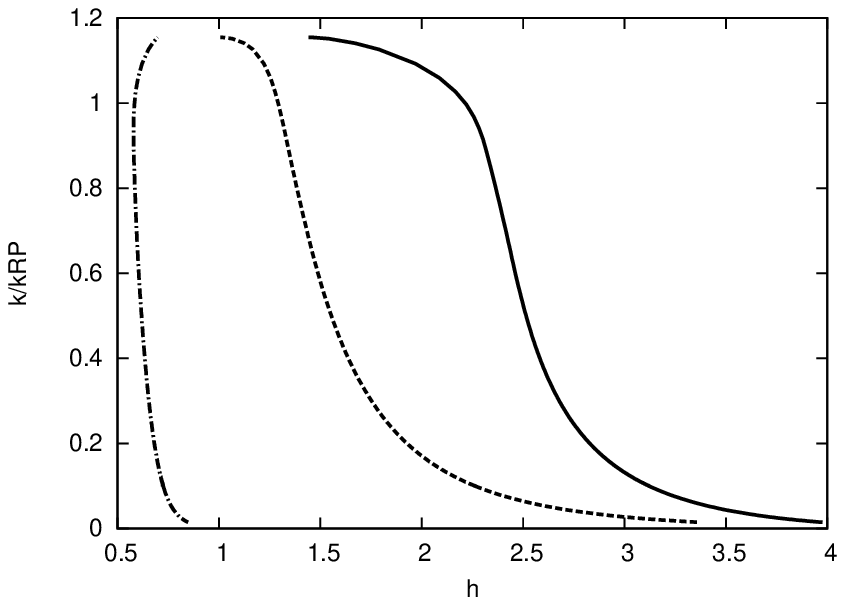}}\hfill%
\subfigure[]{\label{R0.475q0.296Dbs}%
\includegraphics[width=0.45\textwidth]{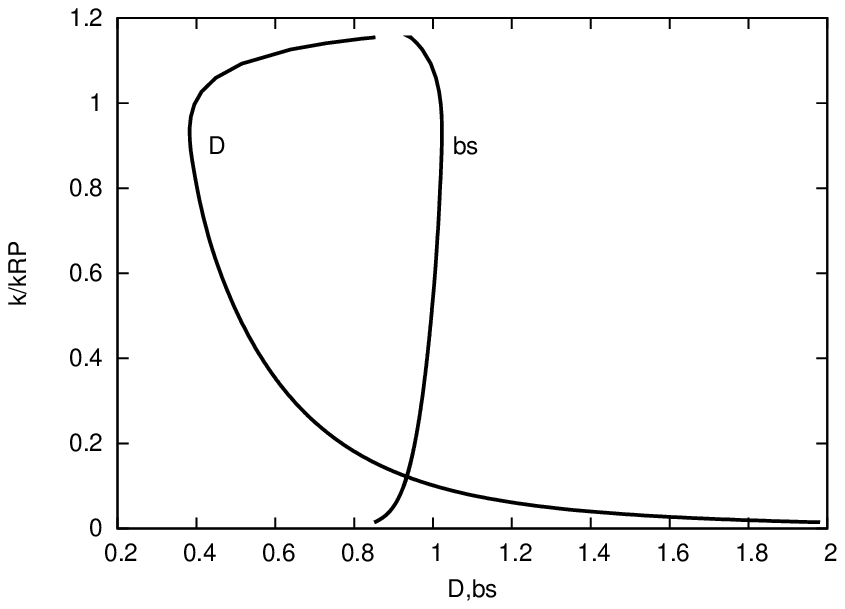}}
\EPS
\end{center}
\caption{(a) Maximum thickness $h_{\rm max}$ (dashed line) and
substrate thickness $h_s$ (dashed-dotted line) versus the normalised
wavenumber $k/k_{\rm RP}$. The solid line refers to the maximum
relative
 thickness $h_{\rm max}/h_s$. (b) Local reduced Reynolds number $\D_s$ and
$\beta^\star_s$ versus $k/k_{\rm RP}$. Parameters correspond to
Rhodorsil v50 silicon oil, $\qN=296$mg/s and $R=0.475$~mm
($\Go=0.32$, $\D=4.1$, $\aN=1.7$ and $\e=0.44$).}
\label{R0.475q0.296}
\end{figure}

We now turn to the experimental conditions considered in
\cite{Dup09a} corresponding again to Rhodorsil silicon oil v50 but
with a larger fibre, $R=0.475$~mm. The chosen flow rate
$\qN=296$~mg/s gives values of the reduced Reynolds number $\D$ and
of the viscous dispersion parameter
 $\e$ which do not vary significantly (compare e.g. $\D=4.1$ and $\e=0.44$ to $\D=3.9$ and $\e=0.36$).
Inertia and viscous dispersion effects are therefore comparable
whereas the RP instability mechanism is significantly lowered
($\beta^\star$ is reduced from $1.30$ to $0.77$). As for $R=0.5$~mm
and $\qN=10$~mg/s, we found a single branch of TW solutions
bifurcating from the marginal stability curve
 and the obtained
bifurcation diagram (not shown) is similar to Fig.~\ref{KliacR0.5}.
 Figure~\ref{R0.475q0.296h} compares the
corresponding substrate thickness $h_s$, the absolute and relative
amplitudes $h_{\rm max}$ and $h_{\rm max}/h_s$. As $k$ is lowered,
all curves exhibit a similar monotonic shape similar to what is
expected in the planar case. Again, this can be understood by
examining the local values of the two governing parameters (cf.
Fig.~\ref{R0.475q0.296Dbs}). The local reduced Reynolds number
$\beta_s^\star$ does not exceed unity whereas $\D_s$ sharply
increases for very long waves, which signals the prevalence of the K
mode over the RP mode.

Interestingly, TWs can be found at a wavenumber $k$ higher than the
critical wavenumber $k_c\approx k_{\rm RP}$ corresponding to the
marginal stability conditions, in which case the Nusselt solution is
linearly stable. This subcritical bifurcation of TW solutions from
the Nusselt uniform film occurs at small radii of the fibre, i.e.
small Goucher number (see figures~\ref{Kliac} and~\ref{KliacR0.5}).
The subcritical onset of TWs has also been reported in the recent
work by Novbari and Oron~\cite{Nov11} based on an `energy
 integral method' though
these authors did not give a physical interpretation of this
phenomenon. Since the RP instability mechanism is strong when
$\Go\ll1$, we can conjecture that the onset of subcriticality is
related to surface tension effects associated with the azimuthal
curvature (hence this effect is absent in the planar case). From a
thermodynamic viewpoint, surface forces tend to reduce the free
energy of the system and equilibrium is reached when the contact
area between the liquid and the surrounding gas is minimum.
Obviously, the flow is out of equilibrium and a thermodynamic
argument must be taken with care \footnote{Indeed, Kapitza himself
erroneously predicted the instability threshold of a falling film
using thermodynamic arguments~\cite[]{Kap48}, a question that was
settled with Benjamin's work a few years later~\cite[]{Ben57}.}.
Yet, when $\D \ll1$, $\e \ll 1$ and $\Go \ll 1$,
model~(\ref{model2shk}) reduces at leading order to the long-wave
Young-Laplace equation and the shape of the TWs thus only slightly
differs from the shape of static drops on fibres as was shown in
\S\,\ref{S-drop} and in Fig.~\ref{fig-prof-tab2}.

\begin{figure}
\begin{center}
\BPS
\psfrag{k/kRP}{$k/k_{\rm RP}$}
\psfrag{Anorm}{$A$}
\psfrag{hM}{$h_{\rm max}$}
\subfigure[]{\label{R0.02aN8-h}
\includegraphics[width=0.45\textwidth]{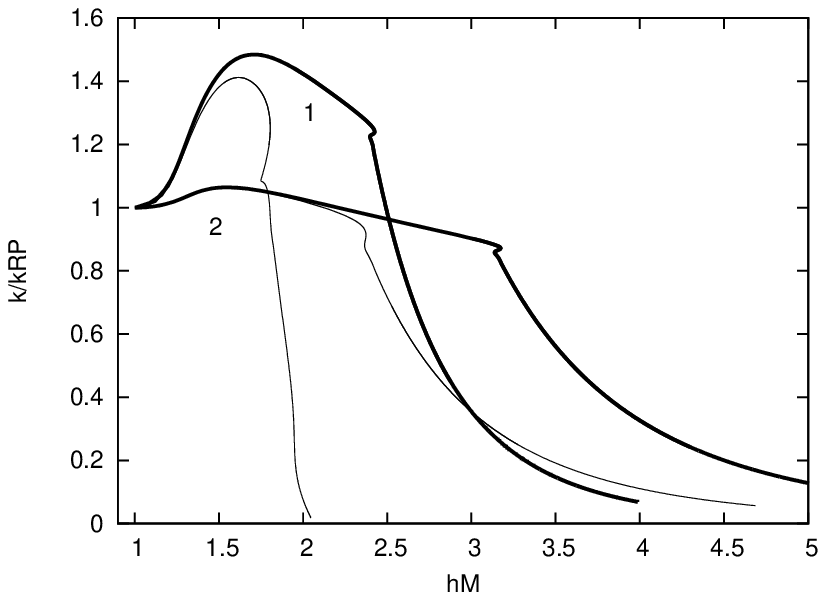}}\hfill%
\subfigure[]{\label{R0.02aN8-Anorm}%
\includegraphics[width=0.45\textwidth]{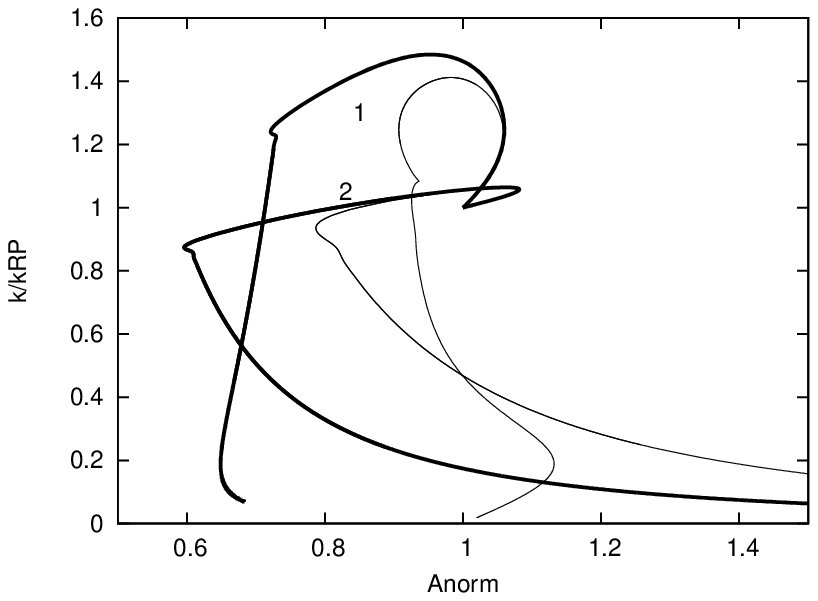}}
\EPS
\end{center}
\caption{\label{R0.02aN8} Traveling wave solutions to the CM
equation (\ref{KDB}) for $\aN=8$ and $\aN=0.5$. Thick and thin solid
lines refer to $\Go=0.02$ and $\Go=0.055$, respectively. Curves
labeled 1 (respectively 2) correspond to $\aN=8$ ($\aN=0.5$).}
\end{figure}

We have computed the TW solutions to the CM equation (\ref{KDB}) for
$\Go=0.02$ and $\Go=0.055$, and for $\aN=0.5$ and $\aN=8$. The
maximum amplitude $h_{\rm max}$, and the interfacial area $A$ of the
waves normalised by the area of the uniform film solution of the
same volume are displayed in Fig.~\ref{R0.02aN8}. The subcritical
onset of the TW solutions is accompanied by a normalized interfacial
area lower than unity, and therefore a lowering of the free surface
energy in comparison to the Nusselt uniform film solutions, which
supports our conjecture that subcriticality is promoted by capillary
effects. We note that the minimum wavenumber at which TWs are
observed seem to depend only on $\aN$ independently of $\Go$.
Conversely, at a given value of $\aN$, the amplitude of the waves is
strongly affected by the value of $\Go$, a small Goucher number
implying larger waves as expected, since the saturation number
$\beta=\aN^{2/3} \Go^{-4/3}$ is also larger.

\begin{figure}
\begin{center}
\BPS
\psfrag{k/kRP}{$k/k_{\rm RP}$}
\psfrag{aN}{$\aN$}
\includegraphics[width=0.45\textwidth]{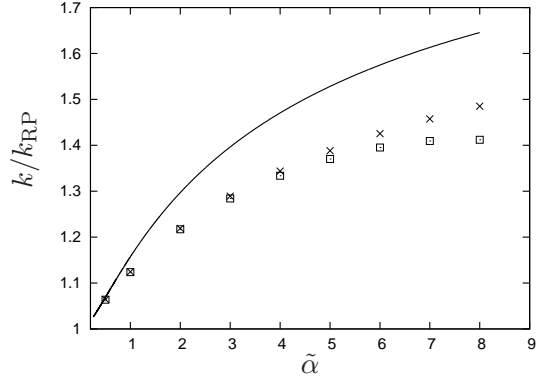}
\EPS
\end{center}
\caption{\label{R0.02LP}  Location of the maximum of the normalised
wavenumber $k/k_{\rm RP}$ as function of the aspect ratio $\aN$ for
the TW solutions of the CM equation (\ref{KDB}). Crosses
(respectively squares) correspond to $\Go=0.02$ ($\Go=0.055$). The
solid line denotes the static drop solution to (\ref{Kum88-LY-eq})
with a normalized area $A=1$ (see text and
Appendix~\ref{S-Static-Drops}). }
\end{figure}

In Fig.~\ref{R0.02LP} the maximum of the normalised wavenumber
$k/k_{\rm RP}$ of the TW solutions of the CM equation is depicted as
function of the aspect ratio $\aN$ for $\Go=0.02$ and $\Go=0.055$.
We note again that the location of this local maximum only weakly
depends on the Goucher number. The shape of the corresponding TW
solutions (not shown) is also nearly symmetrical. We thus conclude
that the subcritical behavior of the TW solutions is not affected by
gravity effects at small values of the Goucher number but is
strongly dependent on the parameter $\aN$ and therefore on the
geometry of the base flow.

By neglecting gravity effects and assuming a wettable substrate, the
shape of a static drop is governed by its length and volume.
Therefore, for a normalized area $A$ set to unity, the shape of a
static drop depends only on the aspect ratio $\aN$ of the uniform
film solution with the same volume. The solid line curve in
figure~\ref{R0.02LP} represents the wavenumber of static drops with
unit normalized areas. Above this curve, $A>1$ and the static drop
solution is not energetically favorable, whereas below it the static
drop solution is favored. Indeed, large drops have a nearly
spherical shape with an interfacial area that is lower than the
uniform film with the same volume, with further lowering of the
interfacial area being forbidden by the presence of the fibre. For a
given amount of liquid, the thicker the uniform film solution is,
the lower the normalized area of the static drop would be.
Therefore, the range of energetically favored wavenumbers of the
static drop solution increases with the aspect ratio $\aN$ as can be
observed from figure~\ref{R0.02LP}.

We can now deduce two more arguments from Fig.~\ref{R0.02LP} which
support a capillary origin of the subcritical behavior of the branch
of TW solutions that emerge from the Hopf bifurcation of the uniform
film solution: (i) Since all points lie below the solid curve, TWs
only exist when an energetically favorable static drop solution is
available; (ii) The trend of the maximum of the TW wavenumber with
respect to the aspect ratio $\aN$ is similar to the trend of the
boundary separating energetically favored and unfavored static drop
solutions.

\section{Phase diagram}
\label{S-Phase} 
\begin{figure}
\begin{center}
\BPS \psfrag{RP}{DL} \psfrag{VIS}{SOL} \psfrag{DI}{DI}
\psfrag{DG}{DG} \psfrag{aN}{$\aN$} \psfrag{R/lc}{$\Go$}
\psfrag{beta=1}{$\beta^\star=1$} \psfrag{D=1}{$\D=1$}
\psfrag{e=1}{$\e=1$} \subfigure[water ($\G=3376$)]{\label{domainev1}
\includegraphics[width=0.45\textwidth]{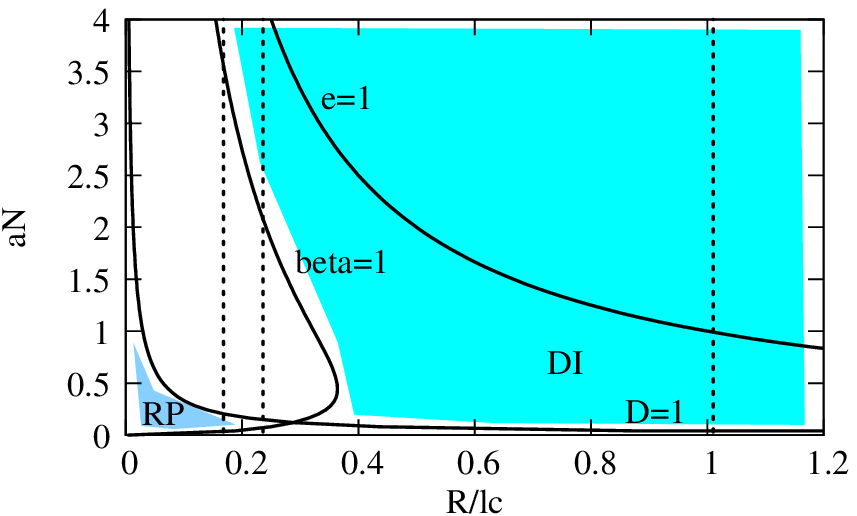}}\hfill%
\subfigure[silicon oil v50 ($\G=5.48$)]{\label{domainev50}
\includegraphics[width=0.45\textwidth]{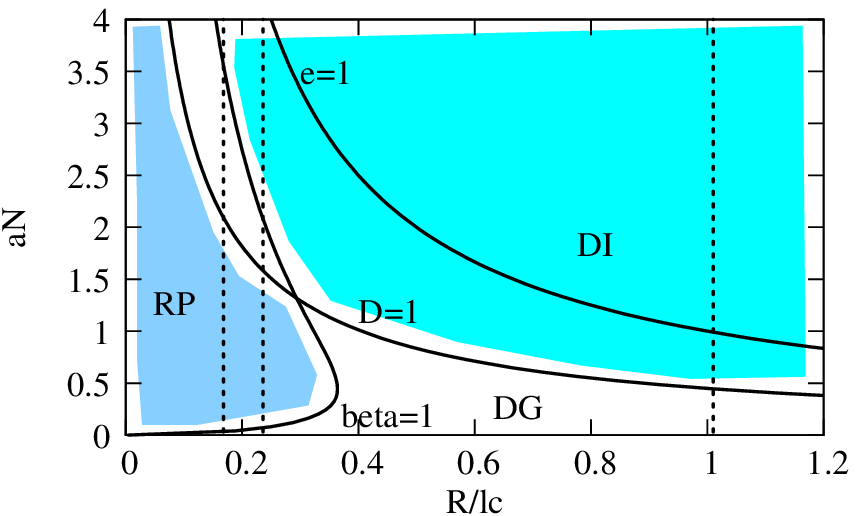}}\\
\subfigure[castor oil ($\G=0.45$)]{\label{domainev440}
\includegraphics[width=0.45\textwidth]{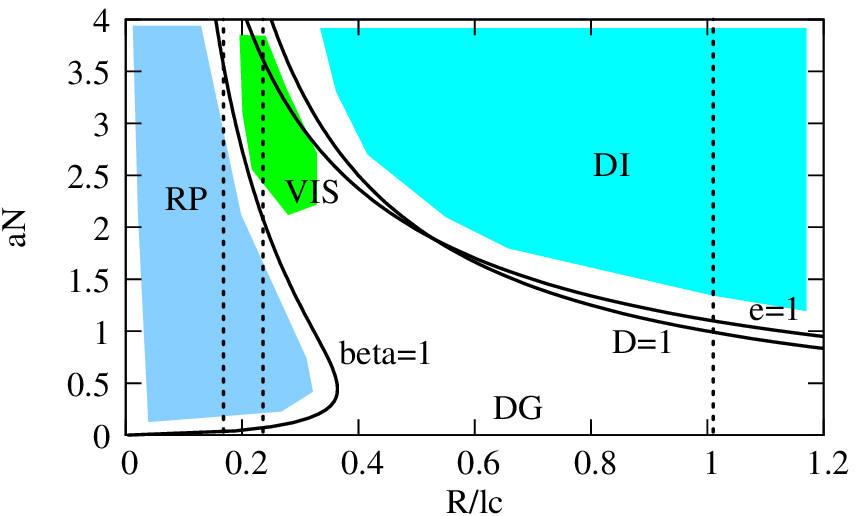}}\hfill%
\subfigure[silicon oil v1000 ($\G=0.10$)]{\label{domainev1000}
\includegraphics[width=0.45\textwidth]{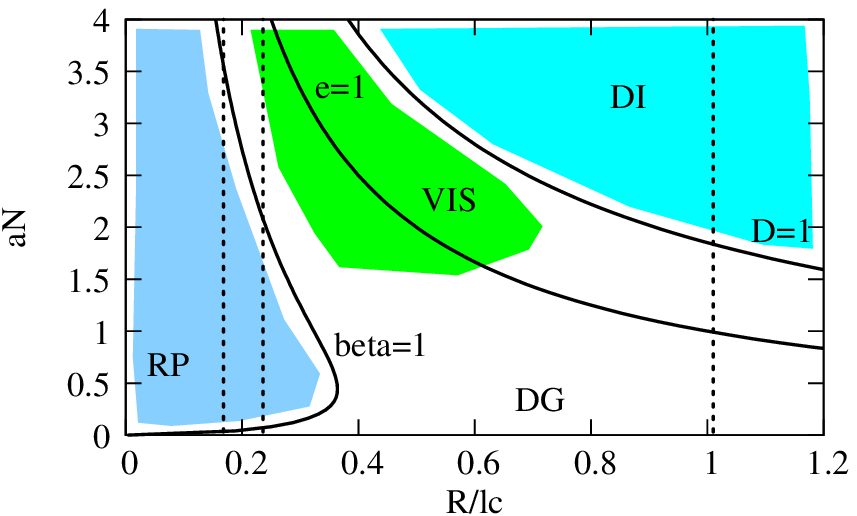}}
\EPS
\end{center}
\caption{Maps of the different regimes in the $\Go-\aN$ parameter
space for fluids of increasing viscosity. The different curves are
the loci of $\D=1$, $\e=1$ and $\beta^\star=1$. `DI' refers to the
`drag-inertia' regime, `DL' to the Rayleigh-Plateau regime, `SOL' to
the `soliton-like' regime and `DG' to the `drag-gravity' regime. }
\label{domaine}
\end{figure}

We are now in a position to give a phase diagram of
the different regimes found for all possible TW solutions on a film
flow down a fibre. The onset of the `drop-like' regime and the
`drag-inertia' regime corresponds roughly to $\beta^\star\approx 1$
and $\D \approx 1$. The `soliton-like' regime arises when the
instability mechanisms are weak ($\D \lessapprox 1$ and
$\beta^\star\lessapprox 1$), the film is thick, $\aN = O(1)$, and
viscous dispersion is strong, $\e = O(1)$. Finally, the
`drag-gravity' regime is observed when all other effects are weak
($\D \lessapprox 1$ and $\beta^\star\lessapprox 1$ and $\e \ll 1$).
Therefore, a phase diagram can be obtained for a given fluid, thus a
given Kapitza number $\G$, by drawing the curves $\D=1$, $\e=1$ and
$\beta^\star=1$ in the plane ($\aN$, $\Go$). Since
$\e=(\aN\Go)^{4/3}$ and $\beta^\star=\{\aN c_k(\aN)/[\Go^2
(1+\aN)^4]\}^{2/3}$ are functions of $\aN$ and the Goucher number
only, the corresponding curves $\beta^\star=1$ and $\e=1$ are
independent of the working fluid considered. Thus, $\D=1$ is the
only boundary that moves in the plane ($\aN$, $\Go$) when $\G$ is
varied. Figure~\ref{domaine} is a tentative representation of the
phase diagrams for the four working fluids considered in this study,
from weakly viscous fluids like water with a high Kapitza number,
$\G=3376$, to highly viscous fluids like silicon oil v1000
corresponding to a small Kapitza number, $\G=0.10$.
In the case of water, a large overlap region exists for the
`drop-like' and the `drag-inertia' regime corresponding to a mutual
reinforcement of the two K and RP instabilities. The `soliton-like'
regime takes over at relatively high viscosities where the curve
$\D=1$ moves below the curve $\e=1$.

\section{Conclusions and discussion}
\label{S-concl} We have presented new results and insights on the
characteristics of axisymmetric waves propagating down a fibre. Our
analysis was based on the two-equation model derived in \cite{Ruy08}
using a weighted-residuals approach, equation (\ref{model2shk}). The
model accounts for all physical effects, namely inertia, azimuthal
curvature and viscous dispersion and has been validated in the
studies by Ruyer-Quil \emph{et al.}~\cite{Ruy08} and Duprat \emph{et
al.}~\cite{Dup09a} through direct comparisons to the experiments by
Kliakhandler \emph{et al.}~\cite{Kli01} and Duprat \emph{et
al.}~\cite{Dup07,Dup09a}.

We first focused on isolated waves running on a constant thickness
substrate, or solitary waves. The dynamics of the film seems to be
dominated by these structures even when the flow becomes disordered
\cite[]{Kli01,Cra06}. We examined in detail, via both asymptotic
analysis and through elements from dynamical systems theory, the
shape, speed and amplitude of the waves for four fluids of
increasing viscosities: Water, Rhodorsil silicon oils v50, v1000 and
the castor oil utilised in the experiments by \cite{Kli01}.

We identified four distinct regimes corresponding to the competition
of the two instability modes, the K and RP modes, prompted by
inertia and azimuthal curvature (the fibre curvature effectively),
respectively, with the viscous dispersion (\ie the second-order
axial viscous diffusion and second-order viscous contributions to
the tangential stress at the free surface) and with the advection of
the structures by the flow, which results from the balance between
gravity and viscous drag. Two of these regimes are similar to what
is found in the planar geometry \cite[]{Oos99}. The `drag-gravity'
regime corresponds to the predominance of the flow advection over
the instability mechanisms, either when inertia effects are weak,
\ie for $\D \lessapprox 1$ or when the azimuthal curvature effects
are non-dominant, $\beta^\star \lessapprox 1$. In both cases it is
possible to interpret the `drag-gravity' regime as one where the
instability growth is arrested  by the flow which determines the
amplitude and speed of the solitary waves as reflected by the
asymptotic relations (\ref{power-laws-bis}). The `drag-inertia'
regime is observed at large reduced Reynolds numbers, $\D \gg 1$,
when the wave characteristics are determined by the balance of
inertia, drag and gravity. We have obtained the asymptotic limit of
the speed and showed that the rate of convergence to this limit is
governed by $\e/\D^2 \propto \R^{-2}$.

The `drop-like' regime corresponding to the predominance of the RP
instability mechanism over the flow advection. It is specific to the
cylindrical geometry and is observed for small fibre radii $R$
compared to the capillary length $l_c$, that is at small Goucher
numbers $\Go$, when the typical time of growth of the RP instability
is greater than the advection time of a wavy structure, \ie for
$\beta^\star \gtrapprox 1$. The maximum reachable amplitude and
speed of the waves in this regime is governed by the radius $R$ of
the fibre and the balance of gravity and viscous drag. Comparisons
to quasi-static drop solutions of the Laplace--Young equation
(\ref{Kum88-LY-eq}) sliding down a fibre with a speed verifying the
Landau--Levich--Derjaguin law (\ref{Landau}) show excellent
agreement, even in the case of  spherical drops where the long-wave
lubrication assumption does not strictly apply. In this regime,
waves have a drop-like nearly symmetrical shape determined by
capillary effects. The thickness of the substrate film on which the
drops slide is governed by the balance of viscosity and capillarity.
`Drop-like' TW solution branches subcritically emerge from the
Nusselt uniform film branch. We have given an explanation for this
subcritical onset based on geometric and thermodynamic arguments and
thus completed its recent investigation in~\cite{Nov11}. This
phenomenon arises from capillary effects and depends only on the
aspect ratio $\aN$ for sufficiently low Goucher numbers.

We have also found a possible fourth regime for very viscous fluids
and thick films $\aN = O(1)$), for which both K and RP instability
mechanisms are weak ($\D\lessapprox 1 $ and $\beta^\star \lessapprox
1$) and viscous dispersion is significant ($\e =O(1)$). This
`soliton-like' regime corresponds to the balance of the
nonlinearities with the dispersion induced by second-order viscous
effects, with the speed and amplitude of the solitary waves being
functions of the logarithm of the aspect ratio $\aN$.

Our study of the solitary-wave solutions has been followed by
construction of the TW branches of solutions corresponding to the
experimental conditions for which the average flow rate is the true
control parameter, with the substrate thickness being determined by
the solution itself. If the substrate thickness $h_s$ and the
maximum amplitude $h_{\rm max}$ grow when TWs approach
homoclinicity, the ratio of the two, $h_{\rm max}/h_s$, evolves in a
manner that strongly depends on which instability mechanism is
dominant. If the RP instability is dominant, $h_{\rm max}/h_s$
decreases as the wavenumber $k$ tends to zero, whereas if the K mode
is dominant $h_{\rm max}/h_s$ has an opposite trend. This picture
can be even more complex since the predominance of the instability
modes can be exchanged by varying the periodicity of the waves and
thus the substrate thickness (cf. Fig.~\ref{q0.151Rb0.23-h-Dbs}).
The selected wave regime depends not only on the properties of the
Nusselt flow at the inlet but also on the periodicity selected by
the system. Indeed, the boundaries separating the different regimes
are not only functions of $\Go$ and $\aN$ but also functions of the
thickness of the substrate, which is determined by the typical
distance separating solitary-like waves. Therefore, the phase
diagrams displayed in Fig.~\ref{domaine} must be taken with caution.
The wave selection process of a noise-driven falling film is the
complex result of the linear amplification of inlet perturbations
and the downstream nonlinear interaction mechanisms
\cite[]{Cha95,Kal07,Pra11}.

\begin{figure}
\begin{center}
\BPS \psfrag{x (mm)}{$x$ (m)} \psfrag{h (m)}{$h$ (mm)}
\psfrag{x}{{\large \CCW $x$}} \psfrag{t}{{\large \CCW $t$}}
\subfigure[$t=0$~s]{
\includegraphics[height=0.45\textwidth,angle=-90]
{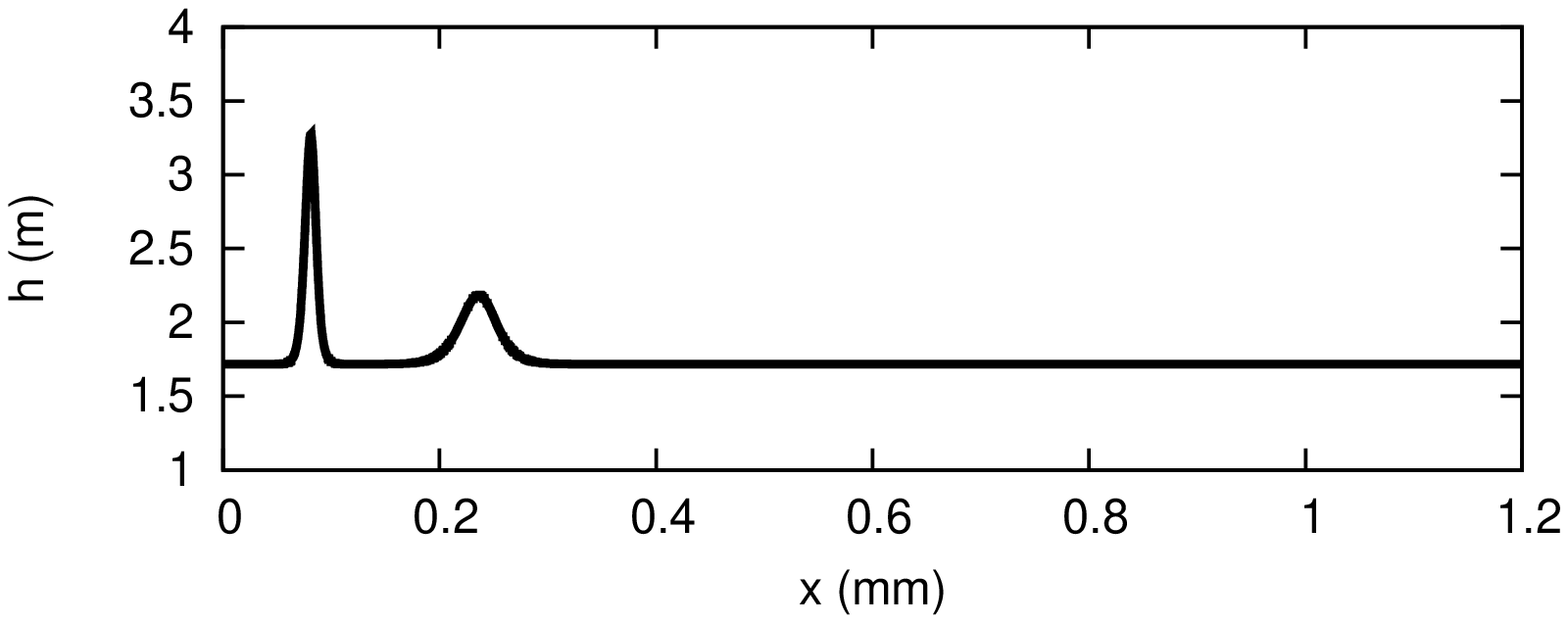}}\hfill%
\subfigure[$t=4$~s]{
\includegraphics[height=0.45\textwidth,angle=-90]
{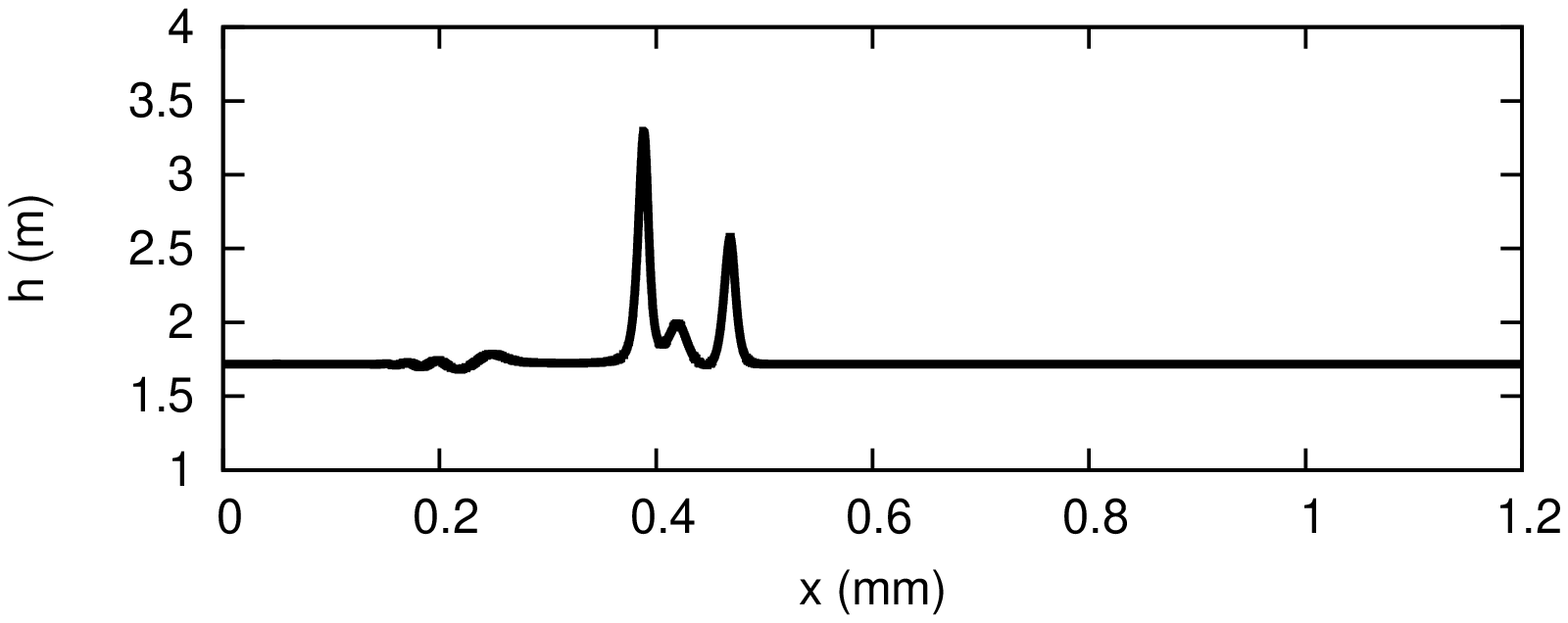}}\\
\subfigure[$t=8$~s]{
\includegraphics[height=0.45\textwidth,angle=-90]
{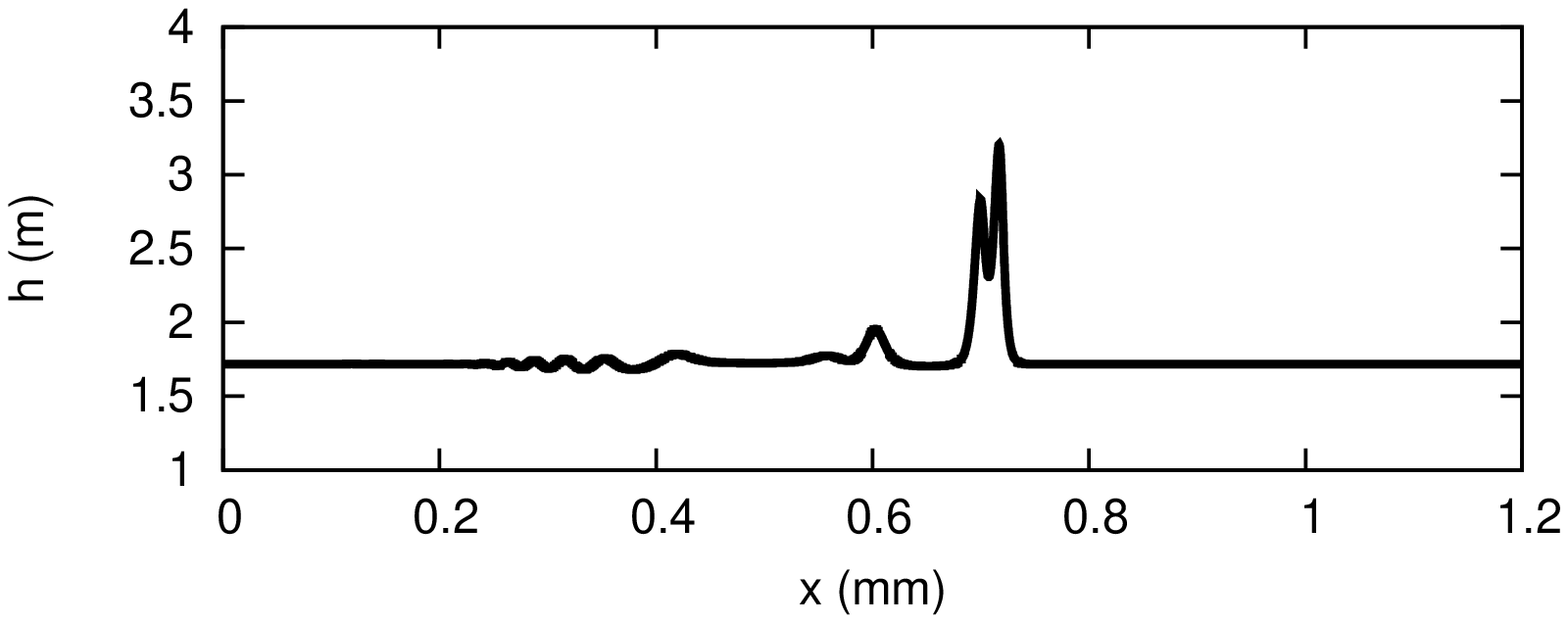}}\hfill%
\subfigure[$t=12$~s]{
\includegraphics[height=0.45\textwidth,angle=-90]
{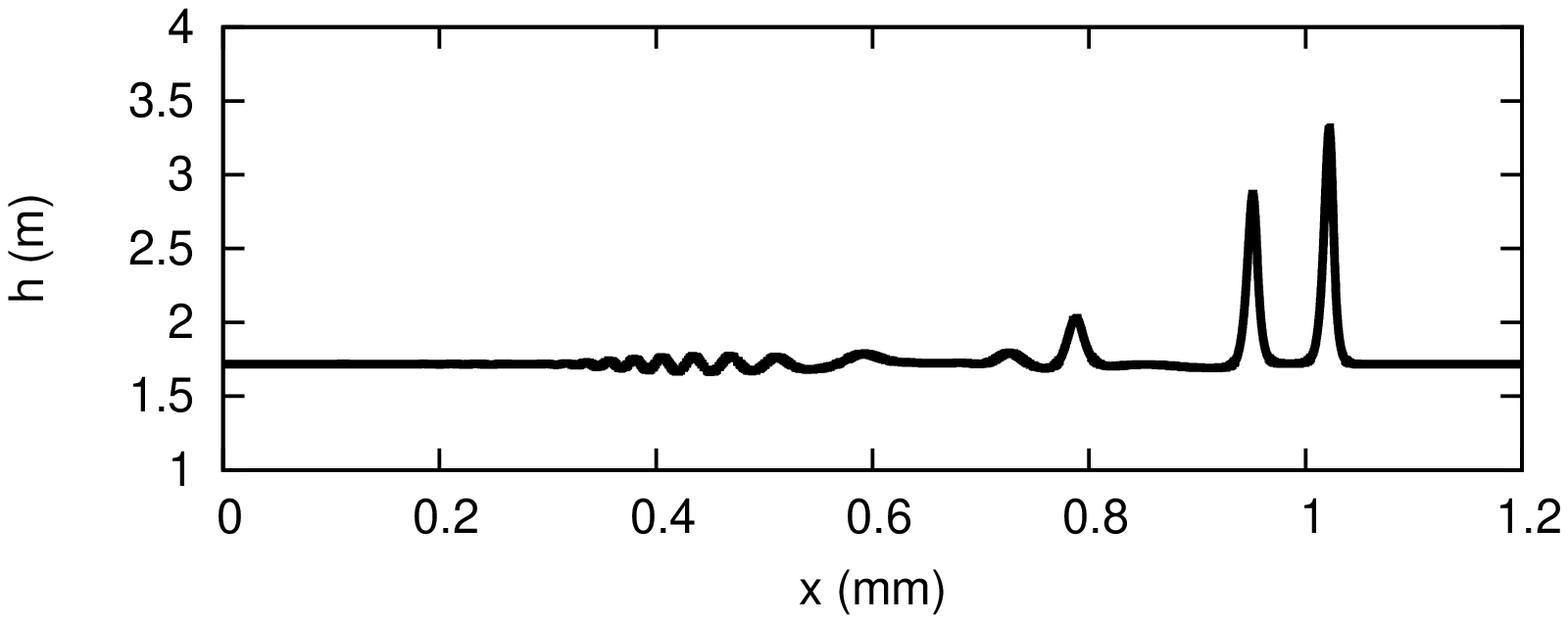}}\\
\subfigure[]{
\includegraphics[width=\textwidth,height=0.3\textwidth]
{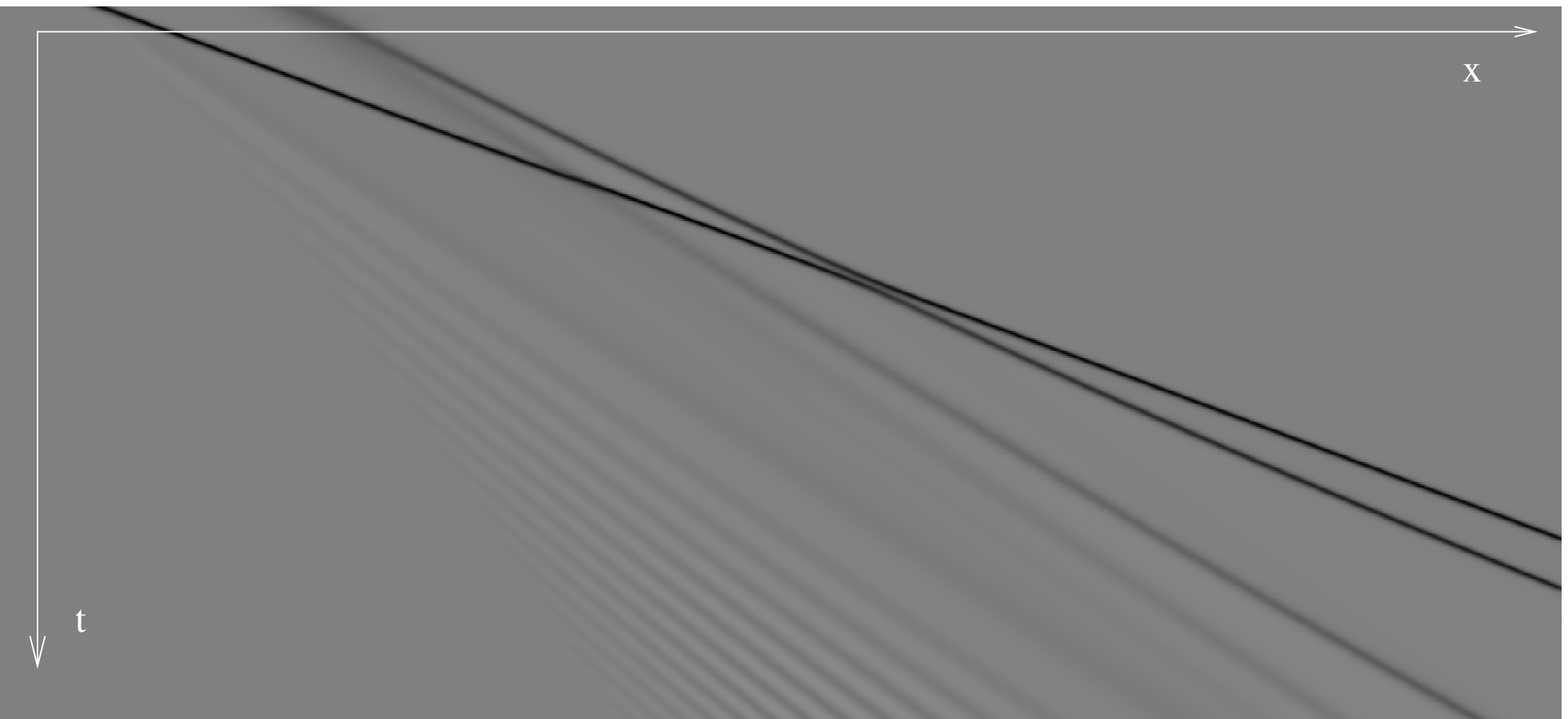}} \EPS
\end{center}
\caption{Simulation of the response of the film in the
`soliton-like' regime. Initial conditions consist of two pulse-like
perturbations of different amplitudes. Parameters are $\D=0.18$,
$\aN=1.9$ and $\e=1.2$ (Rhodorsil silicon oil v1000, $\qN=300$~mg/s
and $R=0.89$~mm). (a-d) Snapshots of the film thickness at
increasing times; (e) Spatio-temporal diagram. Vertical and
horizontal ranges are $1.25$~m and $28$~s respectively. Elevations
(depressions) of the free surface are coded in dark (light) grey.
\label{Rlc06-q03} }
\end{figure}

Noteworthy is that in our previous study on the fibre
problem~\cite[]{Ruy08}, we have shown that the wave selection
process of the noise-driven wave dynamics down the fibre is strongly
affected by viscous dispersion. This effect is generally weak in the
experiments devoted to falling films on planar substrate, where
working fluids were often weakly viscous, i.e.
water~\cite[]{Ale85,Liu94,Tih06}. In the case of films down fibres,
the oils used in the experiments are much more viscous than water
which explains that viscous dispersion can be dominant and can
promote a `soliton-like' regime. In this regime and despite the
dissipative nature of the flow, it is possible to observe the
formation of solitons, \ie solitary waves whose shape and speed are
not altered by collisions with other solitary waves (such waves are
still dissipative but they share several common features with
solitons in conservative systems~\cite[]{Chr95}). We explore such
effects in Fig.~\ref{Rlc06-q03} which depicts a simulation of the
interaction between two solitons. Chosen parameters correspond to
Rhodorsil silicon oil v1000 and $\Go = 0.6$. Two pulse-like
perturbations of different amplitudes are initially placed in the
computational domain (cf. panel~a). The excess mass carried by these
two perturbations are then drained behind the two pulses creating
two wave-packets. The spatial extend of these packets grows in time
due to the convective instability of the film~\cite[]{Dup07,Ruy08}.
Panel~b shows the interaction of the second pulse with the mass
ejected by the first pulse. The spatio-temporal diagram shown in
panel~e indicates that the second pulse first overtakes this excess
mass and next the first pulse without absorbing them (cf. panels~c
and d). The coalescence of the two pulses is accompanied by a phase
shift but no notable modifications of the speeds and amplitudes of
the pulses as in Hamiltonian systems: The initial wave profiles get
superimposed as the waves collide and reappear as the waves move
apart. Yet, a slow evolution of the amplitude and speed of the two
solitons can be observed towards the speed $c= 7.5$~cm/s and
amplitude  $h_{\rm max}=3.3$~mm of the infinite-domain solitary-wave
solution for the given set of parameters.

Our hope is that this new evidence of existence of solitons on a
liquid film flowing down a fibre, a dissipative system as opposed to
a Hamiltonian one, may motivate a renewed interest for the
experimental investigation of the resulting wave dynamics and, in
particular, on the role of viscous dispersion.

\begin{acknowledgments}
We acknowledge financial support through a travel grant supported by
the Franco-British Alliance Research Partnership Programme and from
the Multiflow ITN Marie Curie network funded by the European
Commission  (GA-2008-214919). The authors thank Imperial College and
Laboratoire FAST for hospitality.
\end{acknowledgments}

\begin{appendix}

\section{Coefficients of the model (\ref{model2shk})}
 \label{S-Coeffs} The expressions of
the coefficients entering the averaged momentum balance
(\ref{mom-shk}) consist of ratios of polynomials in $b$ and
 $\log(b)$, where $b=1 +\a h$ is the ratio
of the total radius $R+h$ to the fibre radius $R$:
 \BSE \label{F-I}
\BA
\label{def-phi}
\phi &=& \left[3 \left((4 \log(b)-3) b^4+4 b^2-1\right)\right]/
\left[16 (b-1)^3\right]\,,\\
\F &=& 3 \F_a/[ 16 (b-1)^2 \phi \F_b]\,,\\
\NNM
\F_a &=& -301 b^8+622 b^6-441 b^4+4 \log(b) \left\{197 b^6-234 b^4
+6 \log(b)
\right.\\ && \left.
\times\left[16 \log(b) b^4-36 b^4+22 b^2+3\right] b^2
+78 b^2+4\right\} b^2
+130 b^2-10 \,,\\
\F_b &=& 17 b^6+12 \log(b) \left[2 \log(b) b^2-3 b^2+2\right] b^4
-30 b^4+15 b^2-2\,,\\
\GG &=& \GG_a/[ 64 (b-1)^4 \phi^2 \F_b]\,,
\\ \NNM
\GG_a &=&
9 b \left\{
\vphantom{\left(b^2-1\right)^2}
4 \log(b) \left[-220 b^8+456 b^6-303 b^4
+6 \log(b) \left(61 b^6-69 b^4
\right. \right. \right. \\ &&  \NNM \left. \left. \left.
+4 \log(b) \left(4 \log(b) b^4-12 b^4+7 b^2+2\right)
 b^2+9 b^2+9\right) b^2+58 b^2+9\right]
   b^2
\right. \\ && \left.
+\left(b^2-1\right)^2 \left(153 b^6-145 b^4+53 b^2-1\right)\right\}\,,
\\
\I &=& 64 (b-1)^5 \phi^2/[3 \F_b]\,,\\
\J &=& 3\J_a/[128 (b-1)^4 \phi^2 \F_b]\,,\\
\NNM
\J_a &=& 9 \left\{\left(490 b^8-205 b^6-235 b^4+73 b^2-3\right)
\left(b^2-1\right)^3
\right. \\ \NNM  &&
+4 b^2 \log (b)
\left[2 b^4 \log (b) \left(72 \log (b)
\left(2 \log (b) b^4
 - 6 b^4+b^2+6\right) b^4
\right.\right. \\ && \left. \NNM
+(b-1) (b+1)
 \left(533 b^6-109 b^4-451 b^2+15\right)\right)
\\ && \left.\left.
-3 \left(b^2-1\right)^2
\left(187 b^8-43 b^6-134 b^4+17 b^2+1\right)\right]\right\}\,, \\
\K &=& 3 \K_a/[16 b^3 (b-1)^2 \phi \F_b]\,,\\
\K_a &=& 4 b^4 \log(b) \left(233 b^8-360 b^6+12 \log(b) \left(12 \log(b) b^4-25 b^4+12 b^2
+9\right) b^4
\right. \NNM \\ && \left.
+54 b^4+88 b^2-15\right)-\left(b^2-1\right)^2 \left(211 b^8-134 b^6-56 b^4+30
   b^2-3\right)\,, \\
\L &=& \L_a/[ 8 b(b-1)^2 \phi \F_b]\,,
\\ \NNM
\L_a &=& 4 b^2 \log(b) \left\{ 6 \log(b)
\left(12 \log(b) b^4-23 b^4+18 b^2+3\right)
 b^4+(b-1) (b+1)
\right. \\  && \left.
\times\left(95 b^6-79 b^4-7 b^2+3\right)\right\}
-\left(b^2-1\right)^2
 \left(82 b^6-77 b^4+4
   b^2+3\right)\,,\\
\M &=& 3 + \left[24 \log(b) b^8-25 b^8+48 b^6-36 b^4+16 b^2-3\right]
/[2 b^2 \F_b]\,. \EA \ESE We note that in Appendix~B in \cite{Ruy08}
a small misprint can be found in the definition of the factor $\J$
---a missing factor of three--- that is here corrected.

\section{Static drops on coated fibres: Computations and asymptotic analysis}
 \label{S-Static-Drops}
The shape of an axisymmetric drop sitting on a vertical fibre has
been computed numerically by Kumar and Hartland~\cite{Kum88} and
determined analytically by Carroll~\cite{Car76} by neglecting
gravity effects. When the contact angle of the liquid with the solid
fibre vanishes, the analytical solution corresponds to an unduloid
\cite[]{Del41,Pla73} that can be written parametrically using
elliptic integrals of the first and second kind. However, the
analytical solution is cumbersome to use and requires numerical
evaluation of the different integrals involved. Hence, we choose to
determine the solution by solving numerically the Laplace--Young
equation parametrically rewritten as \cite[]{Kum88}: \BSE
\label{Kum88-LY-eq} \BA \frac{d\phi}{d\s} &=&
\left(\frac{d\phi}{d\s}\right)_t +
 \frac{1}{R_t}
 - \frac{\sin\phi}{\r}\,,\\
\frac{d \r}{d\s} &=& \cos \phi\,,\qquad \hbox{and} \qquad \frac{d
\x}{d\s} = \sin \phi\,, \EA where the length scale is the fibre
radius, ${\bar R}$. The radial and axial coordinates are denoted by
$\r$ and $\x$, respectively. $\s$ denotes the curvilinear arc length
along the drop interface whereas $\phi$ is the angular inclination
of the drop interface to the radial axis whereas
$\left(d\phi/ds\right)_t + 1/R_t$ denotes the mean curvature at the
top of the drop, $s=0$. The set of equations is completed by the
boundary conditions: \BE
 \phi = \pi/2\,,\qquad \r=R_t\,,  \qquad \x=0\,,  \qquad \hbox{at} \qquad
 \s=0.
\EE The contact area $\A$ separating the liquid and gas phases and
the volume $\V$ of the drop has been computed by solving: \BE
\frac{d \A}{d\s} = 2\pi \r \,\qquad \hbox{and} \qquad
\frac{d\V}{d\s} = \pi(\r^2- R_t^2) \EE \ESE We have solved system
(\ref{Kum88-LY-eq}) using the {\sc Auto07p} software~\cite{auto97}.
We have adjusted $R_t$ to the coated fibre radius $1+\aN$, and the
drop volume $V$ to the volume of the corresponding solitary wave,
after subtracting the volume of the residual film.

The speed of the quasi-static drops sliding on a vertical fibre can
be estimated using the Landau-Levich-Derjaguin
theory~\cite[]{LL42,Der43}. For this purpose we divide a large
amplitude sliding drop in two separate regions. The `inner' one
corresponding to the thin films at the upper and lower ends of the
drop where viscosity balances capillary forces, and the `outer' one,
the quasi-static drop itself which is governed by the Laplace-Young
equation given above. In the inner region, $h\ll R$ and the equation
to be solved reduces to (\ref{Serafim}), whose TW solutions are
governed by: \BE \frac{h^3}{3}(1 + \beta h' + h''') - c(h-1) -1 =0
\EE Following~\cite{Kal94}, we introduce the inner coordinates
$X=c^{-1/3}\xi$ where $c\gg1$ and get to leading order \BE
\label{Bretherton} h''' = 3(h-1)/h^3 \EE which is the so-called
`Bretherton equation' \cite[]{Bret61,Kal94}, originating from
Bretherton's work on the motion of a long gas bubble in a capillary
tube. In the upper region, the solution of the Bretherton equation
with boundary conditions $h=1$, $h'=h''=0$ at $X=0$ is monotonic.
The numerical solution of~(\ref{Bretherton}) yields
$\lim_{X\to\infty} h''=1.34$. The speed of the drops is then
selected by asymptotically matching the solutions of
(\ref{Kum88-LY-eq}) and (\ref{Bretherton}) in the upper overlap
region between inner and outer domains. This can be easily done by
imposing that the Laplace pressures corresponding to the two
solutions are equal in this region giving: \BE \label{Landau}
\left(\frac{d\phi}{d\s}\right)_t +
 \frac{1}{R_t} = 1 + 1.34 \frac{\Ca^{2/3}}{\aN},
\EE where $\Ca = c_{\rm drops} \Bo = \mu {\bar c}_{\rm
drops}/\sigma$ in the limit $\aN\ll1$.

\end{appendix}

\bibliographystyle{apsrev4-1}
\bibliography{TW}

\end{document}